\documentclass[11pt]{article}

\usepackage[usenames,dvipsnames,svgnames,table]{xcolor} 
\usepackage[obeyspaces,hyphens,spaces]{url}
\usepackage{jcapmod}
\usepackage{verbatim}

\usepackage{booktabs}
\usepackage[english]{babel}
\usepackage{amsmath, amssymb, amsbsy, amstext, amsthm, simplewick}
\usepackage{hyperref}
\usepackage{graphicx}
\usepackage{amsfonts}
\usepackage{upgreek}
\usepackage{framed}
\usepackage{tensor}
\usepackage{pifont}
\usepackage{latexsym, mathrsfs}
\usepackage{array}
\usepackage{hyperref}
\usepackage{xspace}
\usepackage{longtable}
\usepackage{multirow}
\usepackage{cases}
\usepackage{empheq}

\usepackage{bm}
\usepackage{bbm}

\usepackage{afterpage}
\usepackage{setspace}

\usepackage{braket}

\usepackage[load-configurations=astronomy, range-units=brackets, range-phrase=-, per-mode=reciprocal, mode=math]{siunitx}
\usepackage{threeparttable}
\usepackage{subfig}

\usepackage{ragged2e}

\usepackage{hhline}
\usepackage{array}
\newcolumntype{P}[1]{>{\centering\arraybackslash}p{#1}}
\usepackage{booktabs}
\usepackage{tikz}
\usetikzlibrary{calc,shadings,patterns,tikzmark,fadings}

\definecolor{Blue}{rgb}{0.25, 0.41, 0.88}
\definecolor{Red}{rgb}{0.92,0.,0.}
\definecolor{darkorange}{rgb}{1.0,0.549,0.}
\definecolor{cobalt}{RGB}{44, 98, 120}
\definecolor{Mathematica1}{rgb}{0.368417, 0.506779, 0.709798}
\definecolor{Mathematica2}{rgb}{0.880722, 0.611041, 0.142051}
\definecolor{Mathematica3}{rgb}{0.560181, 0.691569, 0.194885}
\definecolor{Mathematica4}{rgb}{0.922526, 0.385626, 0.209179}
\definecolor{Mathematica5}{rgb}{0.528488, 0.470624, 0.701351}
\definecolor{Mathematica6}{rgb}{0.772079, 0.431554, 0.102387}
\definecolor{Mathematica7}{rgb}{0.363898, 0.618501, 0.782349}
\definecolor{Mathematica8}{rgb}{1, 0.75, 0}
\definecolor{Mathematica9}{rgb}{0.647624, 0.37816, 0.614037}
\definecolor{plotBlue}{RGB}{94, 130, 181}
\definecolor{plotRed}{RGB}{233, 85, 54}
\definecolor{plotGreen}{RGB}{142, 176, 50}
\definecolor{plotPurple}{RGB}{135, 120, 178}

\newcolumntype{C}[1]{>{\centering\let\newline\\\arraybackslash\hspace{0pt}}m{#1}}

\def\d{{\rm d}}

\def\r{{\bf r}}

\makeatletter
\newlength{\apb@width}
\newcommand{\autoparbox}[2][c]{\settowidth{\apb@width}{#2}\parbox[#1]{\apb@width}{#2}}

\makeatother

\makeatletter
\newsavebox\myboxA
\newsavebox\myboxB
\newlength\mylenA

\newcommand*\xoverline[2][0.75]{
    \sbox{\myboxA}{$\m@th#2$}%
    \setbox\myboxB\null
    \ht\myboxB=\ht\myboxA%
    \dp\myboxB=\dp\myboxA%
    \wd\myboxB=#1\wd\myboxA
    \sbox\myboxB{$\m@th\overline{\copy\myboxB}$}
    \setlength\mylenA{\the\wd\myboxA}
    \addtolength\mylenA{-\the\wd\myboxB}%
    \ifdim\wd\myboxB<\wd\myboxA%
       \rlap{\hskip 0.5\mylenA\usebox\myboxB}{\usebox\myboxA}%
    \else
        \hskip -0.5\mylenA\rlap{\usebox\myboxA}{\hskip 0.5\mylenA\usebox\myboxB}%
    \fi}
\makeatother

\numberwithin{equation}{section}

\def\beq{\begin{equation}}
\def\eeq{\end{equation}}

\def\bea{\begin{eqnarray}}
\def\eea{\end{eqnarray}}

\def\d{{\rm d}}

\def\beq{\begin{equation}}
\def\eeq{\end{equation}}
\def\bea{\begin{eqnarray}}
\def\eea{\end{eqnarray}}

\def\d{{\rm d}}

\def\d{{\rm d}}

\newcommand{\ud}{\mathrm{d}}
\newcommand{\lab}[1]{{\mathrm{#1}}}
\newcommand{\mb}[1]{{\mathbf{#1}}}
\newcommand{\minus}{{\scalebox {0.75}[1.0]{$-$}}}
\newcommand{\sminus}{{\scalebox {0.5}[0.85]{$-$}}}
\newcommand{\indlab}[1]{{\scriptscriptstyle{(#1)}}}
\newcommand{\indlac}[1]{{#1}}

\def\r{{\bf{r}}}

\DeclareRobustCommand{\SkipTocEntry}[4]{}

\newcommand{\es}{\hspace{0.5pt}}

\usepackage{colortbl}
\definecolor{blue2}{cmyk}{1, 0.1, 0.1, 0.1}

\definecolor{pyBlue}{RGB}{31, 119, 180}
\definecolor{pyRed}{RGB}{214, 39, 40}
\definecolor{pyGreen}{RGB}{44, 160, 44}
\definecolor{pyBlue2}{RGB}{0, 111, 237}
\definecolor{pyRed2}{RGB}{224, 52, 36}

\newcommand{\red}[1]{\textcolor{pyRed}{#1}}
\newcommand{\blue}[1]{\textcolor{pyBlue}{#1}}

\newcommand{\green}[1]{\textcolor{pyGreen}{#1}}

\DeclareRobustCommand\mySingularity{\tikz[baseline=-3pt]{ \draw[black] (0ex,0ex) circle (0.8ex); \fill[black] (0, 0) circle (0.4ex);}}

\begin{document}

\pagenumbering{roman}
\begin{titlepage}
\baselineskip=15.5pt \thispagestyle{empty}

\bigskip\

\vspace{1cm}
\begin{center}
{\fontsize{22}{24}\selectfont  \bfseries The Spectra of Gravitational Atoms}
\end{center}
\vspace{0.1cm}
\begin{center}
{\fontsize{12}{18}\selectfont Daniel Baumann, Horng Sheng Chia, John Stout and Lotte ter Haar} 
\end{center}

\begin{center}
\vskip8pt
\textit{Institute for Theoretical Physics, University of Amsterdam,\\Science Park 904, Amsterdam, 1098 XH, The Netherlands}


\end{center}

\vspace{1.2cm}
\hrule \vspace{0.3cm}
\noindent {\bf Abstract}\\[0.1cm]
We compute the quasi-bound state spectra of ultralight scalar and vector fields around rotating black holes.
These spectra are determined by the gravitational fine structure constant $\alpha$, 
which is the ratio of the size of the black hole to the Compton wavelength of the field. 
When $\alpha$ is small, the energy eigenvalues and instability rates can be computed analytically. Since the solutions vary rapidly near the black hole horizon, ordinary perturbative approximations fail and we must use matched asymptotic expansions to determine the spectra. 
Our analytical treatment relies on the separability of the equations of motion, and is therefore only applicable to the
scalar field and the electric modes of the vector field. However, for slowly-rotating black holes, the equations for the magnetic modes can be written in a separable form, which we exploit to derive their energy eigenvalues and conjecture an analytic form for their instability rates.
To check our conjecture, and to extend all results to large values of $\alpha$, we 
solve for the spectra numerically. 
We explain how to accurately and efficiently compute these spectra, without relying on separability.
This allows us to obtain reliable results for any $\alpha \gtrsim 0.001$ and black holes of arbitrary spin. 
 Our results provide an essential input to the phenomenology of boson clouds around black holes, especially when these are part of binary systems. 
 
\vskip10pt
\hrule
\vskip10pt

\end{titlepage}

\thispagestyle{empty}
\setcounter{page}{2}
\begin{spacing}{1.03}
\tableofcontents
\end{spacing}

\clearpage
\pagenumbering{arabic}
\setcounter{page}{1}

\clearpage

\section{Introduction}
\label{sec:introduction}

Light particles with very weak couplings to ordinary matter are hard to detect by traditional experimental means. For example, neutrinos---the lightest massive particles of the Standard Model and one of the most abundant particles in the universe---have only been
 detected directly through their rare interactions in enormous water tanks~\cite{Cowan:1992xc, Fukuda:1998mi}. 
New (ultra)light particles with couplings much weaker than those of neutrinos can therefore easily have escaped all of our current detection efforts.  However, if these hypothetical particles are much lighter than neutrinos, and if they are bosons, then they can be produced around rotating black holes through a process called superradiance~\cite{Zeldovich:1971a,Zeldovich:1972spj}.  This effect can extract enough mass and angular momentum from the black hole to form a large condensate of the bosonic field.
The gravitational influence of these superradiantly-generated `boson clouds' is potentially observable, making this an interesting laboratory of physics beyond the Standard Model~\cite{Arvanitaki:2009fg, Arvanitaki:2010sy}.

\begin{figure}[t]
    \centering
    \begin{minipage}{0.45\textwidth}
        \centering
        \includegraphics[scale=1.3]{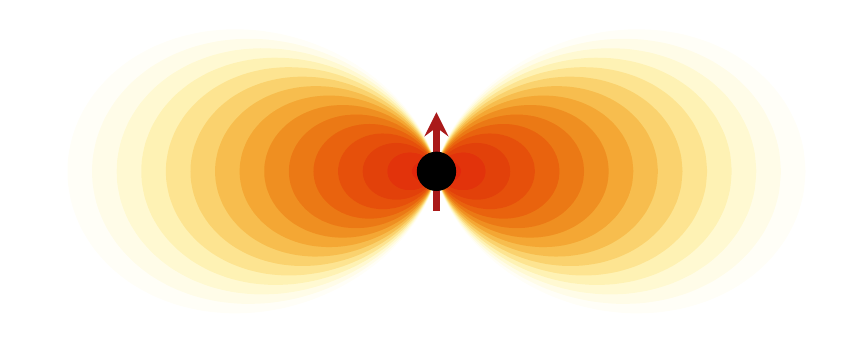}  
        \end{minipage}\hfill
    \begin{minipage}{0.45\textwidth}
        \centering
        \includegraphics[scale=1.3]{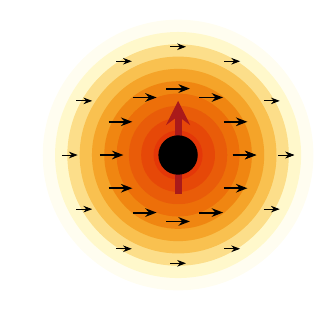} 
    \end{minipage}
    \vspace{-0.25cm}
    \caption{Illustration of the dominant growing modes of the scalar ($2p$) and vector ($1s$) gravitational atoms. The small arrows denote the intrinsic spin of the vector field, which allows for superradiant growth even for states without orbital angular momentum.}
    \label{fig:atom}
\end{figure}

\vskip 4.4pt 
The black hole carrying the boson cloud is sometimes called a {\it gravitational atom} since
 the system closely resembles the proton-electron structure in a hydrogen atom (see Fig.\,\ref{fig:atom}). 
While superradiance with scalar fields has been studied extensively in the past~\cite{Detweiler:1980uk, Dolan:2007mj, Arvanitaki:2010sy, Yoshino:2012kn, Dolan:2012yt, Okawa:2014nda, Yoshino:2014, Baumann:2018vus, Brito:2014wla, Arvanitaki:2014wva, Arvanitaki:2016qwi, Brito:2017wnc, Brito:2017zvb}, the exploration of vector fields is much more recent~\cite{Witek:2012tr, Pani:2012vp, Pani:2012bp, Endlich:2016jgc, Baryakhtar:2017ngi, East:2017ovw, East:2017mrj, East:2018glu, Cardoso:2018tly, Dolan:2018dqv}.  In this paper, we study the spectra of these gravitational atoms for both scalar and vector clouds.  
Our principal goal is to accurately compute the energy splittings between the eigenstates of the cloud---the analog of the fine and hyperfine structure of the hydrogen atom---for both types of fields. The differences in the spectra are potentially observable in future gravitational wave experiments. In particular, if the gravitational atom is part of a binary system \cite{Baumann:2018vus} (see also~\cite{Paper2, Hannuksela:2018izj, Zhang:2018kib, Berti:2019wnn, Zhang:2019eid}), the gravitational perturbation due to the companion can induce transitions among the different states of the cloud. These transitions can then affect the dynamics of the inspiral, leaving an imprint on observed gravitational waveforms.  Detailed knowledge of the spectra will be required to interpret such experimental results and infer the microscopic nature of the cloud.

\vskip 4.4pt 
Both the efficiency of black hole superradiance and the spectrum of the cloud depend on the ratio of the gravitational radius of the black hole, $r_g \equiv GM/c^2$, to the (reduced) Compton wavelength of the field, $\lambda_c \equiv \hbar/(\mu c)$,  where $\mu$ is the mass of the field. This defines the gravitational fine structure constant
\beq
\alpha \equiv \frac{r_g}{\lambda_c} = \frac{G M \mu}{\hbar c}  \, , \label{eqn:alphaDef}
\eeq
which becomes $\alpha = M \mu$ in natural units. 
A bosonic condensate only forms around the black hole if $\alpha$ is smaller than of order unity. In the limit $\alpha \ll 1$, the superradiance phenomenon can be treated perturbatively as an expansion in powers of $\alpha$.  Using such an expansion, the energy spectrum~\cite{Baumann:2018vus} and instability rates~\cite{Detweiler:1980uk} were computed for a massive scalar field, up to the order in $\alpha$ at which the degeneracies between all modes in the spectrum are broken. 
The computation for a massive vector field, on the other hand, is much more involved~\cite{Endlich:2016jgc, Baryakhtar:2017ngi, Pani:2012vp, Pani:2012bp}, and has not yet been treated with complete rigor.  Until recently,
the main technical obstruction in the vector analysis has been the fact that the Proca equation could not be decomposed into separate
 angular and radial equations. However, in~\cite{Krtous:2018bvk, Frolov:2018ezx}, a separable ansatz was discovered for the electric modes  of a massive vector field on the Kerr background.\footnote{In~\cite{Dolan:2018dqv}, it was shown that a special class of magnetic modes are contained within this ansatz (see Appendix~\ref{app:details}). However, it is not yet clear whether it contains every magnetic mode, nor is it clear how to recover them.}  In this paper, we use this ansatz to compute the spectrum and instability rates for a Proca field perturbatively. A special feature of superradiantly-generated vector clouds is that the dominant growing mode is a $1s$ state, which has vanishing orbital angular momentum. In this case, the spin of the black hole is converted into the intrinsic spin of the field, and the cloud has non-negligible support near the black hole, see Fig.~\ref{fig:atom}. These modes are then especially sensitive to the near-horizon geometry of the black hole, and this sensitivity necessitates going beyond ordinary perturbation theory to  determine the spectrum. 

\vskip 4pt
While the magnetic modes of the Proca field are separable on the Schwarzschild background~\cite{Rosa:2011my}, a separable ansatz in the Kerr spacetime remains elusive,
 and thus rigorous analytic results are difficult to achieve for this part of the spectrum. Fortunately, all degeneracies between the states in the spectrum are broken at linear order in the spin of the black hole, and the Schwarzschild ansatz still separates the Proca equation at this order~\cite{Pani:2012bp,Pani:2012vp}. Assuming that the fine and hyperfine structure of the magnetic modes are similar to their electric counterparts, we can use this small-spin expansion to derive the magnetic spectrum  
perturbatively in $\alpha$. 
Furthermore,  we `derive' their leading-order instability rates using educated guesswork, and are thus able to attain perturbative 
results for the most phenomenologically relevant aspects of the spectrum, for all modes of the Proca field.

\vskip 4pt
To test our analytic results, and to move beyond the limit of small $\alpha$,
we compute the spectra of scalar and vector clouds numerically (see also e.g.~\cite{Dolan:2007mj,Pani:2012bp, Cardoso:2018tly, Dolan:2018dqv}). Formulating this as a nonlinear eigenvalue problem allows us to attain highly accurate results for the spectrum with little computational cost. Our method also has the advantage that it does not rely upon a separable ansatz, which allows us to rigorously determine the magnetic spectrum and check our analytic guesswork for its instability rates. 
Our method is reliable even for very small values of $\alpha$, where precise numeric results are typically difficult to achieve without separability~\cite{Pani:2012bp, Cardoso:2018tly}, and we apply it to both the dominant growing modes and other phenomenologically relevant states. We pay special attention to the numerical obstacles that can potentially destroy the accuracy of a solution, and describe how to avoid them. 
This new formulation thus provides a robust and flexible method for finding the quasi-bound state spectra for arbitrary ultralight tensor fields about any stationary black hole, without relying on separability of the equations of motion.

\paragraph{Outline} The plan of the paper is as follows: in Section~\ref{sec:preliminaries}, we set up the basic equations describing massive scalar and vector fields around rotating black holes. We
 outline the computations performed in the rest of the paper and present a summary of key results. 
 In Section~\ref{sec:analytic}, we compute the spectrum of energy eigenvalues and the associated instability rates, for both the scalar field and the electric modes of the vector field, using matched asymptotic expansions. 
 We also compute the energy spectra of the magnetic modes at linear order in the black hole spin $\tilde a$ and motivate their conjectured instability rates. 
 In Section~\ref{sec:numeric}, we describe our numerical methods. Our techniques, in principle, work for ultralight fields of arbitrary spin and do not assume separability of the equations of motion.
In Section~\ref{sec:outlook}, we summarize our results and present an outlook on future applications. The appendices contain additional technical details: In Appendix~\ref{app:harmonics}, we review essential facts on tensor spherical harmonics.   In Appendix~\ref{app:details}, we present material supporting the analytical treatment in Section~\ref{sec:analytic}, while Appendix~\ref{app:numericalDetails} spells out many of the details left out in the description of the numerics in Section~\ref{sec:numeric}. Finally, Appendix~\ref{app:notation} collects the most important variables used in this paper.

\paragraph{Notation and conventions} We will use a metric with `mostly plus' signature $(-,+, +, +)$ and, unless stated otherwise, work in natural units with $G=\hbar = c = 1$. 
 Except when presenting explicit results, we will set $\mu \equiv 1$, so that times and distances are measured relative to the typical oscillation timescale and Compton wavelength of the fields. Greek letters will denote spacetime indices ($\mu,\nu, \ldots$), while Latin letters will either stand for spatial indices ($i, j, \ldots$), label indices ($i, k, \ldots$),
or for vierbein indices ($a,b,\ldots$). To avoid confusion, we will sometimes wrap label indices in parentheses. We will adopt Boyer-Lindquist coordinates for the Kerr black hole with mass $M$ and specific angular momentum $a$. The line element is then
\beq
\d s^2 = - \frac{\Delta}{\Sigma}\left(\d t - a \sin^2 \theta\, \d \phi \right)^2 + \frac{\Sigma}{\Delta} \d r^2 + \Sigma\hskip 1pt \d \theta^2 + \frac{\sin^2 \theta}{\Sigma}  \left(a \hskip 1pt \d t - (r^2 + a^2) \, \d \phi \right)^2\, ,  \label{equ:Kerr}
\eeq
where $\Delta \equiv r^2 -  2 Mr +a^2 $ and $\Sigma \equiv r^2 + a^2 \cos^2 \theta$. The roots of $\Delta$ determine the inner and outer horizons, located at $r_\pm = M \pm \sqrt{M^2 -a^2}$, and the angular velocity of the black hole at the outer horizon is $\Omega_H \equiv a/ 2 M r_+$. Dimensionless quantities, defined with respect to the black hole mass~$M$, are denoted by tildes; e.g.~$\tilde{a} \equiv a/M$ and $\tilde{r}_{\pm} \equiv r_\pm/M$. 

\vskip 4pt
The scalar and vector eigenstates are denoted by $|\es n \es \ell \es m\rangle$ and $| \es n \es \ell \es j \es m\rangle$, with the integers $\{n, \ell, j, m \}$ labeling the principal, orbital angular momentum, total angular momentum, and azimuthal angular momentum numbers, respectively. Following the convention in atomic physics, we have~$n \geq \ell + 1$. We refer to vector modes that acquire a factor of $(-1)^j$ under a parity transformation as `electric modes,' and those that acquire a factor of $(-1)^{j+1}$ as `magnetic modes.' These modes are to be distinguished from odd and even modes, which, by our convention, receive a factor of $-1$ and $+1$ under parity, respectively.

\newpage
\section{Scalars and Vectors around Kerr}
\label{sec:preliminaries}

Massive scalar and vector fields in a general spacetime with metric $g_{\alpha \beta}$ satisfy similar equations of motion,
\begin{align}
\left(g^{\alpha \beta} \nabla_\alpha \nabla_\beta - \mu^2 \right) \Phi \ \, &= 0 \, , \label{equ:KG} \\
\left(g^{\alpha \beta} \nabla_\alpha \nabla_\beta - \mu^2 \right) A_\mu &= 0 \, .  \label{equ:Proca}
\end{align}
These are called the Klein-Gordon and Proca equations, respectively, and the latter must be supplemented by the Lorenz constraint $\nabla^\mu A_\mu = 0$. The goal of this paper is to determine the quasi-bound state solutions of these equations in the Kerr background. 
This section serves as an overview for the rest of the paper and we present a summary of both our analytic and numeric
results in \S\ref{sec:Summary}, relegating their detailed derivations to Sections \ref{sec:analytic} and \ref{sec:numeric}.

\subsection{Tensor Representations} 
\label{sec:TensorKerr}

We begin with a discussion of general tensor fields $T_{\mu \nu \dots \rho}$ on the Kerr background.
We are interested in quasi-bound states solutions, which are `purely ingoing' at the outer horizon and vanish at infinity.
As in atomic physics, these solutions will be characterized by a set of discrete `quantum numbers' describing the energy, the orbital angular momentum and the intrinsic spin of the states.
A precise definition of these quantum numbers is complicated by the fact that the spin of black hole breaks spherical symmetry. We will label our states by the quantum numbers they attain in the flat-space limit, where the spin of the black holes  can be ignored.

\vskip 4pt
In order to define the flat-space limit,
it is convenient to rescale the temporal and radial coordinates, $t \mapsto t \mu^{-1}$ and $r \mapsto r \mu^{-1}$, so that the metric (\ref{equ:Kerr}) becomes
\beq
\mu^2 \d s^2 = - \frac{\Delta}{\Sigma}\left(\d t - \alpha \hskip 1pt \tilde{a} \sin^2 \theta\, \d \phi \right)^2 + \frac{\Sigma}{\Delta} \d r^2 + \Sigma \hskip 1pt \d \theta^2 + \frac{\sin^2 \theta}{\Sigma}  \left(\alpha \hskip 1pt \tilde{a} \hskip 1pt \d t - (r^2 + \alpha^2 \tilde{a}^2) \, \d \phi \right)^2\, , \label{equ:Kerr2}
\eeq
where $\tilde{a} \equiv a / M$ and 
\beq
\begin{aligned}
\Delta &\equiv r^2 -  2 \alpha r +\alpha^2 \tilde{a}^2\, ,  \\ 
\Sigma &\equiv r^2 + \alpha^2 \tilde{a}^2 \cos^2 \theta\, .
\end{aligned}
\eeq
All physical quantities are now measured in units of the Compton wavelength of the field, $\mu^{-1}$, which we henceforth set to $\mu^{-1} \equiv 1$. 
Notice that the spin parameter $\tilde{a}$ in (\ref{equ:Kerr2}) always appears in the combination~$\alpha \hskip 1pt \tilde{a}$, so spherical symmetry is approximately restored when $\alpha \ll 1$. 
In fact, in the limit $\alpha \to 0$, the line element (\ref{equ:Kerr2}) reduces to that of Minkowski spacetime, while at linear order in $\alpha$ it becomes the Schwarzschild solution. Spin-dependent terms, such as those corresponding to the Lense-Thirring effect, only appear at quadratic order in~$\alpha$. 
As far as the dynamics of the field is concerned, we therefore expect the effects of spin to be subleading compared to the gravitational potential sourced by $M$. 

\vskip 4pt
The eigenstates of the field are labeled by a discrete set of quantum numbers that reflect the (approximate) isometries of the background metric. To identify these quantum numbers, we first note that the Kerr geometry has two Killing vectors  
		\begin{equation}
			k_t \equiv - i \frac{\partial}{\partial t} \, , \qquad k_{z} \equiv - i \frac{\partial}{\partial \phi} \, , \label{eq:kerrIsometries}
		\end{equation}
representing the fact that the metric (\ref{equ:Kerr2})  is both stationary and axisymmetric. 
The equations of motion for an arbitrary tensor field $T_{\mu \nu \cdots \rho}$ on the Kerr background will inherit the isometries~(\ref{eq:kerrIsometries}). If these equations are linear, we may decompose their solutions in terms of states with definite frequency and azimuthal angular momentum,\footnote{Our analysis will focus entirely on complex scalar and vector fields, as the complex representations of these isometries are much simpler than for real fields. Because the equations of motion are linear, a real solution can be generated by simply taking the real part of a complex solution, and so the spectra are identical.}
\begin{equation}
    \begin{aligned}
        \pounds_{t} \hskip 2pt T_{\mu \nu \cdots \rho} &= -\omega  \hskip 2ptT_{\mu \nu \cdots \rho}\, , \\  \pounds_{z}  \hskip 2ptT_{\mu \nu \cdots \rho} &= +m  \hskip 2pt T_{\mu \nu \cdots \rho}\,, 
        \end{aligned} \label{eqn:DefiniteEigen}
    \end{equation}
where $\pounds_t \equiv \pounds_{k_t}$ and $\pounds_z \equiv \pounds_{k_z}$ are the Lie derivatives with respect to the isometries (\ref{eq:kerrIsometries}). In the Schwarzschild limit $\tilde{a} \to 0$, the geometry gains two more Killing vector fields $k_x$ and $k_y$ which, together with $k_z$, form an $\lab{SO}(3)$ algebra. 
We can then expand the field into its temporal and  spatial components, which we further expand in representations of this algebra, i.e. eigentensors of the \textit{total angular momentum} operator 
      \begin{equation}
      \pounds^2  \hskip 2pt T_{i k \cdots l} \equiv \left(\pounds_x^2 + \pounds_y^2 + \pounds_z^2 \right) T_{i k \cdots l} = j(j+1)  \hskip 2pt T_{i k \cdots l}\, , \label{eq:quadCasimir}
    \end{equation}
    where $j$ denotes the total angular momentum. Furthermore, in the flat-space limit, $\alpha \to 0$, solutions also have definite orbital angular momentum $\ell$ (cf.~Appendix \ref{app:harmonics}), so that they can be characterized by the quantum numbers $\omega$, $\ell$, $j$ and $m$. We will label such states by $|\es n\es \ell \es j \es m \rangle$, where we have introduced an integer quantum number $n$ that indexes the discrete quasi-bound state frequencies $\omega_n$. Since total and orbital angular momenta are the same for the scalar field, we will label its states by $|\es n \es \ell \es m \rangle$. States still have definite total angular momentum in the Schwarzschild limit, but indefinite orbital angular momentum. At finite $\tilde{a}$, these states no longer have definite total angular momentum.
    Nevertheless, we will still label our states by $|\es n \es \ell \hspace{0.5 pt} j \es m \rangle$
     with the understanding that these quantum numbers regain their physical meaning
    as $\alpha \to 0$.

\vskip 4pt
  Finally, the Kerr metric is invariant under the parity transformation 
  \beq
  {\cal P}:\,\,\, (\theta, \phi) \mapsto (\pi - \theta, \phi + \pi)\, . \label{equ:parity}
  \eeq
  This helps us to further organize the spectrum into states with definite parity. Solutions to the Klein-Gordon and Proca equations with definite parity are invariant under (\ref{equ:parity}) up to a sign, and, with our conventions, we have $\mathcal{P} |\es n \es \ell \es m \rangle = (-1)^{\ell} | \es n \es \ell \es m \rangle$ and $\mathcal{P} |\es n \es \ell \es j \es m \rangle = (-1)^{\ell +1} | \es n \es \ell \es j \es m \rangle$ for scalars and vectors, respectively. See Appendix~\ref{app:harmonics} for a more detailed discussion.

\newpage

\subsection{Massive Scalar Fields} \label{sec:scalarintro}

We now specialize to the case of a massive scalar field around a Kerr black hole. Since the Klein-Gordon equation is relatively simple, this will serve as a useful illustration of our approach without the technical distractions that arise in the vector analysis. 

\begin{figure}[t]
\centering
\includegraphics[scale=1, trim = 0 0 0 0]{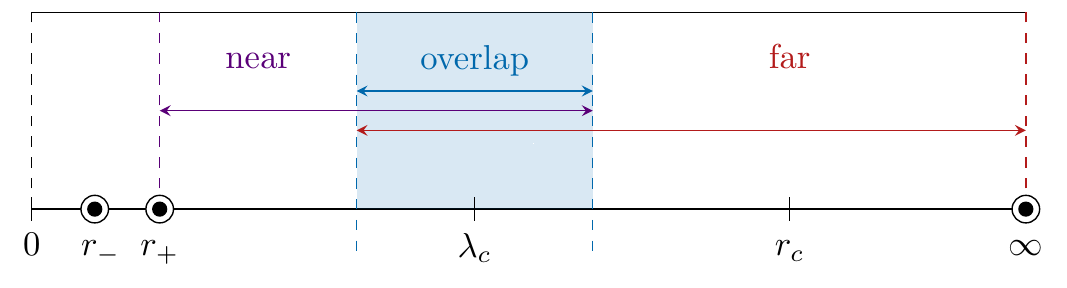}
\caption{ Illustration of the near and far regions used in our perturbative treatment for
the massive scalar field, where $r_\pm$ are the inner and outer horizons of the black hole, $\lambda_c = \mu^{-1}$ is the Compton wavelength of the field, and $r_c = \left(\mu \alpha\right)^{-1}$ is the typical Bohr radius of the quasi-bound state. The two asymptotic solutions are matched in the overlap region.
}
\label{fig:ScalarZones}
\end{figure}

\vskip 4pt
An important feature of the Klein-Gordon equation (\ref{equ:KG}) is that it is separable in Boyer-Lindquist coordinates through the ansatz~\cite{Brill:1972xj, Carter:1968}  
\begin{align}
\Phi \left(t, \textbf{r} \right) = e^{- i \omega t + i m \phi} R(r) S(\theta)  \, . \label{eqn:ScalarAnsatz}
\end{align}
The spheroidal harmonics $S(\theta)$ obey the differential equation\footnote{To avoid clutter, we suppress the dependence on the angular quantum numbers, i.e.~$S\equiv S_{\ell m}$ and $\Lambda \equiv \Lambda_{\ell m}$.} 
\begin{equation}
	\left(-\frac{1}{\sin \theta} \frac{\ud}{\ud \theta}\left(\sin \theta \frac{\ud}{\ud \theta}\right) - c^2 \cos^2 \theta + \frac{m^2}{\sin^2 \theta}\right)S = \Lambda S \, , \label{eqn: spheroidal harmonic equation}
\end{equation}
where $\Lambda$ is the angular eigenvalue and the spheroidicity parameter is 
\beq
c^2 \equiv - \alpha^2 \tilde{a}^2( 1-\omega^2 ) \,.  \label{equ:cc}
\eeq
The radial equation is a {confluent Heun equation},\footnote{The Heun equation is a generalization of the hypergeometric equation with four, instead of three, regular singular points. By merging  
two of these singularities, we arrive at the confluent Heun equation, which is similarly a generalization of the confluent hypergeometric equation (\ref{eqn:scalarRadialLO}) that determines the wavefunctions of the hydrogen atom. The angular equation (\ref{eqn: spheroidal harmonic equation}) is also a confluent Heun equation. For a detailed treatment, see \cite{Andre:1995heun}.} 
\beq
\begin{aligned}
0 \,=\,  \frac{1}{R\, \Delta} \frac{\ud}{\ud r} \!\left(\!\Delta \frac{\ud R}{\ud r}\right)  &\ - \frac{\Lambda}{\Delta} - \left(1 - \omega^2\right) + \ \frac{P_+^2}{(r-r_+)^2} + \frac{P_-^2}{(r - r_-)^2} \\
&\ -\frac{A_+}{(r_+ - r_-) (r -r_+)} +\frac{A_-}{(r_+-r_-)(r-r_-)} \, , \label{eqn: scalar radial equation}
\end{aligned}
\eeq
where  
we have defined the following coefficients 
\beq
\begin{aligned}
A_{\pm}  &\equiv P_+^2 + P_-^2 + \gamma^2 + \gamma_\pm^2   \quad {\rm and} \quad 
P_{\pm}  \equiv \frac{\alpha (\tilde{a} m - 2 \hskip 1pt  r_{\pm} \omega)}{r_+ - r_-} \, , 
\end{aligned}
  \label{equ:AP}
\eeq
with
\beq
\begin{aligned}
\gamma^2 & \equiv \frac{1}{4} (r_+ - r_-)^2 ( 1 - \omega^2) \, , \\
\gamma_\pm^2 & \equiv  \left[ \alpha^2( 1 - 7 \omega^2 )  \pm  \alpha (r_+ - r_-)(1 - 2 \omega^2)  \right]  . \label{eqn:defgamma}
\end{aligned}
\eeq
We are interested in quasi-bound states, i.e.~solutions to (\ref{eqn: scalar radial equation}) 
that are purely ingoing at the horizon and vanish at infinity:
\begin{equation}
		R(r) \propto \begin{cases}
			(r - r_+)^{i P_+} & \text{as } r \to r_+ \\
			\exp(-\sqrt{1-\omega^2}\hskip 1pt r) & \text{as } r \to \infty
		\end{cases}\, . \label{eqn:bc}
	\end{equation}
As is typical for eigenvalue problems, these two boundary conditions can only be simultaneously satisfied
for specific values of~$\omega$. Unfortunately, while solutions to both (\ref{eqn: spheroidal harmonic equation}) and (\ref{eqn: scalar radial equation}) are well known, closed-form expressions for their eigenvalues are not. However, in the limit $\alpha \ll 1$, we can construct a perturbative expansion of the spectrum. This perturbative approach  is complicated by the presence of terms in (\ref{eqn: scalar radial equation}) that diverge as $r \to r_+$. 
These terms represent \emph{singular perturbations} in $\alpha$, i.e.~if we naively expand them in powers of $\alpha$, an infinite number of terms become relevant as $r \to r_+$. Luckily, this is not a disaster. The role of these singular terms is simply to modify the characteristic scale on which $R(r)$ varies.  
As we will see explicitly in \S\ref{sec:MatchedScalar}, the radial function satisfies 
	\begin{equation}
		\frac{1}{R}\frac{\ud R}{\ud r} \sim \begin{cases}
			\alpha^{-1} & \text{as } r \to r_+ \\
			\alpha & \text{as } r \to \infty
		\end{cases}\, .
	\end{equation}
	This motivates splitting the interval $[r_+, \infty)$ into a `near region' and a `far region' (see Fig.\,\ref{fig:ScalarZones}). 
	In the near region, the $(r-r_+)^{-2}$ pole in (\ref{eqn: scalar radial equation}) dominates and an approximate solution to the differential equation for $R(r)$ can be obtained by dropping the subleading terms.  In the far region, on the other hand, the non-derivative terms are dominated by the constant $(1 - \omega^2)$ term. 
By dropping the sub-leading contributions to (\ref{eqn: scalar radial equation}) in each region, we can then construct approximate analytic solutions in the near and far regions, order-by-order in $\alpha$. These solutions are then matched in the `overlap region,' allowing a determination of the
frequency eigenvalue~$\omega$, as an expansion in powers of $\alpha$. The details of this \emph{matched asymptotic expansion} will be presented in Section~\ref{sec:analytic}. For the angular problem, the perturbative solutions are valid over the entire angular domain and no matching is necessary.

\begin{figure}[t]
\centering
\includegraphics[scale=1, trim = 0 0 0 0]{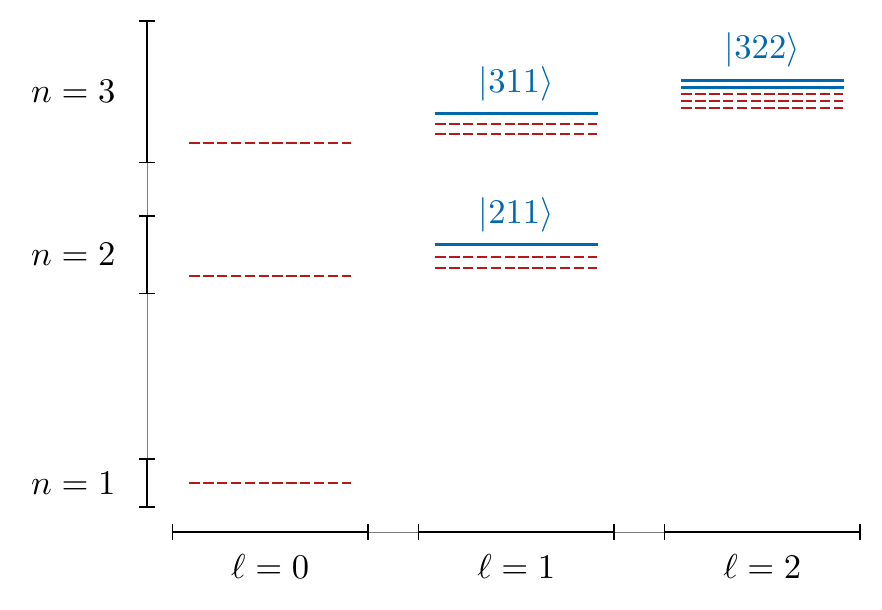}
\caption{Schematic illustration of the energy spectrum for a massive scalar field. Each state is labeled by the  quantum numbers $|\es n \es \ell \es m\rangle$. The solid blue lines are growing modes, while the dashed red lines are decaying modes. } 
\label{fig:ScalarSpectra}
\end{figure}

	\vskip 4pt
To gain an intuitive understanding of the behavior of the solutions at leading order, it is instructive to first consider the equation of motion obeyed by the far-zone radial function, $R^{\rm far}_0$. At leading order, this reads
\beq
\left[ -\frac{1}{2r^2}\frac{\d}{\d r} \left( r^2 \frac{\d }{\d r}\right) - \frac{\alpha}{r} + \frac{\ell(\ell+1)}{2r^2} +\frac{1-\omega^2}{2} \right] R^{\text{far}}_0 = 0 \, . \label{eqn:scalarRadialLO}
\eeq
This is analogous to the time-independent Schr\"odinger equation for the hydrogen atom, where the radial-gradient term arises from the kinetic energy, the $1/r$ term corresponds to a Coulomb-like central potential with gravitational coupling constant $\alpha$, the $1/r^2$ term is its centrifugal barrier with orbital angular momentum number $\ell$, and the constant term captures the exponential behavior of the quasi-bound states at $r \to \infty$. We therefore expect the energy spectrum of all modes to be Bohr-like at leading order, with the radial function
\beq
R^{\text{far}}_0(r) \propto e^{-\sqrt{1-\omega^2} \, r} r^\ell L^{(2\ell+1)}_{n-\ell-1}\big( 2\sqrt{1-\omega^2} \, r\big) \, , \label{eqn: scalar radial far LO}
\eeq
where $L_k^{(\rho)}$ is the associated Laguerre polynomial, and $n$ is the principal quantum number, which satisfies $n \geq \ell+1$. The exponential behavior in (\ref{eqn: scalar radial far LO}) implies that the solution has a typical Bohr radius
\beq
r_c \equiv ( \mu \alpha)^{-1}\,. \label{equ:BohrRadius}
\eeq 
Since (\ref{eqn:scalarRadialLO}) only captures the physics in the far region, it does not describe  
physical effects that depend on the boundary condition at the event horizon, such as the instability rates of the energy eigenstates. To study these effects, we match the near- and far-zone solutions. By pushing this procedure to higher orders in $\alpha$, we determine the complete spectrum of the scalar field. A schematic illustration of this spectrum appears in Fig.~\ref{fig:ScalarSpectra}.

\subsection{Massive Vector Fields} 
\label{sec:vectorintro}

The task of solving for the quasi-bound state solutions of vector fields is more involved because the matching involves three different regions, and the equation of motion is not obviously separable. The latter problem was addressed in~\cite{Frolov:2018ezx} and \cite{Pani:2012bp} where separable ans\"atze were found for the electric modes and magnetic modes (in the limit of small black hole spin), respectively. In this paper, we use these results to derive the spectrum of vector quasi-bound states perturbatively.

\vskip 4pt
Consider the following ansatz for a vector field on the Kerr background~\cite{Krtous:2018bvk, Frolov:2018ezx}
\begin{align}
A^\mu = B^{\mu \nu}\nabla_\nu Z \, , \quad {\rm with} \quad Z(t,\r) = e^{-i\omega t + i m \phi} R(r) S(\theta) \, , \label{eqn: Proca ansatz}
\end{align}
where $R$ and $S$ are the radial and angular functions.\footnote{We emphasize that $R$ and $S$ are not the actual radial and angular profiles
of the vector field, since the tensor~$B^{\mu \nu}$ depends on both $r$ and $\theta$. However, we still retain the terminology `radial' and `angular function,' and denote them by $R$ and $S$, to draw a direct parallel with the scalar. 
 The precise relationships between $R$, $S$ and the profile of the vector field $A_\mu$
 in the far zone will be given in \S\ref{sec:ProcaLO}. \label{footnote:ActualRS}}  The polarization tensor $B^{\mu \nu}$ is defined by
\begin{align}
B^{\mu \nu} \left( g_{\nu \sigma} + i \lambda^{-1} h_{\nu \sigma}\right) = \delta^\mu_\sigma \, , \label{eqn:Bdef}
\end{align}
where $\lambda$ is generally a complex parameter, which we will refer to as the angular eigenvalue, and $h_{\mu \nu}$ is the principal tensor of the Kerr spacetime~\cite{Frolov:2017kze}.\hskip 1pt\footnote{To simplify many of the following equations, we take the angular eigenvalue $\lambda$ to be the inverse of that in~\cite{Krtous:2018bvk,Frolov:2018ezx}.}
An explicit expression for $B^{\mu \nu}$ is given in Appendix~\ref{app:details}.  The three independent degrees of freedom of the vector field can be organized in terms of
$j = \ell \pm 1, \ell$. 
Since the $j = \ell \pm 1$ modes acquire a  factor of $(-1)^j$ under a parity transformation, they are called \emph{electric modes}, while the $j = \ell $ modes are the {\it magnetic modes}~\cite{Thorne:1980ru}. In~\S\ref{sec: analytics vector}, we will show that the ansatz (\ref{eqn: Proca ansatz}) captures all of the electric modes of the vector field.\footnote{In~\cite{Dolan:2018dqv}, it was found that the ansatz (\ref{eqn: Proca ansatz}) restores at least a subset of the magnetic modes in a special limit (see Appendix~\ref{app:details} for further details). It remains unclear, however, how to write all magnetic modes in separable form. In this paper, we will instead utilize a different ansatz, which, in the limit of small black hole spin, provides a separable equation for all of the magnetic modes.}

\begin{figure}[]
\centering
\includegraphics[scale=1, trim = 0 0 0 0]{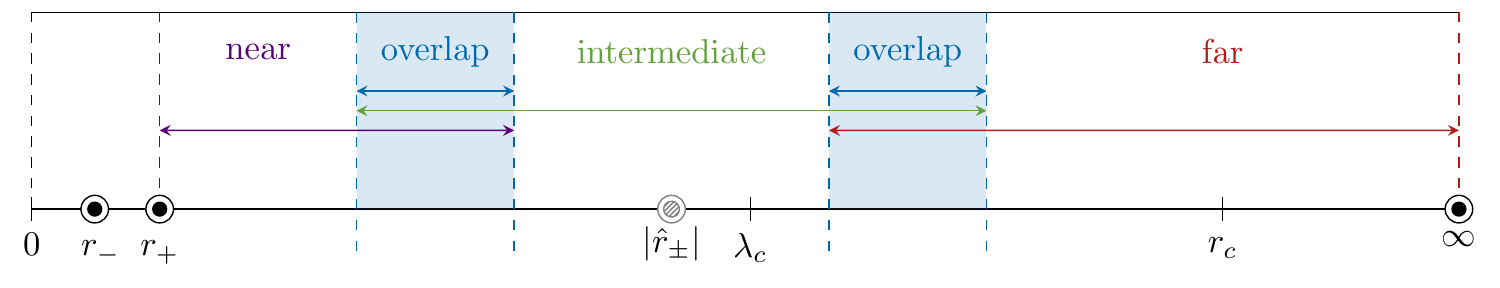}
\caption{
Illustration of the different regions used in our perturbative
 treatment for the massive vector field. The additional poles at $\hat{r}_\pm$ along the imaginary axis reduce the radii of convergence of the near- and far-zone solutions, so that there is no region where they overlap. Matching therefore requires  the intermediate region.} 
\label{fig:Vectorzones}
\end{figure}

\vskip 4pt
Substituting (\ref{eqn: Proca ansatz}) into the Proca equation, we obtain 
differential equations for $S(\theta)$ and $R(r)$. The angular equation reads~\cite{Dolan:2018dqv} 
\beq
\frac{1}{\sin \theta} \frac{\d}{\d\theta} \! \left( \sin\theta \frac{\d S}{\d \theta} \right)  + \left( c^2 \cos^2 \theta - \frac{m^2}{\sin^2 \theta} +  \Lambda  \right) S  = \frac{2 \alpha^2 \tilde{a}^2 \cos \theta}{\lambda^2 q_\theta} \left( \sin \theta \frac{\d }{\d \theta} + \lambda \,\sigma \cos \theta \right)S \, , \label{equ:ProcaS}
\eeq
where the spheroidicity parameter $c^2$ was
 defined in (\ref{equ:cc}), and 
  \beq
\begin{aligned}
q_\theta & \equiv 1 - \alpha^2 \tilde{a}^2  \lambda^{-2} \cos^2 \theta  \, , \\[4pt]
\sigma &\equiv \omega + \alpha \hskip 1pt \tilde{a} \hskip 1pt \lambda^{-2} (m - \alpha \tilde{a} \hskip 1pt \omega)  \, , \\[4pt]
 \Lambda &\equiv \lambda \left( \lambda - \sigma \right) + 2 \alpha \tilde{a} \hskip 1pt m \omega - \alpha^2 \tilde{a}^2 \omega^2 \, . \label{eqn:ProcaSigmaLambda}
\end{aligned}
\eeq
The radial equation becomes
\beq
\begin{aligned}
0 = \frac{\d ^2 R}{\d r^2} & + \left( \frac{1}{r-r_+} + \frac{1}{r-r_-} - \frac{1}{r-\hat{r}_+} - \frac{1}{r-\hat{r}_-} \right) \frac{\d R}{\d r} \\
& + \bigg( - \frac{\Lambda}{\Delta} - (1 - \omega^2) +  \frac{P_+^2}{(r-r_+)^2} + \frac{P_-^2}{(r-r_-)^2}  - \frac{A_+}{(r_+-r_-) (r-r_+) } \\
&  \ \ \ \ \ \ + \frac{A_-}{(r_+-r_-)(r-r_-)} - \frac{ \phantom{\mathcal{B}} \lambda \es \sigma r }{\Delta \left( r- \hat{r}_+ \right)} - \frac{\phantom{\mathcal{B}} \lambda \es \sigma r}{\Delta \left(r- \hat{r}_-\right)}\bigg)\, R \, , \label{eqn:ProcaR}
\end{aligned}
\eeq
where $\hat{r}_\pm  \equiv \pm i \lambda$ depend on the angular eigenvalue  
and the parameters $P_\pm$ and $A_\pm$ were defined in (\ref{equ:AP}). The presence of the additional poles at $r = \hat{r}_\pm$, makes this equation considerably more complicated than the corresponding  equation~(\ref{eqn: scalar radial equation}) in the scalar case.

\vskip 4pt
In principle, the task of determining the spectrum of electric modes of the vector quasi-bound states is the same as for the scalar. However, due to the additional poles at $r=\hat{r}_\pm$,  
the widths of the near and far regions are now much smaller and these regions
 no longer overlap. To match the asymptotic expansions in the near and far regions,
 we must 
 introduce an {\it intermediate region}\footnote{Physically, the necessity of the intermediate region arises because the dynamics in the near and far regions depend on the angular momentum of the vector field in distinct ways.  
As we will see in Section~\ref{sec:analytic}, while the dynamics in the near region is sensitive to the total angular momenta of the field (including its intrinsic spin), the dynamics in the far region depends only on its orbital angular momentum. The intermediate region thus smoothly interpolates between these two behaviors. } that overlaps with both 
 (see Fig.\,\ref{fig:Vectorzones}) and construct a solution that serves as a bridge. The matching is then performed in a two-step procedure, first between the far and intermediate regions, and then between the intermediate and near regions. 

\vskip 4pt
In the Schwarzschild limit, solutions have definite total angular momentum and parity. Because the vector spherical harmonic $Y^{i}_{j,\, j \es m}$ has opposite parity to $Y^i_{j \pm 1,\, j \es m}$ and the scalar harmonic $Y_{j\es m}$, it completely decouples from all other angular modes. This magnetic mode,
\beq
A^i (t, \textbf{r} ) = r^{\minus 1} R(r)\,Y^i_{j, \, j\es m} (\theta, \phi)\,e^{- i \omega t }  \, , \label{eqn:ProcaMagneticAnsatz}
\eeq
is thus completely separable in the Schwarzschild limit~\cite{Rosa:2011my}, and \cite{Pani:2012vp,Pani:2012bp} showed that this persists to linear order in $\tilde{a}$. We can thus use this ansatz to determine the magnetic spectrum in the limit of small spin. 

\begin{figure}[t]
\centering
\makebox[\textwidth][c]{\hspace{0.7cm} \includegraphics[scale=0.99, trim = 0 0 0 0]{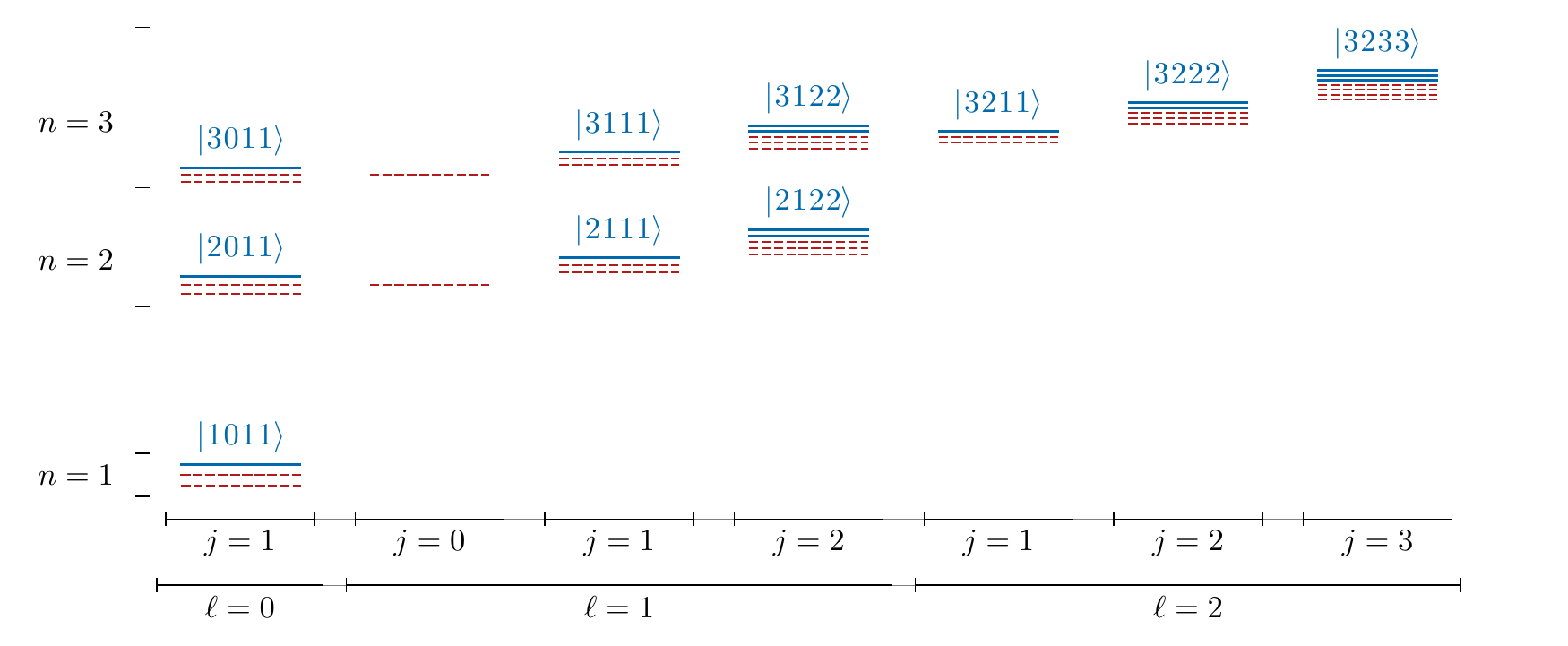}}
\caption{Schematic illustration of the energy spectrum for a massive vector field. Each state is labeled by the quantum numbers $|\es  n \es  \ell \es   j \es   m\rangle$. } 
\label{fig:VectorSpectra}
\end{figure}

\vskip 4pt
Substituting (\ref{eqn:ProcaMagneticAnsatz}) into the Proca equation and expanding 
to first order in $\tilde{a}$, the radial function $R(r)$ satisfies~\cite{Pani:2012vp, Pani:2012bp}
\beq
\begin{aligned}
0 = \frac{\d^2 R}{\d r^2 } + & \left( \frac{1}{r-\check{r}_+} - \frac{1}{r - \check{r}_-} \right) \frac{\d R}{\d r} \ + \bigg( - \frac{\check{\Lambda} }{\Delta} - \left(1 - \omega^2\right)  \\
&\ + \ \frac{\check{P}_+^2}{(r-\check{r}_+)^2} + \frac{\check{P}_-^2}{(r - \check{r}_-)^2} -\frac{\check{A}_+}{(\check{r}_+ - \check{r}_-) (r -\check{r}_+)} +\frac{\check{A}_-}{(\check{r}_+- \check{r}_-)(r- \check{r}_-)} \bigg) R \, ,\label{eqn:MagneticRadial}
\end{aligned}
\eeq
where $\check{\Lambda} = j(j+1)$ and $\check{P}_\pm, \check{A}_\pm$, $\check{r}_\pm$ represent $P_\pm$, $A_\pm$, $r_\pm$ expanded to linear order in $\tilde{a}$, respectively. Specifically, the position of the inner and outer horizons, $\check{r}_- = 0$ and $\check{r}_+ = 2 \alpha$, have shifted in this approximation, and we expect that (\ref{eqn:MagneticRadial}) does not accurately describe the near-horizon behavior of these magnetic modes. When expanded to linear order in $\tilde{a}$, the scalar radial equation~(\ref{eqn: scalar radial equation}) differs from (\ref{eqn:MagneticRadial}) only in its $\ud R/\ud r$ coefficient. In Section~\ref{sec:analytic}, we will discuss the error this small-spin approximation introduces. To compute the magnetic spectra for arbitrary spin, we must still solve the Proca equation (\ref{equ:Proca}) numerically. For a schematic illustration of the vector field spectrum, see Fig.~\ref{fig:VectorSpectra}.

\subsection{Summary of Results} 
\label{sec:Summary}

In Sections \ref{sec:analytic} and \ref{sec:numeric}, we will derive 
the spectra of scalar and vector quasi-bound states around Kerr black holes in detail. Here, we summarize our
main results. 

\vskip 4pt
We write the frequency eigenvalues as 
\beq
\omega \hskip 1pt \equiv \hskip 1pt   E + i \hskip 0.5 pt \Gamma \hskip 2pt  \equiv \hskip 1pt  \mu \, \sqrt{1 - \frac{\alpha^2}{\nu^2}}  \, , \label{eqn:generalspectrum}
\eeq
where the real part represents the energy $E$  and the imaginary part determines the instability rates $\Gamma$. 
At leading order, we expect the energy spectrum to be Bohr-like, so that
the real part of $\nu$ is 
an integer $n$. As we will see below, it is also convenient to work with $\nu$ because the fine and hyperfine structure can be read off directly from its $\alpha$-expansion. 

\begin{figure}[t]
  \begin{center}
    \makebox[\textwidth][c]{\includegraphics{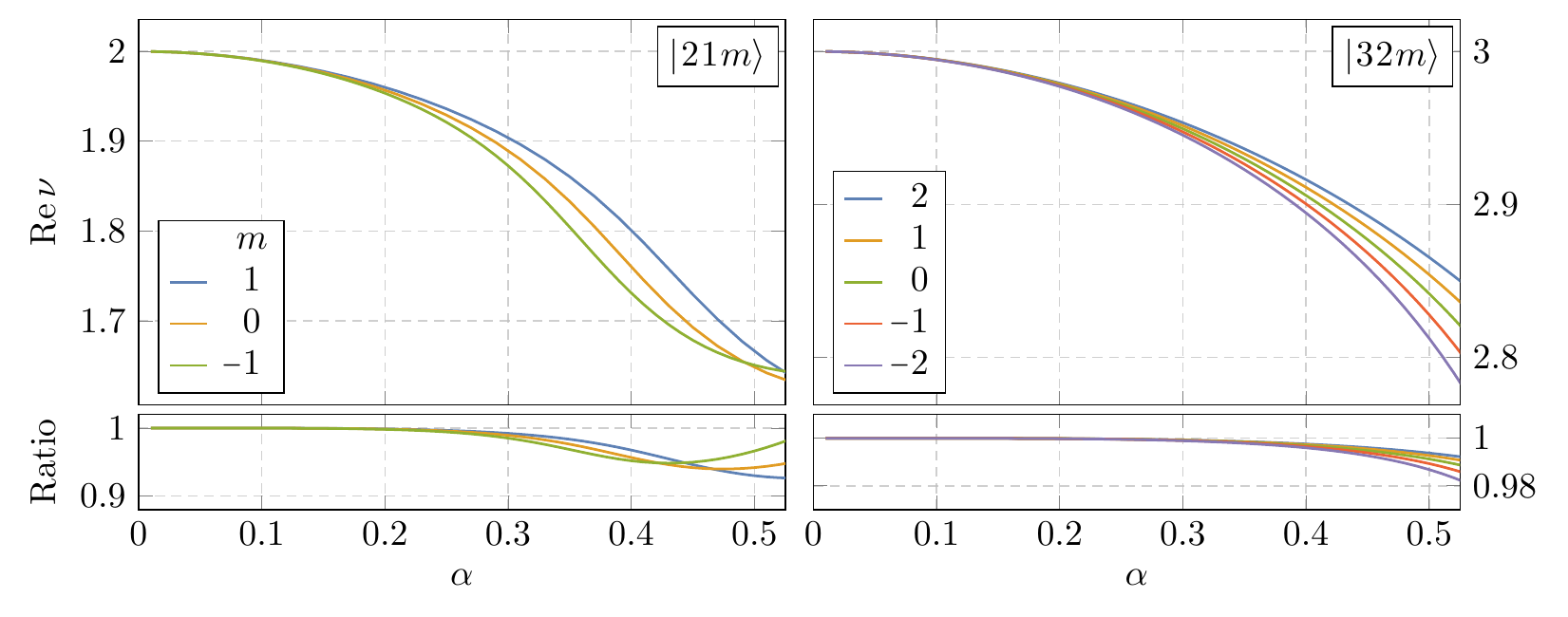}}
    \caption{Numeric results of the energy eigenvalues for scalar field eigenstates $\ket{n \ell m}$, for~$\tilde{a} = 0.5$. The lower panels show the ratio of the numeric results to the analytical predictions in~(\ref{eqn:scalarspectrum}). The fine and hyperfine structures are clearly seen in $\lab{Re}\, \nu$, and we find that (\ref{eqn:scalarspectrum}) is an excellent approximation for $\alpha \lesssim 0.2$.} \label{fig:ScalarSpectrumPlots}
  \end{center}
\end{figure}

\vskip 4pt
For the scalar field, the energy eigenvalues are (see also~\cite{Baumann:2018vus})
\beq
\begin{aligned}
 E_{n \ell m}   &= \mu \left( 1 - \frac{\alpha^2}{2n^2} - \frac{\alpha^4}{8n^4} + \frac{f_{n \ell} }{n^3}\hskip 2pt  \alpha^4 +  \frac{h_{\ell }  }{n^3} \hskip 2pt \tilde a m \hskip 1pt   \alpha^5 + \cdots \right)  , \\[2pt]
\lab{Re}\,(\nu_{n \ell m})  &= n + f_{n \ell} \hskip 1pt \alpha^2 + h_{\ell} \hskip 1pt \tilde a m \hskip 1pt \alpha^3 + \cdots \, , \label{eqn:scalarspectrum}
\end{aligned} 
\eeq
where the principal quantum number $n$ are integers that satisfy $n \geq \ell + 1$, and
\beq
\begin{aligned}
f_{n \ell} &\equiv  - \frac{6}{2\ell + 1} + \frac{2}{n}  \, , \\
h_{\ell} &\equiv \frac{16}{ 2 \ell \left( 2\ell+1 \right) \left( 2\ell+2 \right)}\, . \label{eqn:scalarspectrumfh}
\end{aligned}
\eeq
The first three terms of $E_{n \ell m}$ describe the constant mass term, the hydrogen-like Bohr energy levels and the relativistic corrections to the kinetic energy. The terms proportional to $f_{n\ell}$ and $h_{\ell}$ are the fine-structure $(\Delta \ell \neq 0)$ and hyperfine-structure $(\Delta m \neq 0)$ splittings, respectively.  
As the second line in (\ref{eqn:scalarspectrum}) suggests, these splittings can be more easily extracted from the real part of~$\nu_{n \ell m}$. Numeric results for $\lab{Re}\, \nu$, and their comparison with the perturbative approximations~(\ref{eqn:scalarspectrum}), are shown in Fig.~\ref{fig:ScalarSpectrumPlots} for representative scalar modes; cf.~Fig.~\ref{fig:ScalarSpectra} for a schematic illustration of the scalar spectrum.

\begin{figure}[t]
  \begin{center}
    \makebox[\textwidth][c]{\hspace{-0.25cm}\includegraphics[]{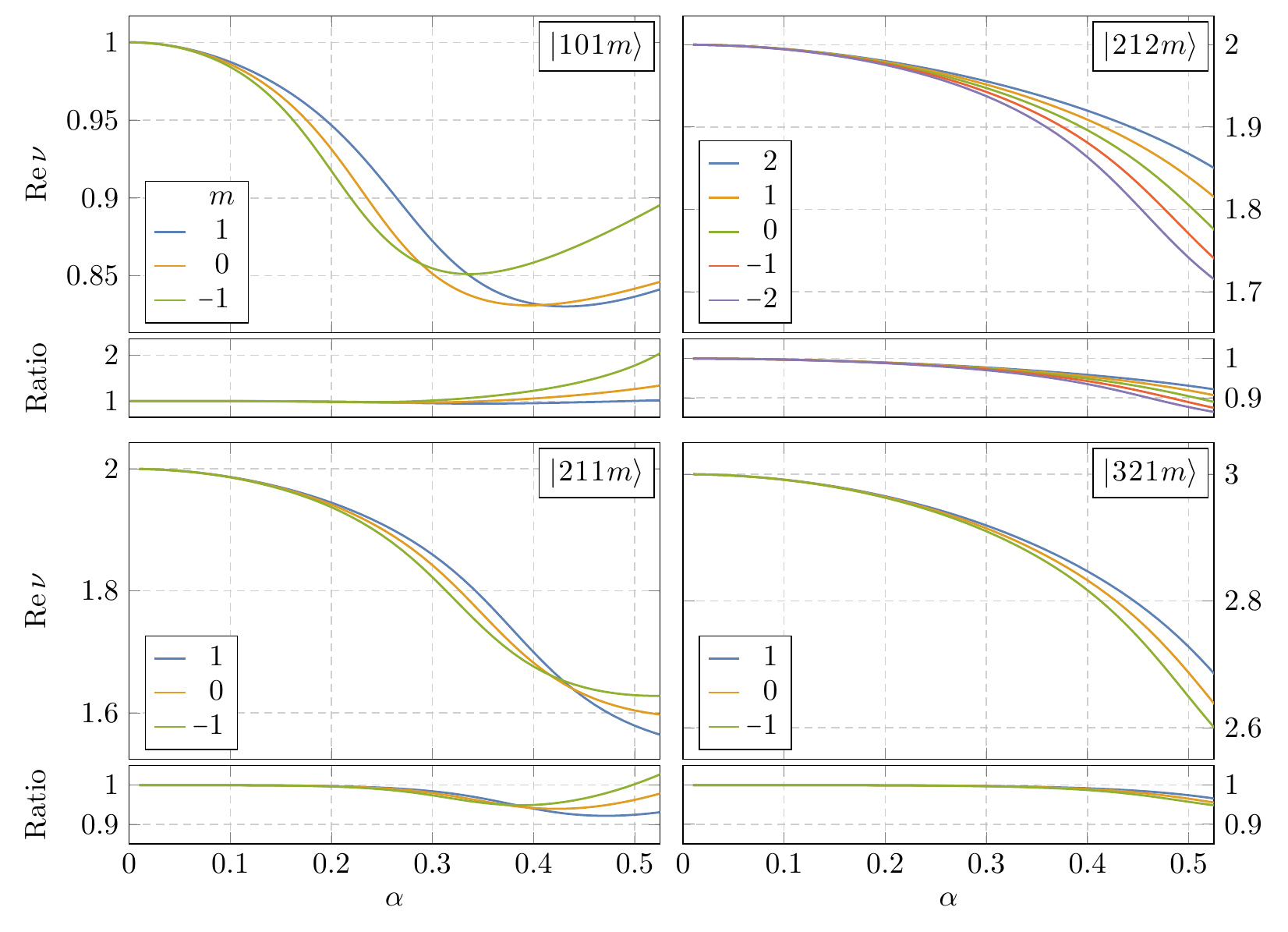}}
    \caption{Numeric results of the energy eigenvalues for vector field eigenstates $\ket{n \ell jm}$, for~$\tilde{a} = 0.5$. The lower panels show the ratio of the numeric results to our perturbative results in~(\ref{eqn:vectorspectrumGeneral}).} 
    \label{fig:VectorSpectrumPlots}
  \end{center}
\end{figure}

\vskip 4pt
For the vector field, we study the electric and magnetic modes separately using the ans{\"a}tze (\ref{eqn: Proca ansatz}) and (\ref{eqn:ProcaMagneticAnsatz}). However, we 
find that the energy eigenvalues for \emph{all} vector modes can be written as
\beq
\begin{aligned}
E_{n \ell j m}  &= \mu \left(  1 -\frac{\alpha^2}{2n^2} - \frac{\alpha^4}{8n^4} + \frac{f_{n \ell j} }{n^3} \hskip 2pt \alpha^4 + \frac{h_{ \ell j } }{n^3}  \hskip 2pt \tilde a  m  \hskip 1pt \alpha^5 + \cdots\right)   , \\[2pt]
\lab{Re}\,(\nu_{n \ell j m})  &= n + f_{n \ell j} \hskip 1pt \alpha^2 + h_{\ell j } \hskip 1pt \tilde a \hskip 1pt m  \alpha^3  + \cdots\, , \label{eqn:vectorspectrumGeneral}
\end{aligned}
\eeq
with the coefficients
\beq
\begin{aligned}
f_{n \ell j} & =  - \frac{4 \left( 6 \hskip 1pt \ell \hskip 1pt j  + 3 \ell + 3 j + 2 \right) }{\left( \ell + j \right)  \left( \ell + j + 1 \right) \left( \ell + j + 2 \right)  }+\frac{2}{n}  \,, \\[-9pt]   \\[-9pt]
h_{\ell j} & = \frac{16}{ \left( \ell + j \right) \left( \ell + j + 1 \right) \left( \ell + j +2 \right) } \, , \label{eqn:vectorspectrumfh}
\end{aligned}
\eeq
for $j = \ell \pm 1, \ell$. Notably, we find that the spectrum of the vector field is qualitatively similar to the scalar case~(\ref{eqn:scalarspectrum}); cf.~Fig.~\ref{fig:VectorSpectra} for a schematic illustration of the vector spectrum. Numeric results for $\lab{Re}\, \nu$ for representative electric and magnetic modes, and their comparison with (\ref{eqn:vectorspectrumGeneral}), 
are shown in Fig.~\ref{fig:VectorSpectrumPlots}.

  \begin{figure}
    \begin{center}
      \hspace{0 pt}\includegraphics[]{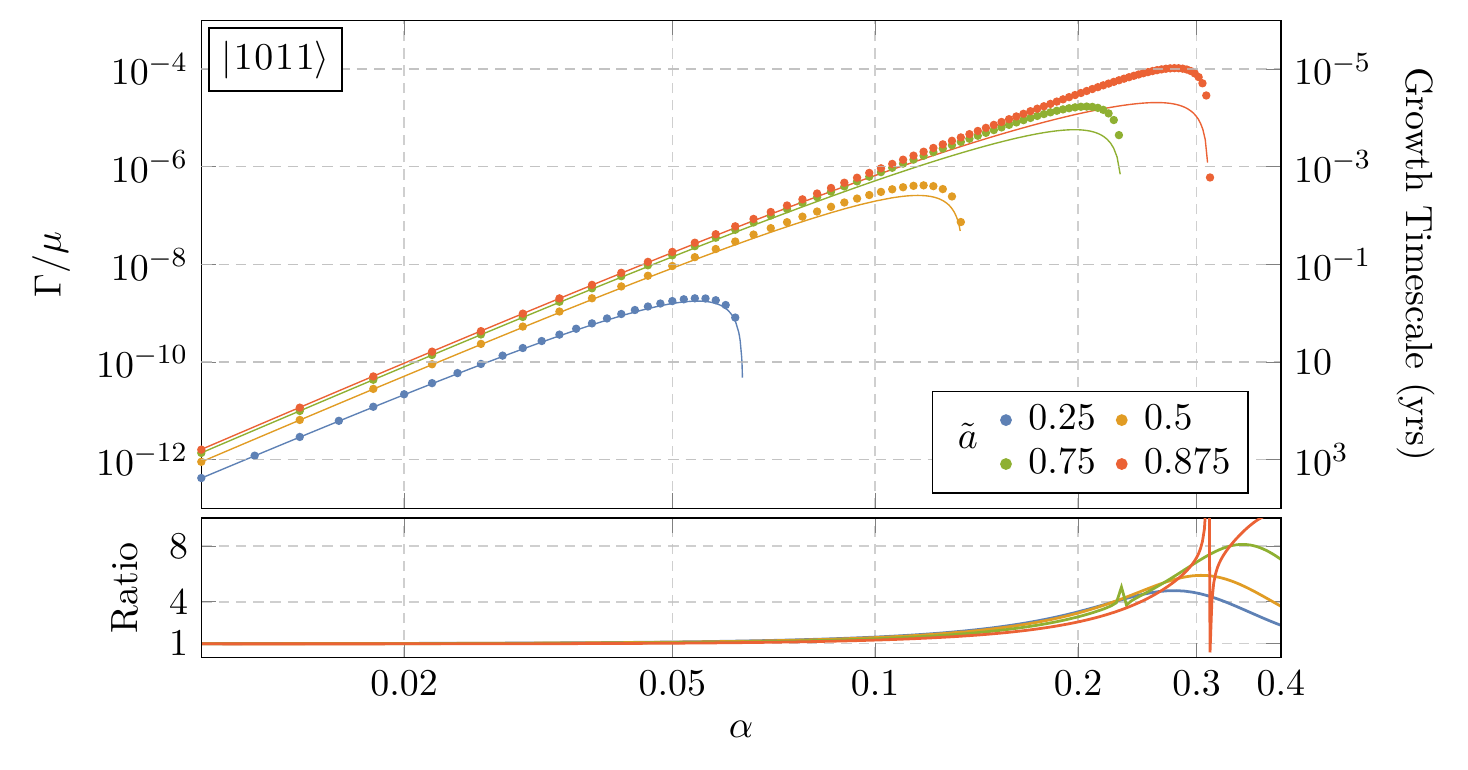}
      \caption{Superradiant growth rates for the dominant vector mode $\ket{1011}$, for different values of the black hole spin $\tilde a$. The growth timescale is for a Proca field of mass $\mu = \hbar/10^{-9}\,\lab{yr}\approx 2.1 \times 10^{-14}\,\lab{eV}$. The points are numeric data, while the solid lines are the approximation 
      (\ref{eqn:VectorRates}). The lower panel shows the ratio of our numerics and analytics, and includes the decaying regime, $\Gamma < 0$, at larger values of $\alpha$.} \label{fig:VectorDominant}
    \end{center}
  \end{figure}
  
\begin{figure}
  \begin{center}
    \makebox[\textwidth][c]{\hspace{-0cm}\includegraphics[]{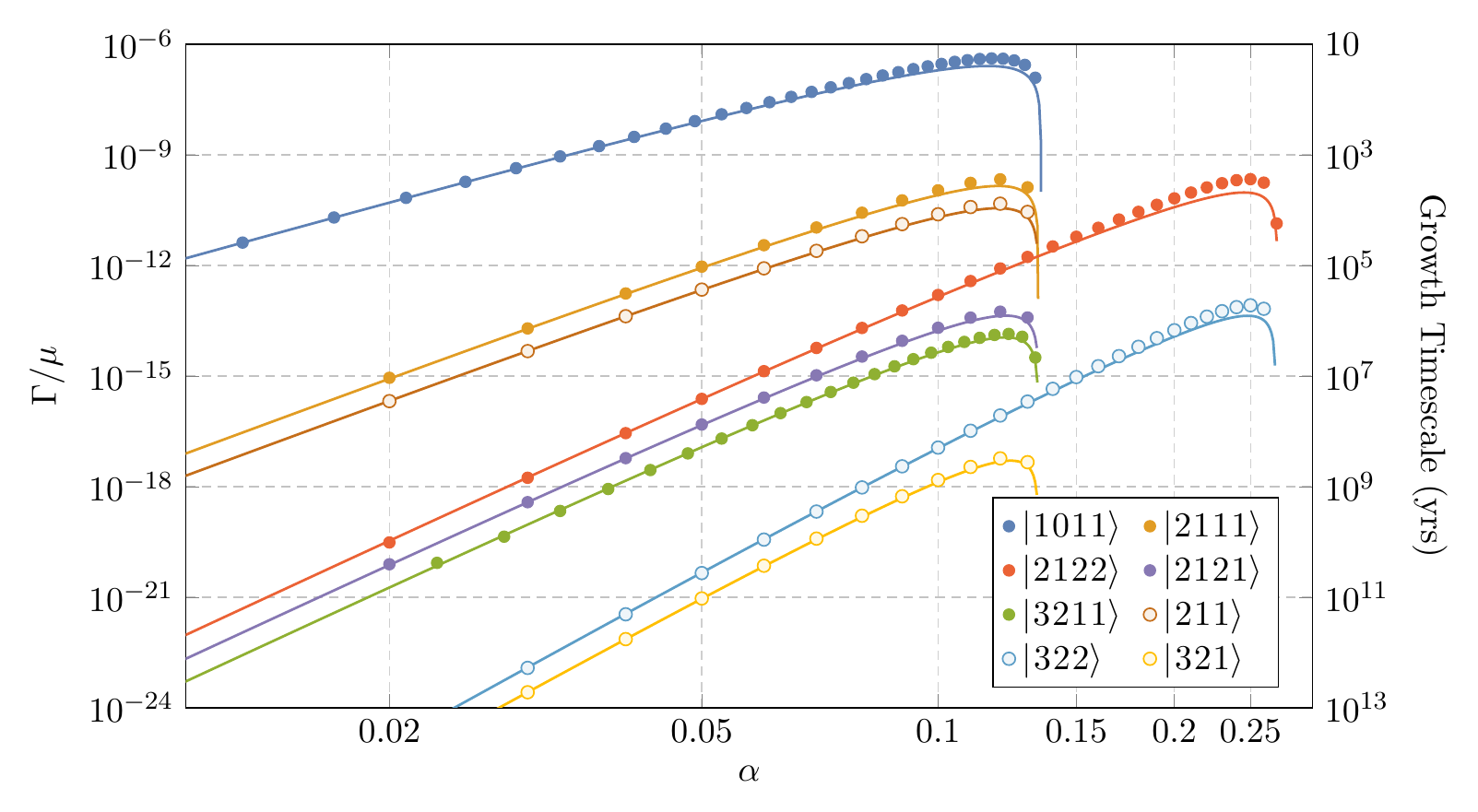}}
    \caption{Superradiant growth rates for selected scalar and vector modes, for $\tilde a =0.5$. The growth timescale is for $\mu = \hbar/10^{-9}\,\lab{yr} \approx 2.1 \times 10^{-14}\,\lab{eV}$. The points represent numeric data, while the solid lines are our perturbative predictions (\ref{eqn:ScalarRate}) and  (\ref{eqn:VectorRates}). \label{fig:growthRates}}
  \end{center}
\end{figure}

\vskip 4pt
We also compute the instability rates for these scalar and vector quasi-bound states. 
For the scalar field, we find (see also~\cite{Detweiler:1980uk}) 
\beq
\Gamma_{n \ell m} = 2 \tilde{r}_+ C_{n \ell} \, g_{\ell m}(\tilde{a}, \alpha,\omega) \, (m \Omega_H - \omega_{n \ell m})\hskip 1pt \alpha^{4\ell+5}  \, , \label{eqn:ScalarRate}
\eeq
where we have defined 
\begin{align}
C_{n \ell} &\equiv  \frac{2 ^{4\ell+1} (n+\ell)!}{  n^{2\ell+4} (n-\ell-1)! } \left[ \frac{\ell !}{(2\ell)! (2\ell+1)!} \right]^2  \, , \label{eqn:ScalarRateCoeff} \\
g_{\ell m}(\tilde{a}, \alpha, \omega) &\equiv \prod^{\ell}_{k=1} \left( k^2  \left( 1-\tilde{a}^2 \right) + \left( \tilde{a} m - 2 r_+ \hskip 1pt  \omega \right)^2  \right)  . \label{eqn:RateCoeffProd}
\end{align}
The dominant growing mode is $|\es n \es \ell \es m \rangle = |\es 2 \es 1 \es 1\rangle$, and hence $\Gamma_{211} \propto \mu \, \alpha^8$~\cite{Dolan:2007mj,Detweiler:1980uk}, where we have used $\Omega_H \sim \mu \alpha^{-1}$. Numeric results for the growth rates $\Gamma$ of the fastest growing scalar states are shown in Fig.~\ref{fig:growthRates}, along with their comparison to the analytic approximations (\ref{eqn:ScalarRate}).

\vskip 4pt 
The instability rates for the vector modes take a very similar form. In particular, we find that they are
\beq
\Gamma_{n\ell jm} = 2 \tilde{r}_+ C_{n \ell j } \, g_{j m}(\tilde{a}, \alpha, \omega)  \left( m \Omega_H - \omega_{n \ell j m} \right) \alpha^{2 \ell + 2 j + 5} \, , \qquad  \text{for $j = \ell \pm 1, \ell$}\, ,  \label{eqn:VectorRates}
\eeq	
where the coefficient, 
\beq
C_{n \ell j}\equiv \frac{2^{2\ell + 2 j +1} (n+\ell)!}{n^{2\ell+4}(n-\ell-1)!} \left[ \frac{(\ell)!}{(\ell + j )!(\ell + j+1)!}\right]^2 \left[ 1 + \frac{ 2  \left( 1+ \ell  - j   \right) \left(1   - \ell  +  j \right)  }{\ell + j}\right]^2 \, , \label{eqn:VectorRatesCoeff}
\eeq
is valid for all $j = \ell \pm 1, \ell$,
and the function $g_{jm}$ is obtained by replacing $\ell$ by $j$ in (\ref{eqn:RateCoeffProd}). The $\alpha$-scaling in (\ref{eqn:VectorRates}) was also found 
in~\cite{Pani:2012vp, Pani:2012bp, Endlich:2016jgc, Baryakhtar:2017ngi}. Notice that (\ref{eqn:VectorRates}) is also valid for the scalar rate~(\ref{eqn:ScalarRate}), if we set $j =\ell$ and multiply $C_{n \ell j}$ by a factor of $j^2/(j+1)^2$.  The dominant vector growing mode is $|\es n \es \ell \es j \es m \rangle = |\es 1 \es 0 \es 1 \es 1\rangle$ and has $\Gamma_{1011} \propto \mu \hskip 1pt \alpha^6$~\cite{Endlich:2016jgc, Baryakhtar:2017ngi, Cardoso:2018tly, East:2017ovw, East:2017mrj, East:2018glu, Pani:2012vp, Pani:2012bp, Dolan:2018dqv}.
This is faster than the dominant growing mode for the scalar field.  In Fig.~\ref{fig:VectorDominant}, we show our numeric results for the growth rate $\Gamma_{1011}$ at different values of the black hole spin $\tilde{a}$, and compare it with (\ref{eqn:VectorRates}). In Fig.~\ref{fig:growthRates}, we compare our numeric results for the growth rates of other electric and magnetic vector modes to their perturbative approximations (\ref{eqn:VectorRates}). For comparison, we also include the dominant scalar modes.

\vskip 4pt
Unfortunately, the separable ansatz (\ref{eqn:ProcaMagneticAnsatz}) for the magnetic mode does not describe the Proca field in the near-horizon geometry of the Kerr black hole, and thus cannot be used to rigorously derive a perturbative expression for the magnetic instability rates. The result (\ref{eqn:VectorRates}) for $j = \ell$ was obtained through an educated guess and, in the absence of numerical evidence, should be taken as purely conjectural.  
Having said that, we find excellent agreement between this guess and our numeric results (c.f.~Figures~\ref{fig:growthRates}~and~\ref{fig:magInstability}). 
and so we include it as a guide for phenomenology.

\newpage
\section{Analytical Computation of the Spectra}
\label{sec:analytic} 

In this section, we compute in detail the energy spectra and instability rates of quasi-bound states of massive scalar and vector fields around rotating black holes, using the method of matched asymptotic expansions~\cite{bender1999advanced, holmes1998introduction}.   The conceptual challenge of the computation can be illustrated using the scalar field, while avoiding the technical heft of the vector. We thus begin with the analysis of a scalar field~\cite{Starobinsky:1973aij,Detweiler:1980uk} in \S\ref{sec:MatchedScalar} before moving to the vector case in \S\ref{sec: analytics vector}.

\subsection{Massive Scalars around Kerr} 
\label{sec:MatchedScalar}

 In \S\ref{sec:scalarintro}, we presented the Klein-Gordon equation of a massive scalar field in the Kerr spacetime. 
 We will now solve for its
spectrum of energy eigenstates. 
It is convenient to write the frequency eigenvalue as 
\beq
\omega \equiv \sqrt{1 - \frac{\alpha^2}{\nu^2}}
 \, , \label{eqn:nudef}
\eeq
where the real part of $\nu$ is an integer for the leading-order Bohr spectrum.
We will compute $\lab{Im}\, \nu$ to leading order in $\alpha$ and $\lab{Re}\, \nu$ to order $\alpha^3$, as all degeneracies are broken at this order.

\subsubsection{Matched Asymptotic Expansion} 

Since perturbative solutions of the scalar field radial equation (\ref{eqn: scalar radial equation}) cannot capture the boundary conditions at $r=r_+$ and $r \to \infty$ simultaneously, a matched asymptotic expansion is necessary to compute the eigenvalues $\nu$. Schematically, we approximate the radial function separately in both the near and far regions (cf.~Fig.~\ref{fig:ScalarZones}), and match these solutions in the region of overlap.
\vskip 4pt
To define the \emph{near region}, it is convenient to introduce the rescaled radial coordinate
\beq
z \equiv \frac{r-r_+}{r_+ - r_-} =  \frac{r- \alpha \hskip 1pt \tilde{r}_+}{\alpha \left( \tilde{r}_+ - \tilde{r}_- \right)} \, , \label{eqn: z variable}
\eeq
where $\tilde{r}_\pm \equiv r_\pm / \alpha$, so that $\tilde{r}_\pm \sim \mathcal{O}(1)$. In this coordinate, the inner and outer horizons are mapped to $z=-1$ and $z=0$, respectively.  As discussed in~\S\ref{sec:scalarintro}, the radial function varies much more rapidly  in the near region than in the far region, $\ud R/\ud r \sim \alpha^{-1}$, and the reason we change coordinates is to accommodate this rapid change in the $\alpha$-expansion. The radial equation  (\ref{eqn: scalar radial equation}) then becomes
\beq
\frac{1}{R \, z(z+1)}\frac{\d}{\d z} \! \left( \! z(z+1) \frac{\d R}{\d z}\right) - \frac{\Lambda}{z(z+1)}  - 4 \gamma^2  + \frac{P_+^2}{z^2} + \frac{P^2_-}{(z+1)^2} - \frac{A_+}{z} + \frac{A_-}{z+1} = 0   \, , \label{eqn: Heun-z}
\eeq
where the coefficients were defined in (\ref{equ:AP}) and (\ref{eqn:defgamma}). 
We then expand (\ref{eqn: Heun-z}) in powers of $\alpha$, with $z$ kept fixed. By comparing the dominant and subdominant terms at each order in $\alpha$, we find that this region covers the range $0 \leq z \lesssim \alpha^{-2}$, cf.~Fig.~\ref{fig:ScalarZones}. 
Near the outer horizon~at $z=0$,  the $1/z^2$ pole dominates and the radial solution scales as 
\beq
R(z) \,\sim\, B_1 \hskip 1pt z^{ i P_{+}} + B_2 \hskip 1pt  z^{- i P_+} \, , \mathrlap{\qquad z \to 0\, ,} \label{eqn:scalar radial outer horizon}
\eeq 
where $B_1$ and  $B_2$ are integration constants. The purely ingoing boundary condition at the event horizon requires that $B_2 = 0$. Similarly, approaching the inner horizon at $z=-1$, we find that $\lim_{z \to \sminus 1} \, R(z) \, \sim \, (z+1)^{\pm i P_-}$. However, unlike the outer horizon, the inner horizon is not physically accessible, and hence no boundary condition has to be imposed on it. 

\vskip 4pt
To identify the \textit{far region}, it is useful to use the alternative coordinate
\beq
x \equiv 2 \sqrt{1 - \omega^2} \, (r-r_+) = 4 \gamma z \, ,  \label{eqn: x variable}
\eeq
where $\gamma$ was defined in (\ref{eqn:defgamma}). In this coordinate, the near horizon region, $z \sim 1$, is mapped to $x \sim \alpha^2$, and the Bohr radius $r_c \sim  \alpha^{-1} $ is at~$x \sim 1$. In the far region, we expand the radial equation in powers of $\alpha$, while keeping $x$ fixed. As illustrated in Fig.~\ref{fig:ScalarZones}, the far region spans~$\alpha^2 \lesssim x < \infty$. 
The radial solution then behaves as 
\beq
R(x) \,\sim\, B_3 \hskip 1pt  e^{- x/2} \, x^{-1 + \nu - 2 \alpha^2/\nu} + B_4 \hskip 1pt  e^{+ x/2} \, x^{-1 -\nu + 2 \alpha^2/\nu} \, ,  \mathrlap{\qquad x \to \infty \, ,}\label{eqn: scalar radial infinity}
\eeq
where $\nu$ was defined in (\ref{eqn:nudef}). Since we are interested in quasi-bound states with $\omega < \mu$, the radial function should vanish at large distances. We therefore set $B_4 = 0$.

\vskip 4pt
To perform matched asymptotic expansions at high orders in $\alpha$, we expand the radial and angular functions\hskip 1pt\footnote{It is essential that $P_+$ is held fixed, since an expansion of $P_+$ in powers of $\alpha$ would change the boundary condition~(\ref{eqn:scalar radial outer horizon}) at the horizon. Indeed, the main reason this matched asymptotic expansion is needed at all is that, regardless of which order in $\alpha$ they appear, \emph{all} terms multiplying the $z^{-2}$ pole of (\ref{eqn: Heun-z}) become arbitrarily important as we approach the outer horizon, $z \to 0$. It is thus crucial that we do not disturb this pole in our $\alpha$-expansion, lest our approximation becomes arbitrarily bad, to
a point where we must impose a different boundary condition. \label{footnote:FixedPp}} 
\beq
X = \sum_k \alpha^k X_{\indlac{k}}\, , \label{eqn:SumX} 
\eeq
where $X = \{R^{\rm near}, R^{\rm far},S\}$. Schematically, the equations of motion become
\beq
\square \, X = \Big[ \square^{\indlab{0}} + \alpha \, \square^{\indlab{1}}  + \alpha^2 \, \square^{\indlab{2}}  +  \cdots \Big] \Big[ X_{\indlac{0}} + \alpha  X_{\indlac{1}}  + \alpha^2  X_{\indlac{2}} +  \cdots \Big] =0 \, , \label{eqn:XExpansion}
\eeq
where the $\alpha$-expansion of the differential operators $\square$ includes expansions of the angular and energy eigenvalues: 
\beq
\Lambda = \sum_k \alpha^k \Lambda_{\indlac{k}}\, , \qquad \nu = \sum_k \alpha^k \nu_{\indlac{k}}\, .
\eeq
In order to lift the degeneracies between all modes of the spectrum, it is 
sufficient to go to order $\alpha^3$ in $\Lambda$ and $\lab{Re}\,\nu$.  We will then solve these equations order by order in $\alpha$, imposing the above boundary conditions at each order. Finally, we match the solutions in the overlap region to determine $\nu$.

\subsubsection{Leading-Order Solution} 
\label{sec:LOscalar} 

We begin with the leading-order expansion of the equations of motion, which we expect to yield the hydrogenic spectrum. However, unlike the hydrogen atom, these states will be quasi-stationary, and we must also compute their instability rates.

\vskip 4pt
At leading order, the angular equation (\ref{eqn: spheroidal harmonic equation}) is
\beq
 \left[\frac{1}{\sin \theta} \frac{\d}{\d \theta}\left( \sin \theta \frac{\d}{\d\theta} \right) - \frac{m^2}{\sin^2 \theta} + \Lambda_\indlac{0} \right]\!S_\indlac{0} = 0 \, .
\eeq
Setting $\Lambda_0 = \ell(\ell+1)$, and imposing regular boundary conditions at the antipodal points $\theta = 0$ and $\pi$, the solutions are given by the associated Legendre polynomials $S_\indlac{0} = P_{\ell m}(\cos \theta)$. This is expected,
since the spheroidal harmonics reduce to the ordinary spherical harmonics in the limit $\alpha \to 0$. 

\vskip 4pt
The leading-order near- and far-zone radial equations are
\begin{align}
\left[  \frac{\d^2}{\d z^2 } + \left( \frac{1}{z} + \frac{1}{z+1} \right)\frac{\d}{\d z} -  \frac{\ell(\ell+1)}{z(z+1)}  + \frac{P_+^2}{z^2} + \frac{P^2_+}{(z+1)^2} - \frac{2P_+^2}{z} + \frac{2P_+^2}{z+1} \right] R^{\rm near}_\indlac{0}  &= 0 \, , \label{eqn:scalarLONear}  \\ 
\left[ \frac{\d^2 }{\d x^2 } + \frac{2}{x}\frac{\d }{\d x } + \frac{\nu_0}{x}   - \frac{\ell(\ell+1)}{x^2} - \frac{1}{4}  \,\right] \! R_\indlac{0}^{\mathrlap{\lab{far}}\hphantom{near}}  &= 0 \, .
\label{eqn:hydrogenic}
\end{align}
Imposing the correct boundary conditions,
we obtain
\begin{align}
R^{\rm near}_\indlac{0}(z)   &= \mathcal{C}^{\rm near}_0 \, \left( \frac{z}{z+1} \right)^{i P_+ } \!\!\!{}_2 F_1(-\ell, \ell+1, 1-2 i P_+, 1+z) \, , \label{eqn:scalarRadialLOnear} \\[4pt]
R^{\rm far}_\indlac{0}(x) &= \mathcal{C}^{\rm far}_{0} \, e^{-x/2} x^{\ell}\, U( \ell+1 -\nu_0, 2+2\ell, x) \, , \label{eqn:scalarRadialLOfar}
\end{align}
where ${}_2 F_1$ is the hypergeometric function and $U$ is the 
confluent hypergeometric function of the second kind. For integer values of $\nu_0 \geq \ell+1$, the functions $U$ become Laguerre polynomials.

\vskip 4pt
The widths of the near and far regions can be determined by comparing the terms in (\ref{eqn: Heun-z}) that have been neglected when writing  (\ref{eqn:scalarLONear}) and (\ref{eqn:hydrogenic}), to those that have been kept.\footnote{The widths we quote for the near, far, and overlap regions are parametric statements, and so will be unaffected by higher-order corrections.}
 We find that the near and far regions are valid for $z \lesssim \alpha^{-2}$ and $x \gtrsim \alpha^2$, respectively. This means that the two regions have an overlapping region than spans the range $ \alpha \lesssim r \lesssim \alpha^{-1}$ (see Fig.~\ref{fig:ScalarZones}).  
Since (\ref{eqn:scalarRadialLOnear}) and (\ref{eqn:scalarRadialLOfar}) are approximations of the same function, they must agree over the entire overlap region. To match the solutions, it is convenient to introduce the following \textit{matching coordinate}\footnote{The $\alpha$-expansions of $R_{0}^{\rm near}(\xi)$ and $R_{0}^{\rm far}(\xi)$ are equivalent to taking the limits $z \to \infty$ and $x \to 0$, respectively. While the latter is more commonly adopted in the literature (see e.g.~\cite{Starobinsky:1973aij, Detweiler:1980uk}), matching in terms of $\xi$ has the advantage that it is organized solely in powers of $\alpha$. As we shall see in \S\ref{sec: Scalar Relativistic Corrections}, this is a particularly useful way of organizing the matching at higher orders. 
}
\beq
\xi \equiv \frac{x}{\alpha^\beta} = \frac{2 (\tilde{r}_+ - \tilde{r}_-) }{\nu } \frac{z}{\alpha^{\beta - 2}} \, , \qquad \text{with} \qquad 0 < \beta < 2 \, . \label{eqn:IntCoord}
\eeq
Requiring the $\alpha$-expansions of the near and far-zone solutions to match, while keeping $\xi$ fixed, will fix the free coefficients of the solutions and thus determine $\nu$.

\vskip 4pt
In terms of the coordinate~$\xi$, the  $\alpha$-expansion of the near-zone solution is\hskip 1pt\footnote{To aid the reader, we group terms that are matched in the overlapping region by color.} 
\beq
\begin{aligned}
R^{\text{near}}_0(\xi)  \sim  \tilde{\mathcal{C}}_{0}^{\rm near} \bigg[ & \red{(\alpha^{\beta-2} \xi)^\ell}  \left( 1  + \cdots  \right) \\ &+ \blue{(\alpha^{\beta-2} \xi )^{-\ell-1}} \, \mathcal{I}_\ell   \left( \frac{2 (\tilde{r}_+ - \tilde{r}_-)}{\nu} \right)^{2\ell+1}  \left( 1  + \cdots  \right) \bigg]\, , \,\, \mathrlap{\,\, \alpha \to 0 \, ,} \label{eqn: scalar matching near2}
\end{aligned}
\eeq
where the ellipses denote expansions in powers of $(\alpha^{\beta-2}\xi)^{-1}$ with real coefficients. In the following, it will only be important that the omitted terms are purely real. We have absorbed an overall coefficient into the rescaling $\mathcal{C}_{0}^{\rm near} \to \tilde{\mathcal{C}}_{ 0}^{\rm near} $ 
and defined
\beq
\begin{aligned}
\mathcal{I}_\ell & \equiv   - i P_+  \frac{(\ell!)^2}{(2\ell)! (2\ell+1)!} \prod_{k=1}^\ell \left( k^2 + 4P_+^2\right)   , \label{eqn:ImI}
\end{aligned}
\eeq
which is purely imaginary. 
Similarly, the $\alpha$-expansion of the far-zone solution~is 
\begin{align}
R^{\text{far}}_0(\xi)  \sim   \mathcal{C}_{0}^{\rm far} \bigg[ & \blue{( \alpha^\beta \xi)^{-\ell-1}}  \, \frac{\Gamma(2\ell+1)}{\Gamma(\ell+1-\nu_0)} \left( 1 +  \cdots \right)  +  \, \red{(\alpha^\beta \xi)^\ell} \, \frac{\mathcal{K}_\ell (\nu_0)}{\Gamma(2\ell+2)\Gamma(-\ell-\nu_0)}  \left( 1+  \cdots \right) \nonumber \\
& \quad +  ( \alpha^\beta \xi)^\ell \log ( \alpha^\beta \xi ) \,  \frac{1}{\Gamma(2\ell+2)\Gamma(-\ell-\nu_0)} \left( 1 + \cdots \right) \bigg]\, , \mathrlap{\qquad \alpha \to 0 \, ,} \label{eqn: scalar matching far} 
\end{align}
where the ellipses denote expansions in powers of $\alpha^{\beta}\xi$, and we have defined the $\nu_0$-dependent constant
\beq
\mathcal{K}_\ell (\nu_0) \equiv \gamma_E - \psi(2\ell+2) + \psi(\ell+1-\nu_0)  +  \sum_{k=1}^{2\ell+1} \frac{(-1)^{k+1} \Gamma(2\ell+2) \Gamma(-\ell + \nu_0)}{ 2^{k} \hskip 1pt k \hskip 1pt \Gamma(\nu_0-\ell+k) \Gamma(2\ell+2-k) } \, , \label{eqn:Kconstant}
\eeq
where $\psi(z) = \Gamma'(z)/\Gamma(z)$ is the digamma function and $\gamma_E$ the Euler-Mascheroni constant. Since logarithmic terms are absent in (\ref{eqn: scalar matching near2}), they do not play a role in the matching at leading order. However, these terms will appear at higher orders in the near-zone expansion and will ultimately match 
the logarithmic term in (\ref{eqn: scalar matching far}). 

\vskip 4pt
Matching the common terms in (\ref{eqn: scalar matching near2}) and (\ref{eqn: scalar matching far}) allows us to solve for the frequency eigenvalue~$\nu_0$. Although there are $2\ell+1$ such terms, 
only the $\red{\xi^\ell}$ and $\blue{\xi^{-\ell-1}}$ terms need to be matched at leading order. This is because they contain the dominant behaviors of $R^{\rm near}_0$ and $R^{\rm far}_0$ in the limit $\alpha \to 0$. The remaining terms are suppressed by powers of $\alpha$ and, like the logarithmic term above, can only be matched consistently when higher-order corrections to $R_0^{\rm near}$ and $R_0^{\rm far}$ are taken into account. Matching the coefficients of $\red{\xi^\ell}$ and $\blue{\xi^{-\ell-1}}$, and taking their ratio,
we obtain the following matching condition 
\beq
\frac{\Gamma(-\ell-\nu_0) \hskip 1pt \nu_0^{2\ell+1}}{\Gamma(\ell +1-\nu_0) \hskip 1pt \mathcal{K}_\ell(\nu_0)} = \alpha^{4\ell+2}\left(   \frac{\big[ 2 (\tilde{r}_+ - \tilde{r}_-)\big]^{2\ell\mathrlap{+1}}}{(2\ell!) (2\ell+1)!} \,{\cal I}_\ell + \cdots \right) , \label{eqn:ScalarMatching}
\eeq
where the ellipses denote terms that are purely real.
We emphasize that, although we have so far only expanded the equations of motion at leading-order,  higher powers of $\alpha$ arise in (\ref{eqn:ScalarMatching}) due to the hierarchy of the coordinates $x/z \sim \alpha^2$.  This is why the leading-order result for $\lab{Im}\,\nu$ can be $\mathcal{O}(\alpha^{4 \ell +2})$.

\vskip 4pt
Solving (\ref{eqn:ScalarMatching}) for $\nu_0$, we get
\beq
\nu_0 = n + i P_+  \big[ 2\alpha^2 (\tilde{r}_+ - \tilde{r}_-)\big]^{2\ell+1} \frac{(n+\ell)!}{n^{2\ell+1}(n-\ell-1)!}\left[ \frac{\ell!}{(2\ell )! (2\ell+1)!}\right]^2 \prod_{k=1}^\ell \left( k^2 + 4P_+^2\right)  ,  \label{eqn:NuLOScalar}
\eeq
where $n$ is an integer, with $ n \geq \ell + 1$. Substituting (\ref{eqn:NuLOScalar}) into (\ref{eqn:nudef}), and restoring a factor of $\mu$, the energy spectrum reads\footnote{From the point of view of the matching procedure, the real and imaginary parts of the 
eigenvalue arise from the matching of the coefficients of $\xi^\ell$ and $\xi^{-\ell-1}$, respectively. This can be seen directly from (\ref{eqn: scalar matching near2}), since the coefficient of $\xi^\ell$ is purely real while that of $\xi^{-\ell-1}$ has an imaginary part.} 
\beq
\omega_{n \ell m} = \mu \left( 1 - \frac{\alpha^2}{2n^2} \right)  + i \Gamma_{n \ell m} \, ,   \label{eqn: Bohr energy1}
\eeq
where the instability rate is given by (\ref{eqn:ScalarRate}). 
The real part of (\ref{eqn: Bohr energy1}) shows that the system has the expected hydrogen-like spectrum. However, due to the nontrivial boundary condition at the horizon, the imaginary part of (\ref{eqn:NuLOScalar}) is non-vanishing and the energy eigenstates are only quasi-stationary. For $m \Omega_H > \omega$, the instability rate is positive and superradiant growth occurs.

\subsubsection{Higher-Order Corrections} \label{sec: Scalar Relativistic Corrections}

Next, we compute higher-order corrections to $\nu$. 
We focus only on the corrections to the real part of the spectrum, since all degeneracies in the imaginary part have already been broken at leading order.   

\vskip 4pt
It is convenient to rearrange the $\alpha$-expansion of the equations of motion (\ref{eqn:XExpansion}) into the following form 
\beq
\square^{\indlab{0}} \hspace{-0.5pt} X_{\indlac{i}} = -  \sum_{k=0}^{i-1} \square^{\indlab{i-k}} X_{\indlac{k}} \equiv J^X_{\indlac{i}} \, , \label{eqn:Jdef}
\eeq
so that the solutions of order $k < i$ are sources for the solution at order $i$.
Expanding the angular equation (\ref{eqn: spheroidal harmonic equation}) in powers of $\alpha$, we obtain 
 $J_{\indlac{i}}^\theta = 0$ and  $\Lambda_{\indlac{i}} = 0$, for $i = 1, 2, 3$, which means that the angular eigenvalue and eigenstate are uncorrected up to third order:
\beq
\Lambda = \ell(\ell+1) + \mathcal{O}\big(\alpha^4\big)  \, , \qquad  S(\theta) = P_{\ell m}(\cos \theta) + \mathcal{O}\big(\alpha^4\big)  \, . \label{eqn:ScalarHigherAngular}
\eeq
This is to be expected, since any deviation between $P_{\ell m}$ and the spheroidal harmonics, which are the exact solutions of (\ref{eqn: spheroidal harmonic equation}), is parametrized by the spheroidicity parameter $c^2 \sim \alpha^4$. 
To the order in $\alpha$ that we are working in, we can therefore use the leading-order angular solutions.

\vskip 4pt 
The asymptotic expansions of the radial functions need to be treated more carefully.
We will distinguish between modes with $\ell=0$ and $\ell \ne 0$.
For the former, the radial function peaks near the horizon and the solution is especially sensitive to the near-horizon geometry.

\paragraph{$\boldsymbol{\ell=0}$ modes} 

While the matching procedure is conceptually the same as before, the system of equations (\ref{eqn:Jdef}) is
more challenging to solve, because the source terms $J_i^X$ do not vanish.
 We solve these inhomogeneous equations via the method of \textit{variation of parameters}, where the general solutions contain integrals over $J_i^X$, with the integration limits appropriately chosen such that the boundary conditions in the respective regions are satisfied.
We relegate all technical details to Appendix~\ref{app:details}, and provide a more qualitative description here.

\vskip 4pt
We perform the matching at higher orders by converting the radial coordinates $x$ and $z$ to the matching coordinate (\ref{eqn:IntCoord}), and expanding the resulting functions in the limit $\alpha \to 0$, just as we did at leading order. However, since $x$ and $z$ also appear in the integration limits of the integrals mentioned above, their asymptotic expansions must be systematically organized such that no spurious divergences appear as $\alpha \to 0$; see Appendix~\ref{app:details}. At finite order in $\alpha$, we expect only a finite number of the terms in $R^\lab{near}(\xi)$ and $R^\lab{far}(\xi)$ to match as $\alpha \to 0$.
For example, for the $\ell=0$ mode, only the terms $\xi$, $\xi^0$, $\log \xi$ and $\xi^{-1}$ are shared between the  near- and far-zone solutions at order $\alpha^2$. We solve for the energy eigenvalues by matching the coefficients of these terms. Although there are generally more terms to match than unknowns, the matching is only consistent if all of these common terms match.
For the $\ell=0$ mode, this procedure yields
\beq
E_{n 0 0} = \mu \left( 1 - \frac{\alpha^2}{2n^2} -\frac{\alpha^4}{8n^4} +  \frac{\left( 2-6n \right)\alpha^4 }{n^4}   \right)  . \label{eqn:l0ScalarSpectrum}
\eeq
Since $\ell=0$ implies $m=0$, there is no hyperfine splitting $\propto m \tilde{a} \hskip 2pt \alpha^5$.

\paragraph{$\boldsymbol{\ell \neq 0}$ modes} 

Although the method of matched asymptotic expansion described so far is robust and correctly captures all boundary conditions of the radial equation, the procedure quickly becomes cumbersome for modes with arbitrary quantum numbers $\{n, \ell, m \}$. 
Fortunately, since the typical Bohr radii of the $\ell \neq 0$ modes are peaked in the far region, most of their support is far from the horizon. The spectra of the $\ell \neq 0$ modes can therefore be obtained by naively extending the radial solution in the far region (\ref{eqn: scalar radial far LO}) towards $x \to 0$ and assuming that it is regular at the origin. While this approximation does not capture the instability rates, it is sufficient for determining the real parts of the energy eigenvalues~$\nu_i$, at least up to the order of interest.\footnote{We have also performed a matched asymptotic expansion for the $\ell=1$ mode, and we found agreement with the stated approximation up to order $\alpha^5$. Since the approximation gets better for larger values of $\ell$, we expect the results obtained through this simplification also hold for all other $\ell \neq 0$ modes.} As shown explicitly in  Appendix~\ref{app:details}, this leads to the expression (\ref{eqn:scalarspectrum}) for the scalar field energy eigenvalue. This result also agrees with (\ref{eqn:l0ScalarSpectrum}) obtained for $\ell=0$ through the matched asymptotic expansion.\footnote{In \eqref{eqn:scalarspectrum}, the hyperfine structure naively diverges for the $\ell=0$ mode. However, since this mode necessarily has $m=0$, it does not actually receive a contribution from the hyperfine splitting.} These higher-order contributions have been computed in \cite{Baumann:2018vus} in the non-relativistic limit. Our results here, which make no such assumption, agrees with the previous result.

\subsection{Massive Vectors around Kerr} 
\label{sec: analytics vector}

Having illustrated our approach for the case of a scalar field, we are now ready to attack the more technically challenging vector field.
We will treat the electric and magnetic modes of the field separately.

\subsubsection{Electric Modes} 
\label{sec:ProcaLO}

Our electric analysis relies on 
the separable ansatz (\ref{eqn: Proca ansatz}) introduced in~\cite{Krtous:2018bvk, Frolov:2018ezx}. Just like for 
the scalar field, we can transform the radial equation (\ref{eqn:ProcaR}) into near and far regions through the variables $z$ and $x$, defined in~(\ref{eqn: z variable}) and~(\ref{eqn: x variable}), respectively. The asymptotic behavior of the radial function as $z \to 0 $ and $ x \to \infty $ is 
\begin{align}
R(z) &\,\sim\, B_1 \hskip 1pt z^{ i P_{+}} + B_2 \hskip 1pt  z^{- i P_+} \,,  \mathrlap{\hspace{4.37cm} z \to 0\, ,} \label{eqn:vector radial outer horizon} \\
R(x) &\,\sim\, B_3 \hskip 1pt  e^{- x/2} \, x^{+\nu - 2 \alpha^2/\nu} + B_4 \hskip 1pt  e^{+ x/2} \, x^{-\nu + 2 \alpha^2/\nu}\,,  \mathrlap{\hspace{0.75cm} x \to \infty \, .} \label{eqn: vector radial infinity}
\end{align}
The near horizon behavior is similar to that of the scalar in (\ref{eqn:scalar radial outer horizon}), since the residues of the $1/z^2$ poles are identical.  The asymptotic behavior in the far region, on the other hand, differs from that of the scalar, c.f.~(\ref{eqn: scalar radial infinity}), because the coefficients of the $\d R/\d r$ term in the radial equations differ as $r \to \infty$.
The purely ingoing boundary condition at the horizon requires that $B_2 = 0$, and for a quasi-bound state solution we must set $B_4=0$.

\vskip 4pt
A crucial difference between the analysis for scalar and vector fields is the presence of additional poles at $r=\hat{r}_{\pm}$ in the radial equation (\ref{eqn:ProcaR}). 
Since these poles are located at an $\mathcal{O}(1)$ distance from the origin of the complex-$r$ plane, the widths of the near and far regions are now much smaller. Indeed, we find that the near and far regions cover $z \lesssim \alpha^{-1}$ and $x \gtrsim \alpha$, and therefore do not overlap. This means that the matching between these regions must be performed indirectly through an intermediate region (see Fig.~\ref{fig:Vectorzones}). To describe this intermediate region, it is convenient to introduce the following coordinate
\beq
y \equiv r - r_+ \, . \label{eqn: y variable}
\eeq
We find that the intermediate region spans the range $\alpha \lesssim y \lesssim \alpha^{-1}$, which overlaps with both the near and far regions and therefore allows the two-step matching procedure.  
Since no boundary conditions need to be imposed in the intermediate region, the coefficients are determined by matching the near- and far-zone solutions.

\vskip 4pt
To solve the coupled differential equations (\ref{equ:ProcaS}) and (\ref{eqn:ProcaR}) order-by-order in $\alpha$, we now expand all relevant quantities in powers of $\alpha$. The system of equations are schematically organized as in~(\ref{eqn:XExpansion}), where $X$ now includes the intermediate radial function $R^{\rm int}$. The parameters appearing in the differential operators, $\lambda$ and $\nu$, are also expanded in powers of $\alpha$.

\subsubsection*{Leading-order solutions}

The leading-order angular equation for the electric modes is
\beq
\left[ \frac{1}{\sin \theta} \frac{\d}{\d \theta} \left( \sin \theta \frac{\d}{\d \theta} \right) - \frac{m^2}{\sin^2 \theta} + \lambda_0 (\lambda_0 - 1) \right] S_0 = 0 \, . \label{eqn: ProcaAngular}
\eeq
Since the vector field has intrinsic spin, the quantum number $m$ now describes the \textit{total angular momentum} projected along the spin of the black hole (see Appendix~\ref{app:harmonics}). The solutions are the associated Legendre polynomials $P_{jm}$, provided that the separation constant obeys $\lambda_0(\lambda_0 - 1) = j(j+1)$. 
In the far region, we may identify $j$ as the total angular momentum of solution.
This yields the following two solutions
\beq
\lambda_0^+   = - j \quad {\rm and} \quad \lambda_0^- = j+1 \, .   \label{eqn:lambda0sol}
\eeq
To understand the physical interpretation of these solutions, it is instructive to consider the leading-order form of the ansatz $A^\mu \equiv B^{\mu \nu} \nabla_\nu Z$. Using the results of Appendix~\ref{app:details}, we find
\begin{align}
A^{\mu}_{\indlab{0}} = \frac{\lambda_0}{\lambda_0^2  + r^2}\begin{pmatrix}
\minus \lambda_0 & \hphantom{\minus}i r  & \hphantom{\,\,}0 \hphantom{\,\,} & \hphantom{\,\,}0 \hphantom{\,\,} \\
 \minus i r & \lambda_0 & 0 & 0 \\
 0 & 0 & 0 & 0 \\
 0 & 0 & 0 & 0 \\
\end{pmatrix}\begin{pmatrix}\hphantom{\,} \partial_t \hphantom{\,} \\ \partial_r \\ \partial_\theta \\ \partial_\phi \end{pmatrix} Z_0 +  \frac{1}{r^2} \begin{pmatrix}
 \hphantom{\,\,}0\hphantom{\,\,} & \hphantom{\,\,}0\hphantom{\,\,} & \hphantom{\,\,}0\hphantom{\,\,} & 0 \\
 0 & 0 & 0 & 0 \\
 0 & 0 & 1 & 0 \\
 0 & 0 & 0 & \lab{csc}^{2}\theta \\
\end{pmatrix}\begin{pmatrix}\hphantom{\,} \partial_t \hphantom{\,} \\ \partial_r \\ \partial_\theta \\ \partial_\phi \end{pmatrix} Z_0 \, ,  \label{eqn: ansatz a0 explicit}
\end{align}
which is diagonal in angular gradients. At large distances $ r \gg \lambda_0$, the spatial components of the vector field simplify to
\beq
A^i_{\indlab{0}} \propto r^{\minus 1}  R_0^{\rm far}(r) \left( r \hskip 1pt \partial^i Y_{jm} -\lambda_0 \,Y_{jm} \hat{r}^i \right) e^{- i \omega t}, \label{eqn:Aisol}
\eeq
where $Y_{jm}$ are the scalar spherical harmonics, and we have normalized the basis vectors with the appropriate scale factors $\{1, r, r \sin \theta \}$ in spherical coordinates. Using the relationships between the `pure-orbital' and `pure-spin' vector spherical harmonics in flat space~\cite{Thorne:1980ru},\footnote{See Appendix~\ref{app:harmonics} for a discussion of vector spherical harmonics in spherically symmetric spacetimes.} 
\beq 
\begin{aligned}
Y^i_{j-1, jm} & = \frac{1}{\sqrt{j(2j+1)}} \left[  r  \hskip 1pt \partial^i Y_{jm} + j Y_{jm} \hat{r}^i \right]  , \\
Y^i_{j+1, jm} & = \frac{1}{\sqrt{(j+1)(2j+1)}} \left[  r  \hskip 1pt \partial^i Y_{jm}  - (j+1) Y_{jm} \hat{r}^i \right] , \label{eqn:PureOrbit-PureSpin}
\end{aligned}
\eeq
we see that the solutions (\ref{eqn:Aisol}) with eigenvalues $\lambda_0^{\pm}$ correspond to the  $j=\ell\pm1$ electric modes of the vector field. Furthermore, (\ref{eqn:Aisol}) relates $R_0^{\rm far}$ to the  actual radial profile of the vector field at large distances, with the additional power of $r^{-1}$ accounting for the different asymptotic behaviors of (\ref{eqn: scalar radial infinity}) and (\ref{eqn: vector radial infinity}). This means that the usual hydrogen-like intuition still applies for all spatial components of the vector field~\cite{Endlich:2016jgc, Baryakhtar:2017ngi}.\hskip 1pt\footnote{This is perhaps most obvious if we rewrite the Proca equation as the system of coupled scalar equations~(\ref{eq:procaEqDecompMain}). In the far-field limit, the mixings vanish, and the Proca equation reduces to a set of uncoupled scalar equations~(\ref{eqn: scalar radial equation}).} 

\vskip 4pt
The near, intermediate and far-zone radial equations are 
\begin{align}
\left[  \frac{\d^2}{\d z^2 } + \left( \frac{1}{z} + \frac{1}{z+1} \right)\frac{\d}{\d z} -  \frac{\lambda_0 (\lambda_0 - 1)}{z(z+1)}  + \frac{P_+^2}{z^2} + \frac{P^2_+}{(z+1)^2} - \frac{2P_+^2}{z} + \frac{2P_+^2}{z+1} \right] R^{\rm near}_0 &  = 0  \, , \label{eqn:ProcaNear} \\
\left[ \frac{\d^2}{\d y^2} +\left(  \frac{2}{y} - \frac{2 y}{  \lambda_0^{2} +  y^2 } \right) \frac{\d}{\d y} - \frac{ \lambda_0 \left( \lambda_0 - 1 \right)}{ y^2} - \frac{2\lambda_0}{ \lambda_0^2 + y^2} \right] \, R_0^{\rm int}\ &= 0 \, , \label{eqn:ProcaInt} \\
\left[ \frac{\d^2 }{\d x^2 } + \frac{\nu_0}{x}   - \frac{\lambda_0(\lambda_0+1)}{x^2} - \frac{1}{4}   \right]\, R_0^{\rm far}\ & = 0\, . \label{eqn:VectorFarF}
\end{align}
We see that the combination of $\lambda_0$ that appears in (\ref{eqn:ProcaNear}) is the same as that in (\ref{eqn: ProcaAngular}). This means that the near-horizon behavior is sensitive to the total angular momentum of the vector field, instead of just its orbital angular momentum. This is to be expected since the near-horizon limit resembles a `massless' limit of the vector field, and Teukolsky's equation for a massless gauge field in this same limit manifestly depends on the spin of the field~\cite{Teukolsky:1973ha}. Furthermore, we find that the dependence on $\lambda_0$ in (\ref{eqn:ProcaNear}) differs from that in (\ref{eqn:VectorFarF}). Since the $\alpha \to 0$ limit in the far region is equivalent to taking the flat-space limit, the far region is therefore only sensitive to the orbital angular momentum of the field. 
Indeed, solving 
$\lambda_0 \left( \lambda_0 + 1 \right) = \ell(\ell+1) $, we obtain
\beq
\lambda_0^+ = -(\ell+1) \quad {\rm and} \quad \lambda_0^- = \ell \, . \label{eqn:lambda0solL}
\eeq
Comparing (\ref{eqn:lambda0sol}) and (\ref{eqn:lambda0solL}), we find that the eigenvalues $\lambda_0^\pm$ correspond to the $j = \ell \pm 1$ modes, which agrees with our analysis above. 

\vskip 4pt
Imposing the correct boundary conditions, we find 
\begin{align}
R^{\rm near}_0(z)   &\,=\, \mathcal{C}_{0}^{\rm near}  \left( \frac{z}{z+1} \right)^{i P_+} \!\!\!\!{}_2 F_1(-j, j+1, 1-2 i P_+, 1+z) \, , \label{equ:Near0} \\[4pt]
R^{\rm int}_0(y)   &\,=\, \mathcal{C}_{0}^{\rm int} y^{- \lambda_0}  + \mathcal{D}_{0}^{\rm int} \, y^{-1 + \lambda_0} \left[ \lambda_0^2 \hskip 1pt (2\lambda_0 +  1 ) + (2 \lambda_0-1) \, y^2 \right]  ,\label{equ:Int0} \\[10pt]
R^{\rm far}_0(x) &\,=\, \mathcal{C}_{0}^{\rm far} \, e^{-x/2} x^{\ell+1}\, U( \ell+1 -\nu_0, 2+2\ell, x)\, . \label{equ:Far0}
\end{align}
Since the intermediate region depends on both $j$ and $\ell$, the powers of the polynomials in (\ref{equ:Int0}) are determined by the angular eigenvalues (\ref{eqn:lambda0sol}). As we shall see, the fact that the term proportional to $\mathcal{D}_{0}^{\rm int}$ contains two distinct $y$-dependences is crucial for the matching with the asymptotic expansions of the near and far-zone solutions. In the following, we will replace $\ell$ by $j$ via the substitution $j = \ell \pm 1$. This is because the azimuthal number $m$ is interpreted as $m_j$, so the eigenstates of the vector field are characterized by $j$ instead of $\ell$.

\vskip 4pt
Since the solutions (\ref{equ:Near0}), (\ref{equ:Int0}) and (\ref{equ:Far0}) are different approximations of the same function, they must agree in the regions of overlap. Although the near and far regions do not overlap with each other, they each overlap with the intermediate region. It is therefore convenient to introduce the following matching coordinates
\beq
\begin{aligned}
\xi_1 & \equiv \frac{x}{\alpha^{\beta}} = \frac{2}{\nu} \frac{y }{\alpha^{\beta-1}} \, , \hskip 65pt \qquad 0 < \beta < 1 \, , \\[4pt]
\xi_2 & \equiv \frac{2(\tilde{r}_+ - \tilde{r}_-)}{\nu} \frac{z}{\alpha^{\beta-2}} = \frac{2}{\nu} \frac{y }{\alpha^{\beta-1}}  \, , \qquad 1 < \beta < 2 \, . \label{eqn:IntCoordVector}
\end{aligned}
\eeq
We match these solutions by demanding that the intermediate-zone solution has the same $\xi_1$ dependence as the far-zone solution, and the same $\xi_2$ dependence as the near-zone solution. This forces the intermediate solution to agree with the solutions in the near and far regions in the regions of overlap. This matching procedure thus fixes the coefficients and determines the frequency eigenvalues. 

\vskip 4pt
In terms of $\xi_2$, the $\alpha$-expansion of the near-zone solution is 
\beq
\begin{aligned}
R^{\text{near}}_0  \,\sim\,  \tilde{\mathcal{C}}_{0}^{\rm near} \bigg[ & \red{(\alpha^{\beta - 2} \xi_2)^j}  \left[ 1  + \cdots  \right] \\
&+\, \blue{(\alpha^{\beta - 2} \xi_2 )^{-j-1}} \, \mathcal{I}_j  \left( \frac{2 (\tilde{r}_+ - \tilde{r}_-)}{\nu} \right)^{2j+1}  \left[ 1  + \cdots  \right] \bigg] \, , \,\, \mathrlap{\,\, \alpha \to 0 \, ,} 
\end{aligned}
\label{eqn: vector matching near2}
\eeq
where $\mathcal{I}_j$ was defined in (\ref{eqn:ImI}), with $\ell$ replaced by $j$. For the far and intermediate-zone solutions, we have to treat the $j = \ell\pm1$ modes separately.

\vskip 4pt
Substituting $\ell = j-1$ into (\ref{equ:Far0}) and expanding in powers of $\alpha$, the far-zone solution becomes 
\begin{align}
R^{\text{far}}_0  \,\sim\,   \mathcal{C}_{0}^{\rm far} \bigg[ & \green{( \alpha^\beta \xi_1)^{-j+1}}  \, \frac{\Gamma(2j-1)}{\Gamma(j-\nu_0)} \left[ 1 +  \cdots \right]  +  \, \red{(\alpha^\beta \xi_1)^{j}} \, \frac{\mathcal{K}_{j-1} (\nu_0)}{\Gamma(2j)\Gamma(1-j-\nu_0)}  \left[ 1+  \cdots \right] \nonumber \\
& +  ( \alpha^\beta \xi_1)^j \log ( \alpha^\beta \xi_1 ) \,  \frac{1}{\Gamma(2j)\Gamma(1-j-\nu_0)} \left[ 1 + \cdots \right] \bigg]\, , \qquad \alpha \to 0 \, , \label{eqn: vector matching far 1} 
\end{align}
where $\mathcal{K}_{j-1}$ was defined in (\ref{eqn:Kconstant}), with the indices appropriately replaced. Although logarithmic terms are absent in (\ref{eqn: vector matching near2}), they will appear at higher orders in the near-zone expansion, such that they can be matched with the logarithmic term in (\ref{eqn: vector matching far 1}). Substituting $\lambda_0 = \lambda_0^+$ into (\ref{equ:Int0}), the intermediate-zone solution reads
\beq
R^{\rm int}_0   \,=\, \mathcal{\tilde{C}}_{0}^{\rm int} \red{(\alpha^{\beta - 1}  \xi )^{j}} + \mathcal{\tilde{D}}_{0}^{\rm int}  \left[ 4 j^2(2j-1) \blue{(\alpha^{\beta - 1}  \xi )^{-j -1}} + (2j+1) \, \nu^2\, \green{(\alpha^{\beta - 1}  \xi )^{-j+1}} \right]  , \label{eqn: vector matching int 1} 
\eeq
where we have rescaled the coefficients $\mathcal{C}_{0}^{\rm int} \to \mathcal{\tilde{C}}_{0}^{\rm int}$ and $\mathcal{D}_{0}^{\rm int} \to \mathcal{\tilde{D}}_{0}^{\rm int}$ for future convenience. For the matchings with the far and near zones, we will use $\xi = \xi_1$ and $\xi = \xi_2$, respectively. 

\vskip 4pt
To determine the correct matching, we consider the behavior of the two asymptotic expansions (\ref{eqn: vector matching near2}) and (\ref{eqn: vector matching far 1}) as $\alpha \to 0$. From the perspective of the near-zone solution (\ref{eqn: vector matching near2}), this limit `zooms' in on the overlap of the near and intermediate regions, as seen `from' the near region. The $\red{\xi_2^j}$ term then dominates (\ref{eqn: vector matching near2}), and it must match the corresponding $\red{\xi^j}$ term in (\ref{eqn: vector matching int 1}). Similarly, if we take $\xi = \xi_1$ in the intermediate region (\ref{eqn: vector matching int 1}), then the $\alpha \to 0$ limit zooms in on the overlap between the intermediate and far regions, as seen from the intermediate region. In this case, $\red{\xi_1^j}$ also dominates, and so it must match the same term in (\ref{eqn: vector matching far 1}). Thus, as we move from the near to intermediate to far regions, the dominant behavior is always $\red{\xi^j}$, and its coefficients in the three different regions must match.

\vskip 4pt
We may also work in reverse. In the far region, the $\alpha \to 0$ limit zooms into its overlap with the intermediate region, but this time as seen from the far region. The dominant behavior now is $\green{\xi_1^{-j+1}}$, and this must match with the corresponding term in (\ref{eqn: vector matching int 1}). Similarly, taking $\xi = \xi_2$ in (\ref{eqn: vector matching int 1}), the $\alpha \to 0$ limit zooms into its overlap with the near region, though this time as seen from the intermediate region. We find that the $\blue{\xi_2^{-j-1}}$ dominates the intermediate solution and so is matched with its partner in (\ref{eqn: vector matching near2}). As we move from the far region, through the intermediate, and into the near region, the dominant behavior changes and thus the resulting matching condition is more nontrivial. 

\vskip 4pt
Taking the ratio of these two matching relations, we obtain
\beq
\frac{\Gamma(1-j-\nu_0) \, \nu_0^{2j-1} }{\Gamma(j-\nu_0) \mathcal{K}_{j-1} (\nu_0)} = \alpha^{4j} \left(    \frac{(2j+1) \, \left[ 2 (\tilde{r}_+ - \tilde{r}_-) \right]^{2j+1}}{\left[ 2 j (2j-1)!\right]^2 }\,\mathcal{I}_j  + \cdots  \right)  .\label{eqn:VectorplusRatio} 
\eeq
We again emphasize that, although we have expanded the equations of motion only to leading order, higher powers of $\alpha$ appear in (\ref{eqn:VectorplusRatio}) from $x/y \sim y/z \sim \alpha$. Solving for the leading-order real and imaginary parts of $\nu_0$, and restoring factors of $\mu$, we find 
\beq
\omega_{n \ell j m} = \mu \left( 1 - \frac{\alpha^2}{2n^2} \right) + i \Gamma_{n \ell j m}\, , \label{equ:omega+}
\eeq
where the instability rate is given by \eqref{eqn:VectorRates}. The dominant $j=\ell+1$ growing mode, $|\es 1 \es 0 \es 1 \es 1 \rangle$, has a growth rate $\Gamma_{1011} \propto \mu \, \alpha^6$ that is much larger than the dominant growing mode in the scalar case.  
The $j = \ell -1$ electric mode has the same energy spectrum (\ref{equ:omega+}), but a different instability rate $\Gamma_{n\ell jm}$;~cf.~(\ref{eqn:VectorRates}). Its dominant growing mode $|\es 3 \es 2 \es 1 \es 1 \rangle$ has a significantly suppressed growth rate, $\Gamma_{3211} \propto \mu \, \alpha^{10}$, compared to the the dominant $j= \ell +1$ mode.

\subsubsection*{Higher-order corrections}

Finally, we compute the higher-order corrections to the real part of the frequency eigenvalues. 
The following is a sketch of the computation, with details relegated to  Appendix~\ref{app:details}.

\vskip 4pt
Due to the presence of additional $\theta$-dependent terms on the right-hand side of (\ref{equ:ProcaS}), the higher-order angular equations are now harder to solve than in the scalar case. In particular, these terms induce new cross couplings in the angular eigenstates
\beq
\begin{aligned}
S(\theta) &= P_{j m}(\cos \theta) +  \Delta S(\theta) +  \mathcal{O}(\alpha^4) \, , \label{eqn: Electric angular eigenstate order2} \\[4pt]
\Delta S(\theta) &= \Big( \alpha^2 \tilde{a}^2  b_{j-2}  + \alpha^3 \tilde{a}^3 c_{j-2}  \Big) \, P_{j-2, m}(\cos \theta) + \Big( \alpha^2 \tilde{a}^2 b_{j+2}  + \alpha^3 \tilde{a}^3  c_{j+2} \Big)  \,  P_{j+2, m}(\cos \theta) \, , 
\end{aligned}
\eeq
where the coefficients $b_{j \pm 2}$ and $c_{j \pm 2}$ are given in Appendix~\ref{app:details}. Strictly speaking, $j$ is no longer a good quantum number at order $\alpha^2$. However, as we discussed in \S\ref{sec:TensorKerr}, an approximate notion of total angular momentum still exists---especially in the $\alpha \to 0$ limit---and we continue to label our states with $j$, even though it has no precise physical meaning. 
The angular eigenvalues $\lambda$ for $j = \ell \pm 1$, expanded up to order $\alpha^3$, are also given in 
Appendix~\ref{app:details}. 
Substituting these results for $\lambda$ into the radial equations allows us solve for the energy eigenvalues at high orders.

\vskip 4pt
It is also instructive to compute the higher-order corrections to the actual vector field configuration (\ref{eqn:Aisol}) in the far zone 
\beq
A^i_{\indlab{1}} \propto r^{-1}R_0^{\rm far} \, i \tilde{a} \alpha   \,  \left(   \cos \theta \,  \epsilon^{ikl} r^k  \partial^l \, Y_{jm} -  \lambda_0\sin \theta \, Y_{jm} \, \hat{\phi}^i \right) e^{- i \omega t} \, . \label{eqn:AisolCorrection}
\eeq
Despite the presence of a magnetic vector spherical harmonic in (\ref{eqn:AisolCorrection}),
the vector field configuration is still of the electric type, since the $\cos \theta$ factor acquires a factor of $(-1)$ under a parity transformation. The presence of the $\hat{\phi}^i$-term in (\ref{eqn:AisolCorrection}) is a manifestation of the fact that spherical symmetry is already broken at this order in $\alpha$.

\paragraph{$\boldsymbol{\ell = 0}$ modes} 
We first consider the dominant growing mode, with $j=1$ and $\ell=0$. Since the radial wavefunctions are peaked near the horizon, they are most sensitive to strong gravity effects. This sensitivity manifests itself through the failure of ordinary perturbation theory, which diverges for this mode. It is not clear how to regulate these divergences, and thus this technology of matched asymptotic expansions is necessary to derive this mode's fine and hyperfine structure.  The matching procedure at higher orders is the same as that illustrated in \S\ref{sec: Scalar Relativistic Corrections}, except that the matching between the near and far-zone solutions must be performed through the intermediate zone. 
 We find that these higher-order intermediate-zone solutions take the form of simple linear combinations of powers and logarithms of $y$, which can be matched with the corresponding terms found in the asymptotic expansions of the near and far-zone solutions.
Performing the matched asymptotic expansion up to third order, we find 
\beq
\begin{aligned}
E_{n01m}  & =   \mu \left( 1 -\frac{\alpha^2}{2n^2} - \frac{\alpha^4}{8n^4} +  \frac{\left(  6 -10n \right) \alpha^4 }{3n^4}  \, + \frac{8  \tilde{a} m \alpha^5}{3n^3} \right)  .\label{eqn:VectorMain}
\end{aligned}
\eeq
Since the vector field has intrinsic spin, the $\ell=0$ modes now have hyperfine splittings $\propto \tilde{a} m \es \alpha^5$.

\paragraph{$\boldsymbol{\ell \neq 0}$ modes}  Since the $\ell \neq 0$ modes peak in the far zone, the calculation of the higher-order corrections of their spectra can be simplified by naively extrapolating the far-zone solution (\ref{equ:Far0}) towards the horizon $x \to 0$ and imposing regular boundary conditions there.
The far-zone radial function and the eigenvalues are then solved as an expansion in powers of $\alpha$. This leads to the result (\ref{eqn:vectorspectrumGeneral}) for the energy eigenvalues. Remarkably, we find that (\ref{eqn:vectorspectrumGeneral}), obtained through this simplified treatment, agrees with (\ref{eqn:VectorMain}), which was derived more rigorously through matching (see Appendix~\ref{app:details} for a more detailed discussion).

\subsubsection{Magnetic Modes} \label{sec:MagneticVector} 

In principle, the task of solving for the magnetic mode spectrum involves a straightforward application of the machinery developed in previous sections to the relevant separated radial and angular equations. Unfortunately, these equations are not yet known, except at linear order in $\tilde{a}$~\cite{Pani:2012vp,Pani:2012bp}. In this section, we discuss the extent to which the 
approximate radial equation (\ref{eqn:MagneticRadial}) can be used to derive the energy spectrum and instability rates. Our discussion will be brief and mostly qualitative, since our solutions are similar to those found in previous sections.

\begin{figure}
  \begin{center}
     \makebox[\textwidth][c]{\hspace{-0.325cm}\includegraphics[]{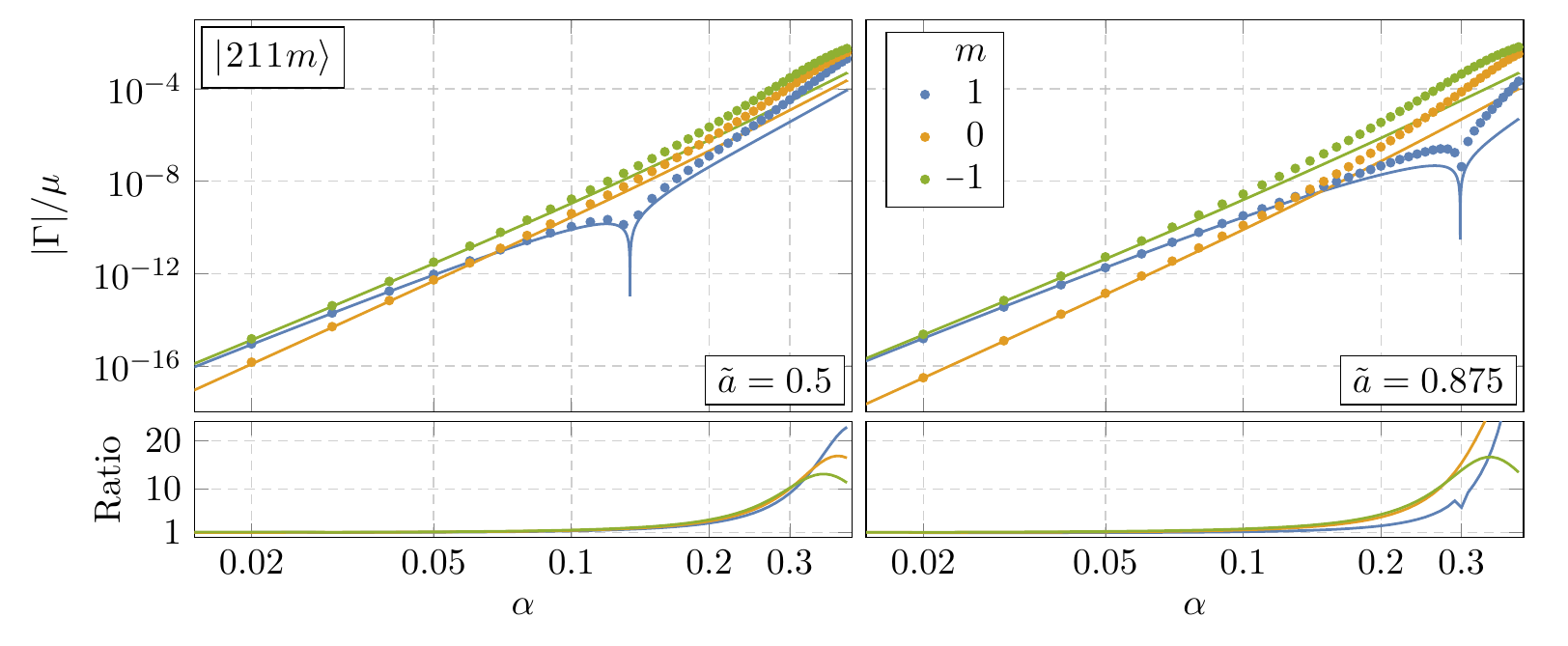}}
    \caption{Instability rates for the magnetic mode $|\es 2 \es 1 \es 1 \es m\rangle$, for $\tilde{a} = 0.5$ (left) and $\tilde{a} = 0.875$ (right). We compare our numeric results (denoted by points) with the conjectured form of the instability rate (solid lines), and plot the ratio of  numerics to analytics in the lower panels.     \label{fig:magInstability}}
  \end{center}
\end{figure}

\subsubsection*{Leading-order solution}

In the far region, the electric  (\ref{eqn:VectorFarF}) and magnetic (\ref{eqn:MagneticRadial}) radial equations are of the same form at leading order,
with $\lambda_0(\lambda_0 + 1)$ replaced with $j(j+1)$~\cite{Rosa:2011my, Pani:2012bp}. The magnetic quasi-bound states therefore take the same form 
as their electric counterparts in the far region (\ref{eqn:Aisol}), and 
\begin{equation}
A^i_\indlab{0} \propto r^{-1} R_0^\lab{far}(r) \, Y_{j,\es j \es m}^i(\theta, \phi) \,e^{-i \omega t}\,.
\end{equation} 
As expected, 
these magnetic modes are therefore also hydrogenic~\cite{Endlich:2016jgc,Baryakhtar:2017ngi} and the spectrum is Bohr-like to leading order in $\alpha$.

\vskip 4pt
In the previous sections, we derived the leading-order instability rates by finding the near- and far-zone solutions, and matched 
them in the overlap region. One might hope to do the same with~(\ref{eqn:MagneticRadial}) to derive the magnetic instability rates to linear order in $\tilde{a}$. However, 
since the radial equation was derived in the far region, extending it into the near region is not as innocuous as it naively seems. This is because the near region---and especially the boundary condition at the outer horizon---is sensitive to the full nonlinear spin-dependence of the metric. As we discussed in Footnote~\ref{footnote:FixedPp}, getting these important terms wrong in our perturbative expansion can cause our approximation 
to deviate strongly from the actual solution, and it is not clear how to infer the correct behavior from a far-zone solution at linear order in $\tilde{a}$.\footnote{The authors of \cite{Pani:2012bp,Pani:2012vp} were able to derive an accurate expression for the magnetic instability rate to linear order in $\tilde{a}$, by matching the solution in the far region to a near-zone solution that is obtained by extrapolating (\ref{eqn:MagneticRadial}) toward the outer horizon $r_+ = 2 \alpha$. However, they needed to discard terms like $\check{P}_+^2$, which contains both constant and linear-in-$\tilde{a}$ pieces, to obtain a finite result. In effect, discarding these divergent terms 
imposes the correct near-horizon behavior.} 

\vskip 4pt
We will instead guess the form of the magnetic instability rate and check this guess against our numerics. 
It is natural to assume that the magnetic instability rates take the same functional form as the rates for the scalar field~(\ref{eqn:ScalarRate}) and the electric modes of the vector field~(\ref{eqn:VectorRates}). The overall normalization and the dominant $\alpha$-scaling can be fixed by demanding that this ansatz matches the Schwarzschild limit~\cite{Rosa:2011my}, and thus we arrive at the conjectural instability rate~(\ref{eqn:VectorRates}) with~(\ref{eqn:VectorRatesCoeff}). 
As we show in Fig.~\ref{fig:magInstability}, we find our guess to be in excellent agreement with our numeric results for small $\alpha$, even at high values of $\tilde{a}$.  However, we emphasize that this formula represents our best educated guess for the magnetic instability rate at arbitrary spin and is {\it not} rigorously derived. A conclusive analytic result can only be found by solving a separable equation valid for all $\tilde{a}$.

\subsubsection*{Higher-order corrections}

\begin{figure}
  \begin{center}
    \includegraphics{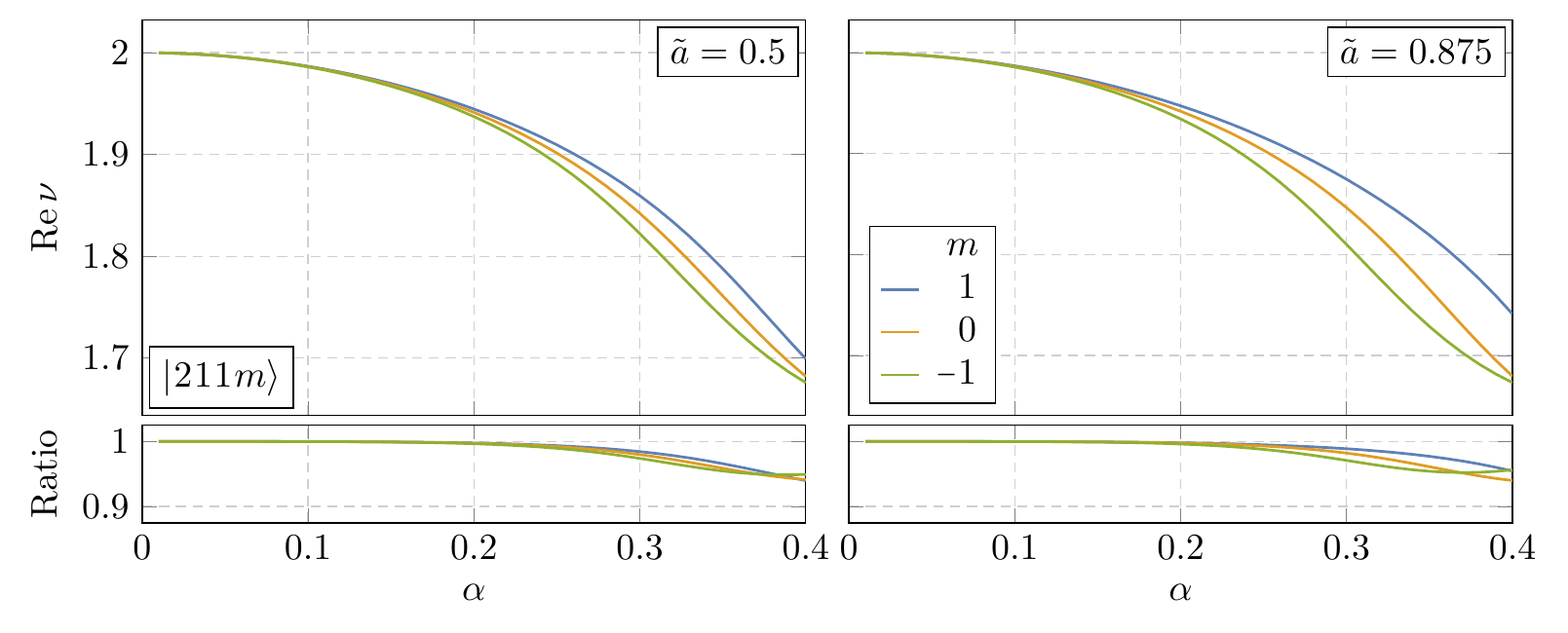}
    \caption{\label{fig:magneticReal} Numeric results for the spectra of the magnetic mode $|\es 2 \es 1 \es 1 \es m \rangle$, for $\tilde{a} = 0.5$ (left) and $\tilde{a} = 0.875$ (right). The ratios of the numeric results to their perturbative predictions (\ref{eqn:vectorspectrumGeneral}) are shown in the panels below.} 
  \end{center}
\end{figure}

Since the linear-spin approximation 
does not fully capture the leading-order behavior of the near region, we cannot use (\ref{eqn:MagneticRadial}) to perform a matched asymptotic expansion at high orders in $\alpha$. Fortunately, the magnetic modes have non-vanishing orbital angular momentum and are thus peaked far away from the horizon. 
 As we have illustrated for both the scalar and the electric modes of the vector, we may derive the energy spectrum of these modes to $\mathcal{O}(\alpha^3)$ by extending the far-zone radial solutions (\ref{equ:Far0}) toward the horizon, imposing regular boundary conditions, and solving for $\nu$ perturbatively in powers of $\alpha$. This then yields the fine and hyperfine structure of the magnetic modes to linear order in $\tilde{a}$. However, following the pattern suggested by the electric modes in the spectrum, we expect the fine structure to be independent of $\tilde{a}$, and the hyperfine structure to be proportional to $m \tilde{a}$,
so that our results for the magnetic energy spectrum should be valid for arbitrary spin. 
In Figure~\ref{fig:magneticReal}, we compare the approximation~(\ref{eqn:vectorspectrumGeneral}) to our numeric results and find that it is very accurate even at large $\tilde{a}$, suggesting that our extrapolation of the fine and hyperfine structure to arbitrary $\tilde{a}$ is correct.

\newpage
\section{Numerical Computation of the Spectra}
\label{sec:numeric}

The analysis of the previous section required $\alpha$ to be small, and we derived rigorous results only for the scalar field and the electric modes of the vector field.
In this section, we numerically solve for the quasi-bound state spectrum, providing results for arbitrary values of $\alpha$ and $\tilde{a}$, including the magnetic modes of the vector field. We will thus be able to use these results to determine when the perturbative approximations summarized in \S\ref{sec:Summary} break down.
In~\S\ref{sec:ScalarNumeric}, we first illustrate our approach using the scalar field. We discuss why precise results for $\alpha \ll 1$ are difficult to achieve numerically and how to surmount these difficulties. In  \S\ref{sec:VectorNumeric}, we then solve the analogous problem for the vector field.

\subsection{Massive Scalars around Kerr}
\label{sec:ScalarNumeric} 
  
The quasi-bound state spectrum of Klein-Gordon fields in the Kerr background has been studied in many previous works; see e.g.~\cite{Dolan:2007mj,Dias:2015nua}.
To make contact with the literature, we will begin by reviewing the continued fraction method for determining the scalar spectrum (\S\ref{sec:contFrac}). We will discuss the main limitations of the method and show how they can be overcome by reformulating the problem as a nonlinear eigenvalue problem (\S\ref{sec:NLEP}).
We then use this to solve for the scalar spectrum (\S\ref{sec:Separable}).

   \subsubsection{Continued Fraction Method} \label{sec:contFrac}
    
    In \S\ref{sec:scalarintro}, we expanded the field $\Phi$ in spheroidal harmonics and obtained the radial differential equation (\ref{eqn: scalar radial equation}). For the numerical analysis, it will be convenient to write the radial function as
      \begin{equation}
        R(r) = \frac{(r - r_+)^{i P_+}}{(r - r_-)^{i P_-}} B(x)\, , \label{eq:contFracAnsatz}
      \end{equation}
      where $x = 2\sqrt{1-\omega^2}(r -r_+)$, as in \S\ref{sec:MatchedScalar} and (\ref{eqn: x variable}). For quasi-bound state solutions, $B(x)$ approaches a constant at the horizon, $x \to 0$, and decays exponentially $B(x) \sim e^{-x/2}$ at spatial infinity, $x \to \infty$. Any function with these properties can be represented by a linear combination of associated Laguerre polynomials $L_{k}^{(\rho)}(x)$ multiplied by $e^{-x/2}$,
      \begin{equation}
        B(x) = \sum_{k = 0}^{\infty} b_k \,e^{-x/2} L_{k}^{(\rho)}(x)\,. \label{eq:lagExp}
      \end{equation}
      In general, the expansion in (\ref{eq:lagExp}) transforms the scalar radial equation (\ref{eqn: scalar radial equation}) into a five-term recursion relation for the coefficients $b_k$. However, by choosing $\rho = 2 i P_+$, 
      one obtains the three-term recursion relation
      \begin{equation} 
        \alpha_{k} b_{k+1} + \beta_{k} b_{k} + \gamma_{k} b_{k - 1} = 0 \,,\label{eq:recurrence}
      \end{equation}
      where,  in the notation of \S\ref{sec:scalarintro}, we have defined
      \beq
    \begin{aligned}
      \alpha_{k} &\equiv (k+1) (k - c_1 + c_2 + 2 + 2 i P_+ - 4 \gamma ) \, , \\
      \beta_{k} &\equiv -2 k^2 + \left(c_1 - 2 (1 + c_2 + 2 i P_+)\right)k - c_2(1 + 2 i P_+) - c_3\, , \\
      \gamma_{k} &\equiv (k+2 i P_+)(k+ c_2 - 1)\, ,
    \end{aligned}
    \eeq
     with 
      \beq
    \begin{aligned}
    \hspace{0.7cm}  c_1 &\equiv 2(1 + i(P_+ - P_-) - 2 \gamma)\, ,\\
      c_2 &\equiv 1 + i(P_+ - P_-) + \frac{1}{4 \gamma}(\gamma_+^2 - \gamma_-^2) \, , \\
      c_3 &\equiv \left(P_+ - P_-\right)^2 - i(P_+ - P_-) + 2 \gamma(1+2 i P_+) + \gamma^2 + \gamma_+^2 + \Lambda\, .
    \end{aligned}
    \eeq
   For any initial data $b_{0}$ and $b_{1}$,\footnote{By convention, we define $b_{-1} \equiv 0$.} we can iteratively solve (\ref{eq:recurrence}) for  $b_{k}$ and thereby find a solution to the radial equation (\ref{eqn: scalar radial equation}).\footnote{An alternative recursion relation was derived in \cite{Dolan:2007mj} by performing a different mode expansion. Although the detailed forms of these coefficient functions change if we use this expansion, our discussion does not.} However, for generic values of $\omega$, solutions to (\ref{eq:recurrence}) diverge as $k \to \infty$, so that $B(x)$ will grow---rather than decay---exponentially at spatial infinity. This is the discrete analog of the fact that the boundary conditions (\ref{eqn:bc}) can only be satisfied simultaneously at special values of $\omega$ and that, for generic $\omega$, the solutions that satisfy the ingoing boundary condition at $r=r_+$ diverge as $r \to \infty$. 

\vskip 4pt
    It is only for special values of $\omega$---the quasi-bound state frequencies---that the recurrence relation (\ref{eq:recurrence}) admits a \emph{minimal solution}, which is finite both as $k \to 0$ and $k \to \infty$ \cite{Gautschi:1967cat}. Denoting the two linearly-independent solutions of (\ref{eq:recurrence}) as $f_k$ and $g_k$, the solution $f_k$ is  minimal  if
    \begin{equation}
      \lim_{k \to \infty} \frac{f_k}{g_k} = 0\, .
    \end{equation}
    A theorem by Pincherle \cite{Gautschi:1967cat} states that the recurrence relation (\ref{eq:recurrence}) admits a minimal solution if and only if $\omega$ solves the continued fraction equation 
    \begin{equation}
      \frac{\beta_0}{\alpha_0} = \cfrac{\gamma_1}{\beta_1 - \cfrac{\alpha_1 \gamma_{2}}{\beta_{2} - \cfrac{\alpha_{2} \gamma_{3}}{\beta_{3} - \cdots}}}\, . \label{eq:contFrac}
    \end{equation}
Determining the quasi-bound state spectrum is therefore reduced to the much simpler problem of finding the roots of a transcendental equation. This method has been used to find the quasi-normal modes of massless perturbations about Kerr black holes~\cite{Leaver:1985ax} and the quasi-normal modes and quasi-bound state spectra of massive fields~\cite{Dolan:2007mj, Rosa:2011my, Pani:2012bp, Dolan:2015eua}.  

\begin{figure}
      \begin{center}
        \includegraphics[scale=1, trim = 0 0 0 0]{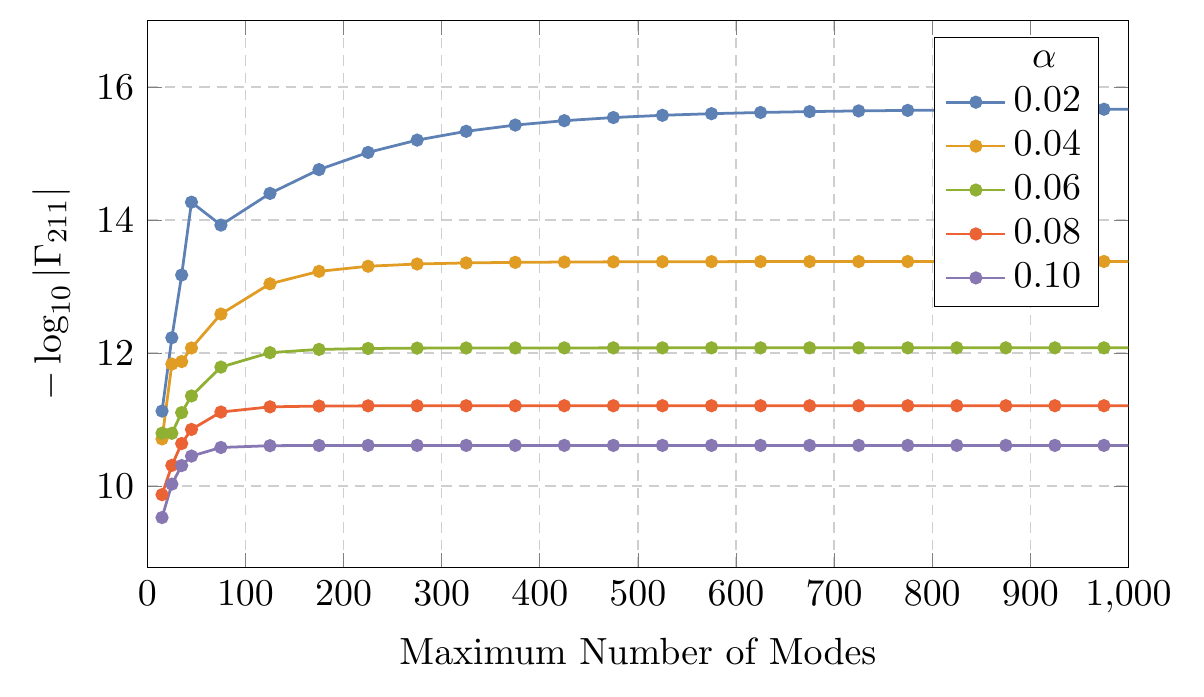}
        \caption{The imaginary part of $\omega$ slowly converges for small $\alpha$, requiring about $1000$ modes (terms in the continued fraction) to achieve accurate results for $\alpha \sim 0.02$. The results shown in the figure are for the $\ell = 1$, $m = 1$ mode, with $\tilde{a} = 0.5$, but similar conclusions apply to other states. Identical results apply to the recursion relation used in \cite{Dolan:2007mj}. \label{fig:contFracResults} }
      \end{center}
    \end{figure}

\vskip 4pt
    As will become apparent, sensitivity to numerical errors is a common challenge in finding the 
     spectrum for a massive field about Kerr, and it is important to use methods that tamp down numerical errors as much as possible. For instance, the fact that the coefficients $b_k$ diverge exponentially quickly when $\omega$ is not exactly a quasi-bound state frequency means that, unless properly accounted for, numerical errors will grow at the same rate. Solving for $\omega$ typically involves starting with an initial seed and stepping towards the solution through a sequence of intermediate frequencies, and large errors introduced during these intermediate steps can impact the accuracy of the `solution.'
    As seen in Fig.\,\ref{fig:contFracResults}, one needs to include several hundred of the coefficients $b_k$ to accurately compute the decay rates $\Gamma_{n\ell m}$, so this exponential growth in error---which scales with the number of coefficients---can be an enormous problem. The major advantage of phrasing this as the continued fraction (\ref{eq:contFrac}) is that there exist numerically robust methods---for instance, the modified Lentz method \cite{Press:2007nr}---for efficiently and accurately evaluating the continued fraction to a specified precision, which can then be passed to standard root-finding methods (i.e.~Newton-Raphson) to find the eigenfrequencies.

   \vskip 4pt
  The main issue with the continued fraction method is its rigidity---one must first separate the partial differential equation and then find a basis of functions, like $e^{-x/2} L_{k}^{(2 i P_+)}$, that reduces the scalar radial equation (\ref{eqn: scalar radial equation}) to a three-term recurrence relation.\footnote{One can sometimes convert a five-term recurrence relation into a three-term relation using Gaussian elimination~\cite{Cardoso:2005vk}. However, it will be more convenient to bypass (\ref{eq:contFrac}) entirely. } This is problematic for two reasons. The first reason is simply that it is not always possible to find such a basis---for instance, it is not clear that there exists a separable ansatz for the magnetic modes of a vector field, or that there exists a basis of functions that reduces (\ref{eqn:ProcaR}) to a relation like (\ref{eq:recurrence}).  The second reason is more subtle. 
    We have seen, in Fig.\,\ref{fig:contFracResults}, that several hundred modes are needed to achieve accurate results for the growth rate of the main superradiant mode $|\es2\es1\es1\rangle$. But the radial profile of this mode does not oscillate wildly and is instead quite smooth. So, why are all of these modes needed? The problem is that, as we discussed in \S\ref{sec:scalarintro}, the solution in the near region varies on extremely short scales, $\Delta x \sim \alpha^2$, while the basis functions $e^{-x/2} L_{k}^{(2 i P_+)}(x)$ naturally vary on much larger scales, $\Delta x \sim 1$. A large number of modes are therefore needed to approximate this (relatively) rapid behavior in the near region, and the number of required modes increases the smaller this region gets.  As we will discuss shortly, the numerical error often scales with the size of a problem, and thus it can be computationally difficult to access reliable results at small $\alpha$ with these methods.

\subsubsection{Nonlinear Eigenvalue Problem}  
\label{sec:NLEP}
    
The rigidity of the continued fraction method presents a serious obstacle to efficiently and accurately computing the quasi-bound state spectrum of scalar and vector fields. However, we can make progress by recognizing that the recurrence relation (\ref{eq:recurrence}) is analogous to the infinite-dimensional matrix equation
        \begin{equation}
          \mathcal{M}(\omega)\, \mb{b} = 
          \begin{pmatrix} 
            \beta_0 & \alpha_0 & 0 & 0 & 0 &\dots\\
            \gamma_1 & \beta_1 & \alpha_1 & 0 & 0 &\dots\\
            0 & \gamma_2 & \beta_2 & \alpha_2 & 0 &\dots\\
            0 & 0 & \gamma_3 & \beta_3 & \alpha_3 &\dots\\
            \vdots & \vdots & \vdots & \vdots & \vdots & \ddots
          \end{pmatrix}
          \begin{pmatrix}
            \,b_0 \,\\
            b_1 \\
            b_2 \\
            b_3 \\
            \vdots
          \end{pmatrix} = 0\,, \label{eq:nlev}
        \end{equation}
        whose elements are nonlinear functions of $\omega$. When we decompose $B(x)$ onto a basis of functions that individually satisfy the boundary conditions (\ref{eqn:bc}) and then truncate to a finite set of size $N+1$, this becomes a \emph{nonlinear eigenvalue problem}, i.e. we must find pairs $\omega$ and $\mb{b}$ such that (\ref{eq:nlev}) is satisfied~\cite{Dias:2015nua}. For a recent review of the numerical techniques developed to attack these problems, see \cite{Guttel:2017cup}. 

\vskip 4pt

       There are two sources of error in using (\ref{eq:nlev}) to solve for $\omega$. The first is truncation error---in order to actually solve this matrix equation on a computer, we must truncate the representation of $B(x)$ to a finite number of coefficients. We similarly needed to truncate the number of `levels' in the continued fraction (\ref{eq:contFrac}) to actually evaluate it. This truncation introduces an error, as $B(x)$ is now \emph{approximated} by a smaller collection of functions. As we increase the size of this collection, we are able to more faithfully represent $B(x)$ and, as seen in Fig.\,\ref{fig:contFracResults}, this truncation error will decrease. We can then estimate the accuracy of our solution by how sensitive it is to changes in~$N$.
      The second source of error is numerical error. Floating point arithmetic is inherently noisy, as there are round off errors incurred after every operation (e.g.~addition and multiplication), and this noise can be amplified by careless numerics. Unfortunately, this error grows with the number of operations performed, and thus with the number of coefficients we include in (\ref{eq:nlev}). This is potentially disastrous, since, if we try suppress truncation error by increasing $N$, we could be overrun by this numerical error. We will specifically choose a representation of (\ref{eq:nlev}) to help sooth these numeric problems. In the following, we will motivate and explain these choices, and discuss them in more detail in \S\ref{app:cheb}.
        
\vskip 4pt
        Regardless of representation, numerical errors can also creep in when we try to numerically solve (\ref{eq:nlev}), and it is important to use methods that avoid such instabilities. For instance, perhaps the simplest way to determine the quasi-bound state spectrum is to find the roots of the equation $\lab{det}\,  \mathcal{M}(\omega) = 0$. Unfortunately, this is not feasible for large $N$. In order for the determinant to vanish, there must be sensitive cancellations between a large number of operations, and accumulated roundoff errors can totally destroy the accuracy of our solution. This is a common problem for any nearly-singular matrix---a rule of thumb is that relative errors are amplified by the so-called condition number, the ratio of the matrix's largest and smallest eigenvalues, and this condition number diverges for exactly the frequencies we are solving for. For small $\alpha$ and large matrices, this numerical error easily dominates the instability rates and energy splittings. 

\vskip 4pt
        We will instead use nonlinear inverse iteration \cite{Guttel:2017cup,Dias:2015nua}, a form of Newton's method applied directly to $\mathcal{M}(\omega)\mb{b} = 0$ that iteratively solves for both $\omega$ and $\mb{b}$. This method circumvents the numerical instability caused by these nearly-singular matrices because this error is amplified much more along the singular direction---i.e. the one we are interested in---than any other. That is, these errors only change the length of the solution $\mb{b}$ and not its direction, and so their effect is nullified. In practice, this method converges both quickly and accurately as long as one has a good initial guess for the pair $(\omega, \mb{b})$.

      \subsubsection{Chebyshev Interpolation}
\label{sec:Separable}

We will now apply the algorithm sketched in the previous subsection to the case of the scalar field in the Kerr background. As shown in \S\ref{eqn:ScalarAnsatz}, the scalar modes are separable into radial and angular functions.
It is convenient to write the radial function as
        \begin{equation}
        R(r) = \left(\frac{r - r_+}{r - r_-}\right)^{i P_+} \!\!\!(r - r_-)^{-1+ \nu - 2 \alpha^2/\nu} e^{-\alpha(r-r_+)/\nu} B(\zeta)\,, \label{eq:pseudoSpecPeelOff}
        \end{equation}
      where the asymptotic behavior shown in  (\ref{eqn:scalar radial outer horizon}) and (\ref{eqn: scalar radial infinity}) has been extracted explicitly. The remaining function $B(\zeta)$ is defined on the finite interval $\zeta \in [-1, 1]$ via a map $\zeta(r)$. Different choices of $\zeta(r)$ will be discussed below. We will work with $\nu$, defined in (\ref{eqn:nudef}), instead of $\omega$. The function $B(\zeta)$ satisfies a linear differential equation of the form
      \begin{equation}
        \mathcal{D}_{\nu}[B(\zeta)] \equiv \left(\frac{\partial^2}{\partial \zeta^2} + \mathcal{C}_1(\nu, \zeta) \frac{\partial}{\partial \zeta} + \mathcal{C}_2(\nu, \zeta) \right) B(\zeta) = 0\,, \label{eq:zetaEq}
      \end{equation}
     where the precise form of the functions\footnote{These functions also depend on $\alpha$, $\tilde{a}$, and the orbital and azimuthal quantum numbers
     $\ell$ and $m$, respectively. We will suppress this dependence, as it does not play a crucial role in our story.} $\mathcal{C}_i(\nu, \zeta)$ depend on our choice of $\zeta(r)$. 
 The function $B(\zeta)$ will satisfy the correct boundary conditions if it approaches a constant 
 at both the outer horizon ($\zeta = -1$) and spatial infinity ($\zeta = 1$). Because~(\ref{eqn: scalar radial equation}) has no singularities between the outer horizon and spatial infinity, $B(\zeta)$ is necessarily a smooth function for all $\zeta \in [-1, 1]$, and we may represent it in a variety of ways. For computational simplicity, we will use Chebyshev polynomials of the first kind---defined by $T_{n}(\cos t) = \cos n t$ and described at length in \S\ref{app:cheb}---but our conclusions will largely be independent of this choice.

\vskip 4pt
    Above, we derived the recurrence relation (\ref{eq:recurrence}) by projecting both $B(x)$ and the radial equation~(\ref{eqn: scalar radial equation}) onto Laguerre functions.  We could now mimic this procedure 
    by first constructing a \emph{polynomial projection} of $B(\zeta)$ in terms of the Chebyshev polynomials,
    \begin{equation}
      B_N(\zeta) = \sum_{k = 0}^{N} b_k T_k(\zeta)\,, \label{eq:bWrong}
    \end{equation}
    which is a degree-$N$ polynomial that 
     is guaranteed to converge $\lim_{N \to \infty} B_{N}(\zeta) = B(\zeta)$, since $B$ is smooth on $\zeta \in [-1, 1]$ and the Chebyshev polynomials form a complete set. 
  Projecting the radial equation (\ref{eq:zetaEq}) onto the Chebyshev polynomials, we would obtain  the matrix equation $\sum_k \mathcal{M}_{n k}(\nu) b_k = 0$,
where\hskip 1pt\footnote{This formula must be multiplied by a factor of $1/2$ when $n = 0$.}
    \begin{equation}
      \mathcal{M}_{n k}(\nu) \equiv \frac{2}{\pi} \int_{-1}^{1}\!\ud \zeta\, \frac{T_{n}(\zeta)\, \mathcal{D}_\nu[T_{k}(\zeta)]}{\sqrt{1- \zeta^2}}\,. \label{eq:specMatrix}
    \end{equation}
    The matrix given by \eqref{eq:specMatrix} is similar to that in (\ref{eq:nlev}), though it has many more non-zero elements. We could again pass this to a solver to determine the quasi-bound state spectrum, but unfortunately this formulation of the problem proves (cf.~\S\ref{app:cheb}) to be numerically unstable, it is difficult to reduce truncation errors without numerical errors biting back. 

   \vskip 4pt
   To circumvent this problem, we will represent $B(\zeta)$ not by its Chebyshev coefficients, but by its values at the \emph{Chebyshev nodes}
    \begin{equation}
      \qquad\qquad\qquad \zeta_n = \cos\left(\frac{\pi(2 n + 1)}{2 N + 2}\right) , \quad \text{where} \quad n = 0, 1, \dots, N\,. \label{eq:chebPoints}
    \end{equation}
Introducing the associated \emph{cardinal polynomials}  $p_k(\zeta)$---degree-$N$ polynomials that are defined by $p_k(\zeta_n) = \delta_{nk}$ (cf.~\S\ref{app:cheb})---we may rewrite (\ref{eq:bWrong}) as 
    \begin{equation}
      B_N(\zeta) = \sum_{k = 0}^{N} B(\zeta_k) \,p_k(\zeta)\,. \label{eq:bRight}
    \end{equation}
Note that this is nothing more than a reorganization of (\ref{eq:bWrong}) in a different basis of degree $N$ polynomials such that the values of $B(\zeta)$ appear explicitly. Furthermore, because the operator $\mathcal{D}_\nu[B(\zeta)]$ appearing in (\ref{eq:zetaEq}) also defines a smooth function on $\zeta \in [-1, 1]$, we can also represent it by its values at these specified points. This means that we can approximate (\ref{eq:zetaEq}) by the matrix equation 
    \beq
    \sum_{k = 0}^{N} \mathcal{M}_{n k}(\nu) B(\zeta_k) = 0\,,
    \eeq
     where
    \begin{equation}
       \mathcal{M}_{nk}(\nu) \equiv p_k^{\hskip 1pt \prime \prime}(\zeta_n) + \mathcal{C}_1(\nu, \zeta_n)\,p_k^{\hskip 1pt \prime}(\zeta_n) + \mathcal{C}_{2}(\nu, \zeta_n) \,\delta_{nk}\,.\label{eq:scalarMatrix} 
    \end{equation}
    In practice, this formulation proves to be both simpler and more numerically robust than (\ref{eq:specMatrix}), as long as $p_k^{\hskip 1pt \prime \prime}(\zeta_n)$ and $p_k^{\hskip 1pt \prime}(\zeta_n)$ are computed using (\ref{eq:cardDeriv}) and (\ref{eq:cardDDeriv}), respectively. See \S\ref{app:cheb} for more details.
    
\vskip 4pt
    The specific form of the mapping $\zeta(r)$ can dramatically affect the truncation error. Just like the convergence of a Laurent series about a point is set by the largest \emph{circular} domain of analyticity, the convergence of the interpolation (\ref{eq:bRight}) is set by the largest \emph{ellipsoidal} domain of analyticity about the interval $\zeta \in [-1, 1]$. Specifically, it can be shown (cf.~Appendix~\ref{app:cheb}) that the largest disagreement between the function and its Chebyshev interpolation anywhere on the interval scales as
    \begin{equation}
      \|B- B_N\| \sim \mathcal{O}\big(\rho^{-N}\big)\,. \label{equ:conv}
    \end{equation}
The parameter $\rho$ measures the size\footnote{Concretely, $\rho$ is the sum of the semi-major and semi-minor axes of the ellipse with foci at $\zeta = \pm 1$. The parametric form of this ellipse is given by (\ref{eq:bernsteinEllipse}) and it is depicted in Fig.~\ref{fig:bernstein}.} of the largest ellipse with foci at $\zeta = \pm 1$ inside which $B(\zeta)$ is analytic. The interpolation thus converges more slowly the closer a singularity of $B(\zeta)$ is to the interval $\zeta\in[-1,1]$, as measured by these ellipses.

\vskip 4pt
     The radial equation (\ref{eqn: scalar radial equation}) has an additional singularity at $r = r_-$, which 
     maps to a singularity in $B(\zeta)$ at $\zeta_- = \zeta(r_-)$. This is why our choice of $\zeta(r)$ can affect the truncation error. If this mapping does not place $\zeta_-$ far from the interval $\zeta \in [-1, 1]$, this singularity can drastically increase the number of modes $N$ needed to accurately approximate the solution. This is also why we avoided introducing factors of $r$ in the ansatz (\ref{eq:pseudoSpecPeelOff}), as they could introduce inessential singularities into $B(\zeta)$ and potentially affect convergence.
     
\vskip 4pt
     Our choice of mapping $\zeta(r)$ will also determine how the interpolation points (\ref{eq:chebPoints}) sample the radial domain $[r_+, \infty)$. As we have argued in \S\ref{sec:contFrac}, one of the reasons the Laguerre basis performed so poorly was that it naturally varies on scales much larger than the width of the near region and so it has trouble approximating the behavior there. This is easier to see if we note that this Laguerre basis samples the function $B(x)$ at the zeros of $L_{N+1}^{(2 i P_+)}(x)$ \cite{NIST:DLMF}, whose smallest root scales as $\mathcal{O}(N^{-1})$. We would thus need to include $\mathcal{O}(\alpha^{- 2})$ modes to accurately sample the boundary layer $x \sim \mathcal{O}(\alpha^2)$, in good agreement with the numerical experiments shown in Fig.\,\ref{fig:contFracResults}. One of the benefits of the Chebyshev basis is that the interpolation points (\ref{eq:chebPoints}) have a much higher density $\Delta \zeta \sim \mathcal{O}(N^{-2})$ near the boundary than in the interior $\Delta \zeta \sim \mathcal{O}(N^{-1})$, which makes it easier for the basis to resolve phenomena in the near region. Still, we must be careful that the mapping we choose does not obstruct this useful behavior.

\vskip 4pt
    The simplest choice that maps the triplet $r =(r_-, r_+, \infty)$ to $\zeta = (-\infty, -1, 1)$ is 
    \begin{equation}
      \zeta_1(r) = \frac{r - 2 r_+ + r_-}{r - r_-} \, . \label{equ:z1}
    \end{equation} 
A drawback of this mapping is that the Bohr peak at $r_\lab{c} \sim \nu/\alpha$ is mapped to $\zeta_1(r_\lab{c}) \sim 1 - \mathcal{O}(\alpha^2)$,
    while the middle of the interval $\zeta = 0$ corresponds to $(2 r_+ - r_-) = \mathcal{O}(\alpha)$. This creates the opposite problem as with the expansion into Laguerre polynomials---we are now sampling the near-horizon region very well, but at the expense of the far region. To ensure the far region is also well-sampled, we again require a relatively large number of modes $N \sim \alpha^{-1}$, although the properties of the Chebyshev nodes help alleviate this problem somewhat compared to the expansion (\ref{eq:lagExp}). In practice, this map works fairly well for the $n = \ell+1$ quasi-bound states, as they have very little structure in the far region.

\vskip 4pt
    An alternative mapping that avoids this problem is
    \begin{equation}
      \zeta_2(r) = \frac{r - \sqrt{4 r_+(r-r_-) + r_-^2}}{r- r_-}\,. 
      \label{equ:z2}
    \end{equation}
This maps the Bohr radius to $\zeta_2(r_\lab{c}) \sim 1 - \mathcal{O}(\alpha)$, which means that we only need $N \sim \mathcal{O}(\alpha^{-1/2})$ modes to resolve the far region.  While the singularity at $r_-$ is only mapped to the finite point ${\zeta_- = -(2 r_+ - r_-)/r_-}$, it is still displaced far enough from the interval that it does not dramatically affect convergence. For instance, $\rho > 2$ in (\ref{equ:conv}) as long as $\tilde{a} \lesssim 0.96$ and so $\zeta_2$ is an effective map except in the extremal limit. In principle, there are better maps that send $r_-$ infinitely far from the interval and more equally distribute the interpolation points (\ref{eq:chebPoints}) across the near, intermediate, and far regions. However, the map $\zeta_2$ works well enough in practice that we will not pursue others. 
    
\begin{figure}
      \begin{center}
        \includegraphics[]{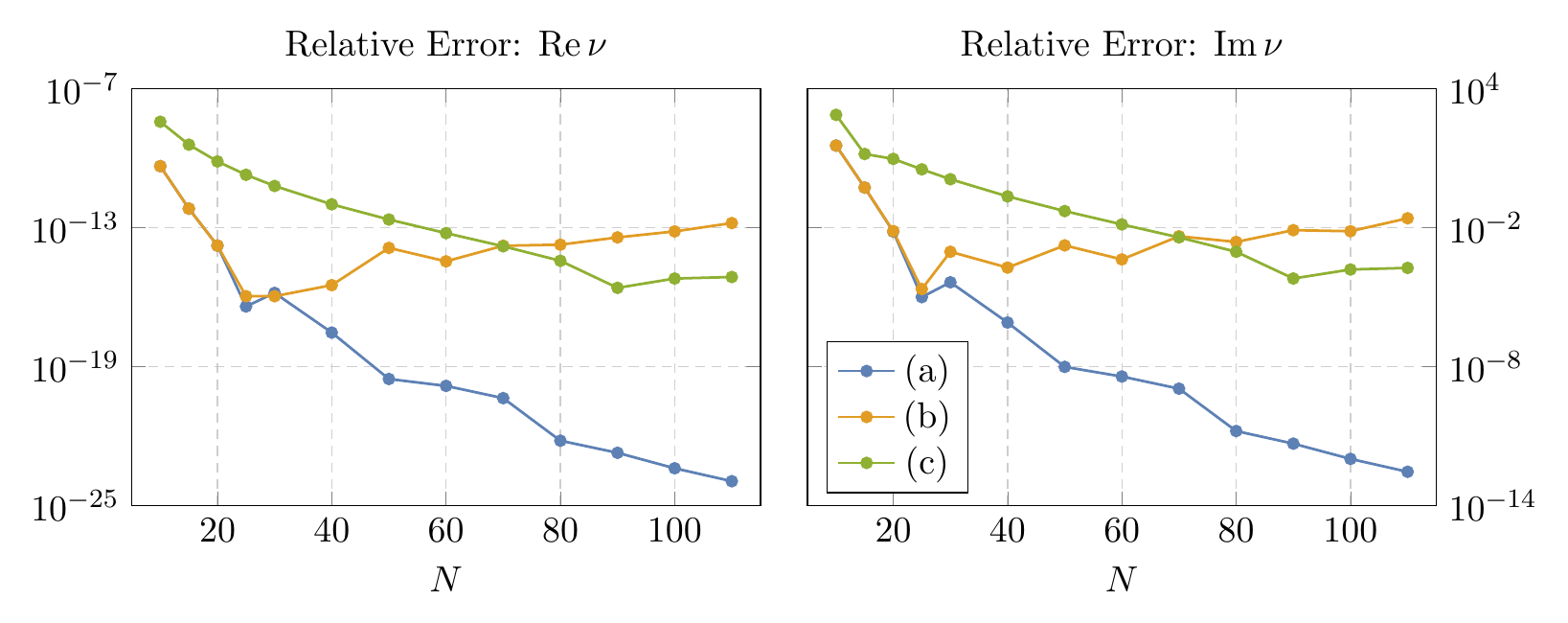}
        \caption{Comparison between the two mappings $\zeta_1(r)$ and $\zeta_2(r)$ defined in (\ref{equ:z1}) and (\ref{equ:z2}), as a function of the number of modes $N$. The displayed lines are:
    {(a)}\,{\color{Mathematica1}[blue]} $\zeta_2$ with 60 digits of precision, {(b)}\,{\color{Mathematica2}[orange]} $\zeta_2$ with machine precision, {(c)}\,{\color{Mathematica3}[green]} $\zeta_1$ with either 60 digits of precision or with machine precision. The data shown in the figures is for $\alpha = 0.01$, $\tilde{a} = 0.5$, and $|n \es \ell \es m \rangle = |2 \es 1 \es 1\rangle$. All relative errors are measured with respect to the high precision $\zeta_2$ result with $N = 120$.
\label{fig:errorsScalar}}
      \end{center}
    \end{figure}

\vskip 4pt
A comparison between the two maps $\zeta_1$ and $\zeta_2$ is shown in Fig.\,\ref{fig:errorsScalar}. We see that we can achieve very accurate results for the parameter $\nu$ using only $N \sim 30$ modes.  For comparison, reaching  a similar level of accuracy using the continued fraction method requires $N \sim 10^4$ modes. Clearly, the map $\zeta_2$ converges much more quickly than $\zeta_1$, and becomes limited by the resolution $\epsilon \sim 10^{-16}$ of machine precision numbers (the `machine epsilon') very quickly. 

\vskip 4pt
This method was used to determine the spectra of the $|\es2\es 1 \es m\rangle$ and $|\es 3 \es 2 \es m \rangle$ modes in Figures~\ref{fig:ScalarSpectrumPlots}~and~\ref{fig:growthRates}. In particular, the extremely small decay rates $\Gamma/\mu \sim 10^{-24}$ of the $|\es 3 \es 2 \es m\rangle$ modes are accurately computed using only $N= 60$. In contrast, the continued fraction method requires $N \sim 2\times 10^{4}$ modes to achieve comparable precision! With the scalar now solved, we can turn our attention to the vector.

\subsection{Massive Vectors around Kerr} 
\label{sec:VectorNumeric}  
  
The quasi-bound state spectrum of Proca fields in the Kerr background has been studied only relatively recently; see e.g.~\cite{Pani:2012vp, Pani:2012bp,East:2017mrj,Dolan:2018dqv,Baryakhtar:2017ngi,East:2017ovw,East:2018glu,Frolov:2018ezx,Endlich:2016jgc,Cardoso:2018tly}. Progress has been impeded by the fact that a separable ansatz is only known \cite{Frolov:2018ezx} for the electric modes of the vector field. 
We used this ansatz, in~\S\ref{sec: analytics vector},  to find perturbative results for both the energy splittings and growth rates for these electric modes. In \S\ref{sec:Sep}, we will show how to modify the techniques developed above for the separated equations (\ref{equ:ProcaS}) and (\ref{eqn:ProcaR}). This yields an efficient, accurate, and robust method for computing the spectra at arbitrary $\alpha$ and $\tilde{a}$. 
To attack the magnetic spectrum, we cannot rely on a separable ansatz. Fortunately, the techniques of the previous section are flexible enough not to rely on one. In \S\ref{sec:nosep}, we describe how to formulate the non-separated equations of motion for both the scalar and vector fields as nonlinear eigenvalue problems. This allows us to accurately and robustly determine the entire quasi-bound state spectrum for a Proca field around a Kerr black hole.

  \subsubsection{Using Separability } 
\label{sec:Sep}

     Clearly, the methods used in \S\ref{sec:Separable} also apply to the separated vector equation~(\ref{eqn:ProcaR}), albeit with one small wrinkle. Our discussion there relied on our ability to accurately compute the eigenvalues of the scalar angular equation (\ref{eqn: spheroidal harmonic equation}), and we skipped past this complication because various software packages already natively compute these quantities.\footnote{For instance, the eigenvalues of the spheroidal harmonics and their derivatives can be computed to arbitrary precision using Mathematica's \texttt{SpheroidalEigenvalue}.} 
     This is not the case for the Proca angular equation (\ref{equ:ProcaS}). 

     \begin{figure}
        \begin{center}
          \includegraphics{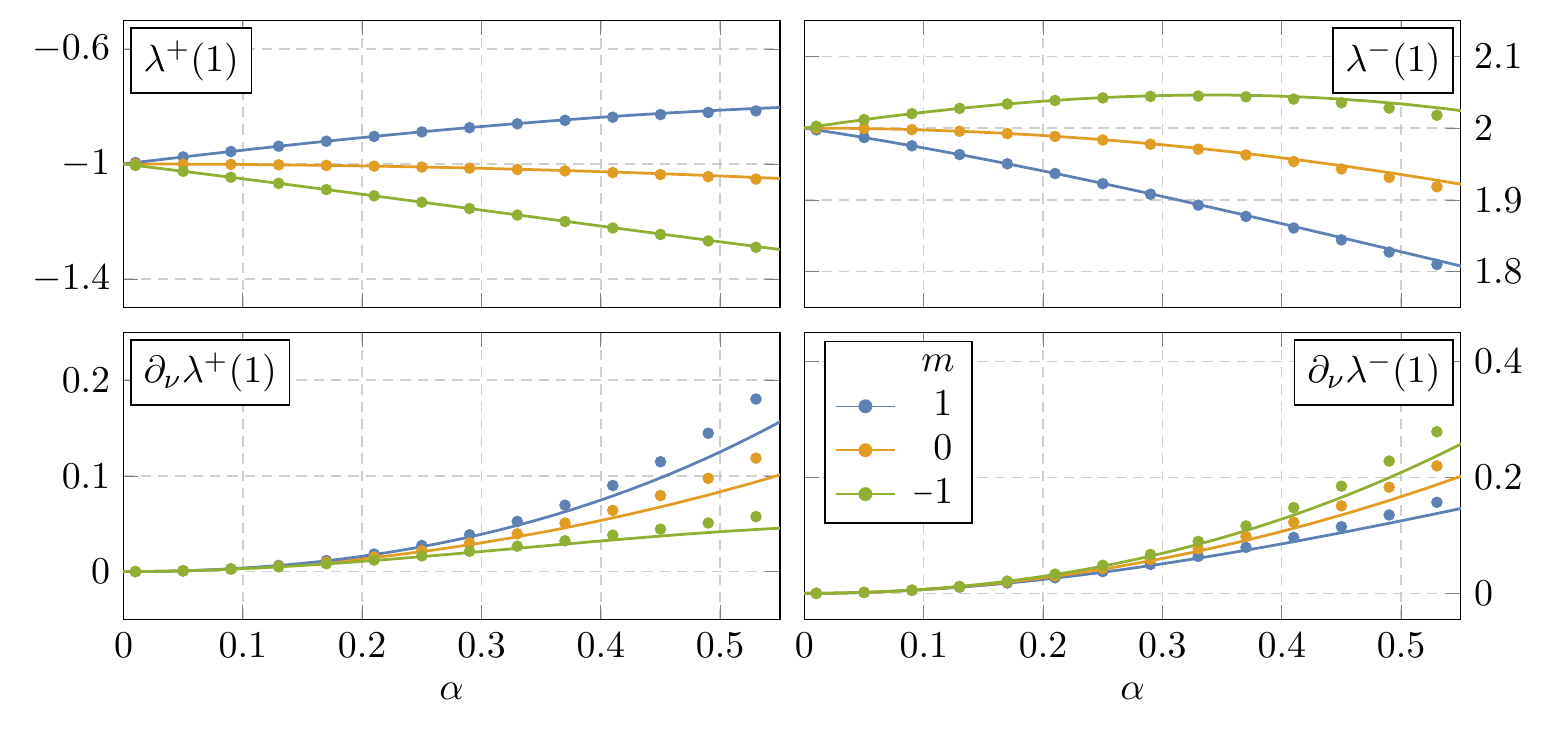}
          \caption{Comparison between the numeric (dots) and perturbative (solid) results for the angular eigenvalues $\lambda^{\pm}$ (top row) and their derivatives with respect to $\nu$ (bottom row) for $\tilde{a} = 0.5$, $j=1$, and $\nu =1$. The perturbative expressions (\ref{eqn:lambdaPlusMinus}) only compute the derivatives $\partial_\nu \lambda^{\pm}$ to next-to-leading order and thus deviate from the numeric results significantly at large $\alpha$. \label{fig:vectorAngular} }
        \end{center}
      \end{figure}

\vskip 4pt
     Fortunately, (\ref{equ:ProcaS}) is also a nonlinear eigenvalue problem which we can solve using the techniques of the previous section. Defining $\zeta = \cos \theta$ and expanding the angular function in the Chebyshev cardinal polynomials,
     \begin{equation}
        S(\zeta) = \sum_{k = 0}^{N} S(\zeta_k)\, p_k(\zeta)\,,
     \end{equation}
     we convert the angular equation (\ref{equ:ProcaS}) into another finite-dimensional matrix equation
     \begin{equation}
        \sum_{k = 0}^{N} \mathcal{A}_{n \es k}(\lambda, \nu) S(\zeta_k) \equiv \mathcal{A}(\lambda, \nu) \,\mb{S} = 0\,. \label{eq:angMatSol}
     \end{equation}
   For a given $\nu$, we can then pass the matrix $\mathcal{A}_{n \es k}(\lambda, \nu)$ to a solver to find~$\lambda$ as a function of~$\nu$. Furthermore, since nonlinear inverse iteration will require the derivative $\partial_\nu \mathcal{M}_{n\es k}(\nu)$ of the radial matrix, we must also find the derivative of the angular eigenvalues $\lambda'(\nu)$. By differentiating~$\lab{det}\, \mathcal{A}_{n\es k}(\lambda(\nu), \nu) = 0$ with respect to $\nu$, we can rewrite this derivative in terms of derivatives of the matrix $\mathcal{A}$,
     \begin{equation}
        \lambda'(\nu) = -\frac{\lab{tr}\left(\mathcal{A}^{-1}\, \partial_\nu \mathcal{A}\right)}{\lab{tr}\left(\mathcal{A}^{-1}\partial_\lambda \mathcal{A}\right)} = -\frac{\mb{S}_\lab{L}^\top \!(\partial_\nu \mathcal{A})\mb{S}^{\phantom{\top}}_\lab{R}}{\mb{S}_\lab{L}^\top \!(\partial_\lambda \mathcal{A})\mb{S}^{\phantom{\top}}_\lab{R}}\, ,
      \end{equation}
     where the second equality follows since the trace is dominated by the zero left and right eigenvectors, $\mb{S}_\lab{L}$ and $\mb{S}_\lab{R}$, when evaluated at the eigenvalue $\lambda(\nu)$. We can thus calculate $\lambda'(\nu)$ for very little additional computational cost. In Fig.\,\ref{fig:vectorAngular}, we compare our numeric results with their perturbative approximations (\ref{eqn:lambdaPlusMinus}).

\vskip 4pt
      With the angular eigenvalue in hand, the techniques discussed in \S\ref{sec:Separable} apply almost unchanged. We introduce a map $\zeta(r)$ from $r\in[r_+, \infty)$ to the finite interval $\zeta \in [-1, 1]$ and rewrite the radial function as
      \begin{equation}
        R(r) = \left(\frac{r - r_+}{r - r_-}\right)^{i P_+} \!\!\!\left(r - r_-\right)^{\nu - 2 \alpha^2/\nu} e^{-\alpha(r - r_+)/\nu} B(\zeta)\,,
      \end{equation} 
      so that $B(\zeta)$ approaches a constant at the endpoints of the interval $\zeta = \pm 1$. We need not worry about the poles at $r = \hat{r}_\pm$ in (\ref{eqn:ProcaR}), which will not affect convergence because $R(r) \sim C_\pm + D_\pm (r - \hat{r}_\pm)^2$ is analytic for $r \to \hat{r}_\pm$. Expanding in the Chebyshev cardinal polynomials $p_k(\zeta)$ and sampling the radial equation (\ref{eqn:ProcaR}) at the Chebyshev nodes, we find a finite-dimensional matrix $\mathcal{M}_{n \es k}(\nu)$ that can then be passed to a solver to determine the quasi-bound state spectrum.

\subsubsection{Without Separability } 
\label{sec:nosep}
  
Having rephrased the problem as a nonlinear eigenvalue problem affords us the flexibility to use a nonseparable ansatz. While this is mainly relevant for the Proca equation, for which a separable ansatz is not known in general, we will first illustrate the technique using the scalar field. It can then be applied to the vector case with minimal complication.

\subsubsection*{Scalar}

    In the Schwarzschild limit, $\tilde{a} \to 0$, spherical symmetry is restored in the Kerr geometry. Moreover, since $\tilde{a}$ always appear dressed by factors of $\alpha$, spherical symmetry is only weakly broken when $\alpha \ll 1$. This mean that, for $\alpha \ll 1$, we can use the weakly broken symmetry to organize our ansatz for the scalar field $\Phi$ and decompose $\Phi$ into scalar spherical harmonics. The general idea is to first decompose the Klein-Gordon equation into operators that act naturally on the scalar spherical harmonics, and then to use the methods of the previous section to convert it into a finite-dimensional matrix equation.

\vskip 4pt
     Our first step is to rewrite the Klein-Gordon equation using the isometries (\ref{eq:kerrIsometries}) and the total angular momentum operator $\pounds^2$, described in detail in Appendix \ref{app:harmonics}. One can show that the operator 
    \begin{equation}
      \left[\Sigma \nabla^2 + \pounds^2 - \left(\Sigma + \frac{2 \alpha r(r^2 + \alpha^2 \tilde{a}^2)}{\Delta}\right) \pounds_t^2 - \frac{\alpha^2 \tilde{a}^2}{\Delta} \pounds_z^2 - \frac{4 \alpha^2 \tilde{a} r}{\Delta} \pounds_t \pounds_z \right]\Phi = \partial_r\left(\Delta \partial_r \Phi\right) \label{eq:purelyRadial}
    \end{equation}
    is purely radial when acting on a scalar field $\Phi$. By assuming that $\Phi$ has definite frequency, $\pounds_t \Phi = - \omega \Phi$, and azimuthal angular momentum, $\pounds_z \Phi =  m \Phi$, the Klein-Gordon equation reduces to 
    \begin{align}
        0 \, =\ \, &\frac{1}{\Delta}\partial_r \left(\Delta \partial_r \Phi \right) - \frac{1}{\Delta}\left(\pounds^2 + \alpha^2 \tilde{a}^2 (1 - \omega^2) \cos^2 \theta\right) \Phi  \label{eq:startingPoint}\\
        &+ \left(-(1 - \omega^2) + \frac{P_+^2}{(r - r_+)^2} + \frac{P_-^2}{(r - r_-)^2} + \frac{A_{-}}{(r_+ - r_-) (r - r_-)} - \frac{A_+}{(r_+ - r_-)(r - r_+)}\right)\Phi\, \nonumber 
    \end{align}
    with minimal effort.
    We recognize the spheroidal harmonic equation (\ref{eqn: spheroidal harmonic equation}) 
    \begin{equation}
      \left(\pounds^2 + \alpha^2 \tilde{a}^2(1 - \omega^2) \cos^2 \theta\right) S = \Lambda S \,,
    \end{equation}
    and so (\ref{eq:startingPoint}) is of the same form as~(\ref{eqn: scalar radial equation}), yet without needing the separable ansatz~(\ref{eqn:ScalarAnsatz}). 
    Mimicking (\ref{eq:pseudoSpecPeelOff}), we then strip $\Phi$ of its asymptotic behavior and  decompose it into scalar spherical harmonics,
    \begin{equation}
      \Phi= e^{-i \omega t} \left(\frac{r - r_+}{r - r_-}\right)^{i P_+}\!\!\! (r - r_-)^{-1 + \nu - 2 \alpha^2/\nu} e^{-\alpha(r-r_+)/\nu} \sum_{\ell} B_\ell(\zeta) \,Y_{\ell m}(\theta, \phi)\,, \label{eq:scalarSHDecomp}
    \end{equation}
    where the sum ranges over even or odd values of $\ell$ for parity even or odd modes, respectively. 

    \vskip 4pt
    With this ansatz, we can project the Klein-Gordon equation (\ref{eq:startingPoint}) onto the scalar spherical harmonics to find 
    \begin{equation}
      \left(\frac{\partial^2}{\partial \zeta^2} + \mathcal{C}_1(\nu, \zeta) \frac{\partial}{\partial \zeta} + \mathcal{C}_2(\nu, \zeta) \right) \hskip 1pt B_{\ell}(\zeta)  + \mathcal{C}_{\ell, \ell+2}(\nu, \zeta)\,B_{\ell+2}(\zeta) + \mathcal{C}_{\ell, \ell-2}(\nu, \zeta) B_{\ell-2}(\zeta) = 0\,, \label{eq:decompScalar}
    \end{equation}
    where the functions $\mathcal{C}_i$ again depend on our choice of $\zeta(r)$, but now have explicit $\ell$ dependence. Importantly, the $\cos^2 \theta$ term in (\ref{eq:startingPoint}) breaks spherical symmetry and thus couples different spherical harmonics to one another, through the function $\mathcal{C}_{\ell,\ell'}$. By expanding the radial functions $B_\ell(\zeta)$ in cardinal polynomials, as in (\ref{eq:bRight}), and sampling at the interpolation points (\ref{eq:chebPoints}), this system of equations can be rewritten as the matrix, 
    \begin{equation}
      \begin{aligned}
      \mathcal{M}_{n \ell; k \ell'}(\nu) = \ &\big(p_{k}^{\hskip 1pt \prime \prime}(\zeta_n) + \mathcal{C}_1(\nu, \zeta_n) \hskip 1pt p^{\hskip 1pt \prime}_k(\zeta_n) + \mathcal{C}_2(\nu, \zeta_n) \hskip 1pt  \delta_{nk}\big) \delta_{\ell \ell'} \\
      &+ \mathcal{C}_{\ell, \ell+2}(\nu, \zeta_n) \hskip 1pt  \delta_{nk} + \mathcal{C}_{\ell, \ell-2} (\nu, \zeta_n) \hskip 1pt  \delta_{nk}\,,
      \end{aligned} \label{eq:scalarNonSepMat}
    \end{equation}
    which can then be passed to a solver to find the bound state spectrum. 

\vskip 4pt
To fit (\ref{eq:scalarNonSepMat}) on a computer, it is necessary both to sample a finite number of radial points and to include only a finite number of angular modes. This angular truncation is an additional source of error that must be controlled. Fortunately, both the Klein-Gordon and Proca equations enjoy an approximate spherical symmetry that is restored in the Schwarzschild limit---this is why we expanded (\ref{eq:scalarSHDecomp}) in terms of scalar spherical harmonics, as opposed to another complete set of functions. In this basis, the $\mathcal{C}_{\ell,\ell\pm 2}$ mixings in (\ref{eq:scalarNonSepMat}) are proportional to $\alpha^2 \tilde{a}^2$. For the main superradiant state $|\es 2 \es 1 \es 1 \rangle$, the angular truncation error roughly scales as $(\alpha \tilde{a})^{2 L}$ if we include only $\ell = 1, 3, \dots, 2L + 1$ in the expansion (\ref{eq:scalarSHDecomp}). However, we must emphasize that this estimate of the truncation error is only accurate at small $\alpha$ or $\tilde{a}$. Of course, the scalar (or vector) spherical harmonics are a complete set and so may represent an arbitrary scalar (or vector) field configuration. As long as we include `enough' of these angular modes, we are guaranteed to faithfully represent any field configuration. In fact, one can show from their recurrence relation \cite{Abramowitz:1965} that the coefficients of the spherical harmonic decomposition of the spheroidal harmonics decay faster than exponentially, independent of $\alpha \tilde{a}$. This reflects the intuition that low-energy solutions to the Klein-Gordon equation are relatively smooth, and the same will be true for the Proca field. In practice, this method is as fast\footnote{Though the matrix (\ref{eq:scalarNonSepMat}) passed to the nonlinear eigenvalue solver is generally much larger than (\ref{eq:scalarMatrix}), this is generally balanced by no longer needing to compute the spheroidal harmonic eigenvalue and (more importantly) its derivative with respect to $\nu$.} as that using the separable ansatz in \S\ref{sec:Separable} and can be made just as accurate. Applying this algorithm reproduces the results presented in \S\ref{sec:Summary}.

  \subsubsection*{Vector}
  
  A similar strategy works for the Proca equation, although we will encounter
   additional technical challenges. As for the case of the scalar, we will first rewrite the equations of motion using operators that act simply on either the radial or angular directions. We will then decompose the temporal and spatial components of the vector field into scalar and vector spherical harmonics, and use this decomposition to convert the equations of motion into a matrix equation. In the following, we provide a mostly qualitative overview of our techniques, keeping the many technical details confined to Appendix~\ref{app:numericalDetails}.

\vskip 4pt
  One complication, compared with the scalar case, is that the Proca equation (\ref{equ:Proca})~needs to be supplemented by the Lorenz condition, $\nabla^\mu A_\mu = 0$. We can do this either by first solving $\nabla^\mu A_\mu = 0$ for the temporal component 
  $A_t$ and then substituting it into the Proca equation to find an equation purely in terms of the spatial components $A_i$, or by including the Lorenz constraint as an additional `block' of the matrix equation. We will choose the latter, since it is more flexible and generally easier to implement.

\vskip 4pt
  With this in mind, our first step is to rewrite both the Proca equation and Lorenz constraint using operators that act simply in either the radial or angular directions. The operator (\ref{eq:purelyRadial}) we used in the previous section is, unfortunately, only radial when acting on a scalar. It will thus be extremely convenient to expand the vector field along a carefully chosen tetrad\hskip 1pt\footnote{We will use tetrad indices $a, b, \dots$ to run from $0$ to $3$, and indices $i,j,\dots$ to run from $1$ to $3$. By convention, repeated indices are summed over.}  $A_\mu = A_a f\indices{^a_\mu}$, so that we may instead work with the four scalar fields $A_a$ instead of the four-vector $A_\mu$. As detailed in Appendix~\ref{app:numericalDetails}, we require that this tetrad is stationary, $\pounds_t f\indices{^a_\mu} = 0$.  We will take $f\indices{^0_\mu}\ud x^\mu \propto \ud t$ to be purely temporal with no angular dependence and the $f\indices{^i_\mu}$ to have definite total and azimuthal angular momentum---their explicit form is given in (\ref{eq:formFieldDefs}).

  \vskip 4pt
   We may then use this decomposition with  (\ref{eq:purelyRadial}) to rewrite the Proca equation as
  \begin{align}
         0  =\  &\bigg(\frac{1}{\Delta} \partial_r(\Delta \partial_r) - \frac{1}{\Delta} \left(\pounds^2 + \alpha^2 \tilde{a}^2 \left(1 - \omega^2\right) \cos^2 \theta \right)\!\bigg)A_b  \label{eq:procaEqDecompMain} \\
        & + \bigg(- (1 - \omega^2) + \frac{P_+^2}{(r - r_+)^2} + \frac{P_-^2}{(r -r_-)^2} - \frac{A_+}{(r - r_+)(r_+ - r_-)} + \frac{A_-}{(r -r_-)(r_+ - r_-)}\bigg) A_b \nonumber \\[6pt]
        & + \mathcal{S}\indices{_b^a} A_a + \mathcal{Q}\indices{_b^a}\, \pounds_z A_a + \mathcal{R}\indices{_b^a} \,\partial_r A_a + \mathcal{P}\indices{_b^a} \,\mathcal{D}_+ A_a + \mathcal{Z}\indices{_b^a} \,\mathcal{D}_0 A_a + \mathcal{M}\indices{_b^a} \,\mathcal{D}_- A_a\,,\nonumber
  \end{align} 
  where we have assumed that the vector field has definite frequency $\pounds_t A_\mu = -\omega A_\mu$ and azimuthal angular momentum $\pounds_z A_\mu = m A_\mu$. We recognize that (\ref{eq:procaEqDecompMain}) is simply four copies of the scalar equation (\ref{eq:startingPoint}), coupled together through the mixing matrices $\mathcal{S}$, $\mathcal{Q}$, $\mathcal{R}$, $\mathcal{P}$, $\mathcal{Z}$ and $\mathcal{M}$, whose precise form can be found in (\ref{eq:mixingMatrices}) and depend on our choice of tetrad. These mixing matrices encode the `vector-ness' of the scalars $A_a$, while $\mathcal{D}_{\pm}$ and $\mathcal{D}_0$---cf.~(\ref{eq:dpm}) and~(\ref{eq:d0})---are purely angular operators that act simply on the scalar spherical harmonics. The Lorenz constraint can be similarly decomposed,
  \begin{equation}
  0 = \mathcal{T}^0 A_0 + \mathcal{S}^i A_i + \mathcal{R}^i \,\partial_r A_i + \mathcal{P}^i \,\mathcal{D}_+ A_i + \mathcal{Z}^i\, \mathcal{D}_0 A_i + \mathcal{M}^i \,\mathcal{D}_- A_i\,, \label{eq:lorenzEqDecompMain}
  \end{equation}
  in terms of the mixing vectors defined in (\ref{eq:mixingVectors}).

  \vskip 4pt
  As we explain in \S\ref{app:boundaryConditions}, we also choose the $f\indices{^a_\mu}$ so that the scalars $A_a$ have the same asymptotic behavior,  (\ref{eqn:scalar radial outer horizon}) and (\ref{eqn: scalar radial infinity}), as the scalar field $\Phi$. Then, by mimicking the scalar decomposition (\ref{eq:scalarSHDecomp}), we strip the $A_a$ of their asymptotic behavior and decompose  the temporal component $A_0$ into scalar spherical harmonics and the spatial components $A_i$ into one-form harmonics
  \begin{align}
    A_0 &= e^{-i \omega t} \left(\frac{r - r_+}{r - r_-}\right)^{i P_+}\!\!(r- r_-)^{-1 + \nu - 2 \alpha^2/\nu} e^{-\alpha(r - r_+)/\nu} \sum_{j} B_{0,j}(\zeta)\, Y_{j m}(\theta, \phi)\,, \label{eq:pseudoVecTemp} \\
    A_i &= e^{-i \omega t} \left(\frac{r - r_+}{r - r_-}\right)^{i P_+}\!\!(r -r_-)^{-1+\nu- 2\alpha^2/\nu} e^{-\alpha(r - r_+)/\nu} \sum_{\ell,j} B_{\ell j}(\zeta)\, Y_{i}^{\ell, j m}(\theta,\phi)\,. \label{eq:pseudoVecSpat}
  \end{align}
   In the scalar case, we found that only components with even (odd) parity could couple to one another. This is a reflection of the fact that the Kerr metric is invariant under parity transformations, and so only scalar harmonics with the same parity couple to one another. This also applies to the Proca field. As discussed in Appendix \ref{app:harmonics}, the parity of the scalar and vector spherical harmonics in (\ref{eq:pseudoVecTemp}) and (\ref{eq:pseudoVecSpat}) are $(-1)^j$ and $(-1)^{\ell+1}$, respectively. When we solve for the spectrum of a parity even or odd mode, we can restrict the expansions (\ref{eq:pseudoVecTemp}) and (\ref{eq:pseudoVecSpat}), halving the number of angular terms we must include to faithfully represent the Proca field.

   \begin{figure}
        \begin{center}
          \makebox[\textwidth][c]{\includegraphics{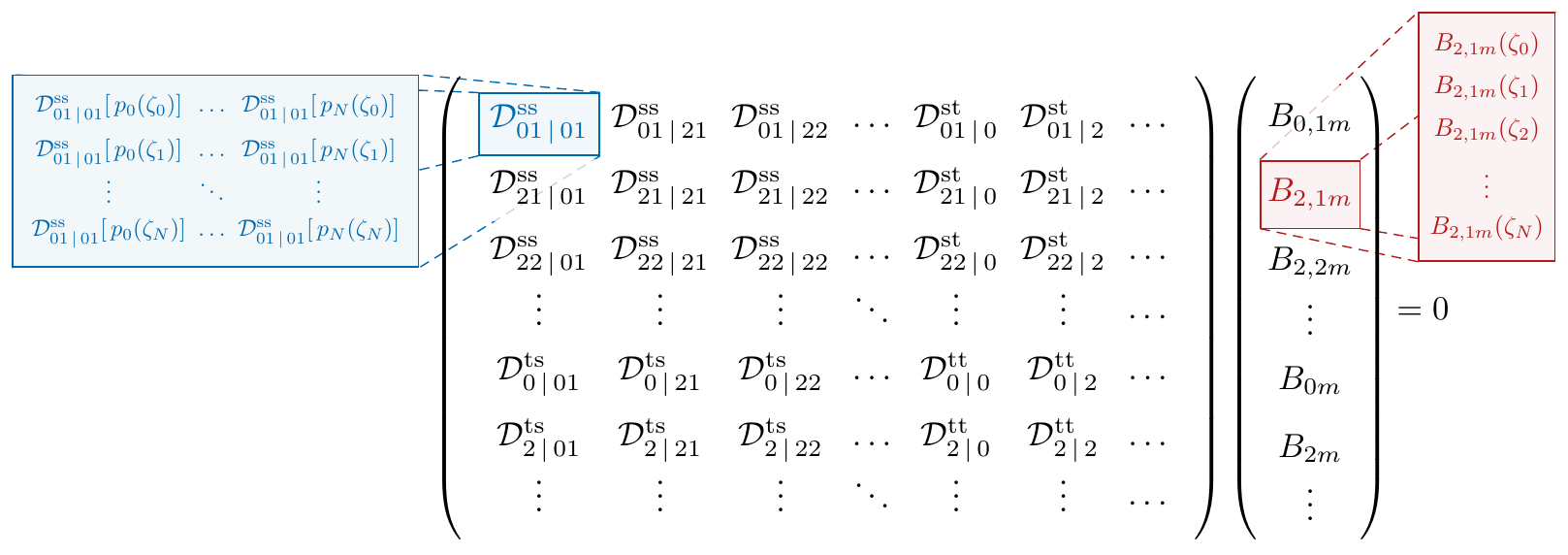}}
          \caption{Structure of the matrix equation for the parity even modes. Each angular block forms a separate radial sector, which is sampled at $N+1$ points $\{\zeta_k\}$.\label{fig:vectorMatrixStructure}}
        \end{center}
      \end{figure}

   \newpage
   Following the scalar case, we project both (\ref{eq:procaEqDecompMain}) and (\ref{eq:lorenzEqDecompMain}) onto the different scalar and vector harmonics to find the following system of equations,
   \begin{align}
        \sum_{\ell',j'} \mathcal{D}^{\lab{ss}}_{\ell j\,|\,\ell' j'}[B_{\ell', j' m}(\zeta)] + \sum_{j'} \mathcal{D}^{\lab{st}}_{\ell j\, |\, j'}[B_{j'm}(\zeta)] &= 0\,, \label{eq:procaProject} \\
          \sum_{\ell',j'} \mathcal{D}^{\lab{ts}}_{j\, |\, \ell' j'}[B_{\ell', j' m}(\zeta)] + \sum_{j'} \mathcal{D}^{\lab{tt}}_{j\, | \, j'}[B_{j' m}(\zeta)] &= 0\,, \label{eq:lorenzProjectMain}
    \end{align}
    which can be written as the matrix equation schematically depicted in Fig.~\ref{fig:vectorMatrixStructure}, where we have introduced a collection of operators defined in (\ref{eq:sphericalMatrices}). This system of equations is the vector analog of (\ref{eq:decompScalar}), though it appears more complicated because we have explicitly separated the spatial and temporal components. As before, this matrix may then be passed to a nonlinear eigenvalue solver to determine the quasi-bound state spectrum. This method reproduces the quasi-bound state frequencies of the electric modes computed using the separable ansatz in~\S\ref{sec:Sep}. More importantly, it accurately computes the energy spectrum of the magnetic modes $|\es 2 \es 1 \es 1 \es m \rangle$, shown in Figures~\ref{fig:VectorSpectrumPlots}~and~\ref{fig:magneticReal}, and provides a crucial check of the magnetic instability rate (\ref{eqn:VectorRates}), displayed in Figures~\ref{fig:growthRates}~and~\ref{fig:magInstability}.

   \newpage 

\section{Summary and Outlook}
\label{sec:outlook}

In this work, we have computed the quasi-bound state spectra of massive scalar and vector fields around rotating black holes, both analytically and numerically. 
The main challenge has been the fact that the fields vary rapidly in the  near-horizon region of the black hole, which causes ordinary perturbative approximations to fail and many numerical methods to become unreliable.
To address this issue in our analytical treatment, we
constructed independent solutions in different asymptotic regions  of the black hole spacetime and determined the spectrum by demanding that these solutions match in their regions of shared validity. For the scalar field, the asymptotic expansions in the near and far regions were matched directly, while, for the vector field, these expansions could only be matched indirectly, via a solution in an additional intermediate region. 
This reflected the fact that the vector field in the near and far regions depend on different types of angular momentum, which a matched solution must smoothly interpolate between. 
We presented results for the energies and instabilities as perturbative expansions in powers of the gravitational fine structure constant $\alpha$.

\vskip 4pt
Our perturbative analysis relied on the separability of the equations of motion, and thus did not apply to all magnetic modes of the vector field. However, by working at linear order in the black hole spin $\tilde{a}$, we derived perturbative results for the energy spectrum of the magnetic modes, which we argued plausibly extend to arbitrary $\tilde{a}$. Furthermore, we provided an educated guess for the magnetic instability rate by demanding that it takes the same functional form as the electric instability rates, and fixing its undetermined overall coefficient and dominant $\alpha$-scaling
using its known Schwarzschild limit.

\vskip 4pt
As a check of our conjectural magnetic results,
and to study all modes 
at large values of $\alpha$, we computed the spectra numerically. The rapid variation of the fields in the near-horizon region presents serious numerical obstacles which can potentially destroy the accuracy of a solution. We described how to avoid these pitfalls and presented a formulation of the problem that accurately computes the quasi-bound state frequencies without relying on a separable ansatz. In principle, this formulation can be extended to ultralight fields of arbitrary spin about any stationary spacetime, although it works best if spherical symmetry is approximately restored at large distances. These numeric results provided a valuable check of our analytic approximations, which we found to accurately predict the energy eigenvalues and instability rates for both electric and magnetic modes as long as  $\alpha \lesssim 0.2$, even at large $\tilde{a}$. 

\vskip 4pt
As shown in Figures~\ref{fig:ScalarSpectra} and \ref{fig:VectorSpectra}, the scalar and vector gravitational atoms
 have qualitatively distinct spectra. In particular, the vector field's intrinsic spin allows for many more nearly degenerate states than the scalar. This suggests that we could differentiate between the scalar and vector atoms by studying induced transitions between these different energy levels. 
In fact, such transitions naturally occur when the gravitational atom is part of a binary system, in which case the gravitational interaction with the binary companion leads to mixings of the energy levels~\cite{Baumann:2018vus}.  The results of this paper therefore provide an essential input for the phenomenology of such systems and, in a forthcoming work~\cite{Paper2}, we study how these spectral differences manifest themselves in gravitational wave signals.

	\subsection*{Acknowledgements}
	We thank Alex Alemi, Will East, Thomas Helfer, Tanja Hinderer, David Nichols, Samaya Nissanke, Wilke van der Schee, and Helvi Witek for useful discussions, and especially Sam Dolan for a fruitful exchange on the separability of the magnetic modes. 
	We are particularly grateful to Rafael Porto for collaboration and many enlightening discussions on the subject of `precision gravity.'
	DB~and JS are supported by a Vidi grant of the Netherlands Organisation for Scientific Research~(NWO) that is funded by the Dutch Ministry of Education, Culture and Science~(OCW). The work of DB and JS is part of the Delta-ITP consortium, and the work of HSC is supported by NWO. DB, HSC and JS thank the Henri Poincar\'e Institute in Paris for its hospitality while some of this work was performed. HSC and JS would like to thank the Perimeter Institute and Utrecht University, respectively, for their hospitality while some of this work was performed.  Portions of this work were completed by JS at the Aspen Center for Physics, which is supported by National Science Foundation grant PHY-1607611.

\clearpage
	\appendix
	\section{Tensor Spherical Harmonics}
 	\label{app:harmonics}
	
In this appendix, we describe the spherical harmonic decomposition of tensor fields in black hole backgrounds, focusing in particular on the construction of vector spherical harmonics.

\subsection{Tensor Representations of SO(3)}

Besides the time-like Killing vector $k_t = -i \partial_t$, the spherically symmetric Schwarzschild black hole enjoys three additional Killing vectors, which in Boyer-Lindquist 
coordinates take the form
	 \begin{equation}
			\begin{aligned}
				k_x &= -i \left(y \frac{\partial}{\partial z} -  z \frac{\partial}{\partial y}\right)= i \left(\sin \phi\, \frac{\partial}{\partial \theta} + \cot \theta \cos \phi\, \frac{\partial}{\partial \phi}\right) , \\
				k_y &= -i \left(z \frac{\partial}{\partial x} - x \frac{\partial}{\partial z}\right) = i \left(-\cos \phi \,\frac{\partial}{\partial \theta} + \cot \theta \sin \phi\, \frac{\partial}{\partial \phi}\right) , \\
				k_z &= -i \left(x \frac{\partial}{\partial y} - y \frac{\partial}{\partial x}\right) = -i \frac{\partial}{\partial \phi}\, .
				\label{eq:so3algebra}
			\end{aligned}
	\end{equation}
	These Killing vectors are the
	generators of the group $\lab{SO}(3)$ and therefore satisfy the commutation relations
	\begin{equation}
		\left[k_i, k_j\right] = \pounds_{i} k_j = i \epsilon_{ijk} k_k\,,
	\end{equation}
  where we use $i,j,k = 1, 2, 3$ to represent $x, y, z$, respectively, and have employed the shorthand $\pounds_{i} = \pounds_{k_i}$ for the Lie derivative along $k_i$. This algebra has a quadratic Casimir,
	\begin{equation}
		\pounds^2 = \pounds_x^2 + \pounds_y^2 + \pounds_z^2 \, ,
	\end{equation}
	that commutes with the generators $k_i$. Irreducible representations of this group can then be labeled by their eigenvalues under $\pounds^2$ and one of the generators $\pounds_i$.
It is conventional to define the representation  $|j, m \rangle$ by
	\begin{equation}
	\begin{aligned}
		\pounds^2 |j, m \rangle &= j(j+1)|j, m \rangle\, , \\
		 \pounds_z | j, m \rangle &= m |j, m \rangle \,.
		 \end{aligned}
	\end{equation}
  For now, $|j, m\rangle$ denotes an arbitrary (tensor) representation of this algebra with total angular momentum $j$ and total azimuthal angular momentum $m$.

\vskip 4pt
We may also define the vector fields
	\begin{equation}
		k_{\pm} = k_{x} \pm i k_y = e^{\pm i\phi} \left(\pm \frac{\partial}{\partial \theta} + i \cot \theta \frac{\partial}{\partial \phi}\right)  , 
	\end{equation}
	and the corresponding operators $\pounds_{\pm} = \pounds_{k_\pm}$. These operators raise/lower the azimuthal angular momentum
	\begin{equation}
	\begin{aligned}
		\pounds_{\pm} k_{\mp} &= [k_{\pm}, k_{\mp}] = \pm 2 k_{z} \,, \\
		 \pounds_z k_\pm &= [k_z, k_{\pm}] = \pm k_{\pm} \,.
		\end{aligned}
	\end{equation}	
  A physical, finite-dimensional representation requires that 
	\begin{equation}
		\pounds_\pm | j, \pm j \rangle = 0\,, \label{eq:loweringDef}
	\end{equation}
	since we could otherwise violate the requirement that $\pounds^2 - \pounds_z^2 \geq 0$. With the normalization condition $\langle j, \pm j | j, \pm j \rangle = 1$, (\ref{eq:loweringDef}) defines the states $|j, \pm j\rangle$.  The other states are then defined in the \emph{Condon-Shortley phase convention}\footnote{Which convention should be used depends on the convention for the Clebsch-Gordan coefficients. We use this convention since it is the one used by Mathematica.} by 
	\begin{equation}
	\begin{aligned}
		\pounds_+ |j,  m \rangle &= \sqrt{(j - m)(j + m + 1)}\, |j, m+1 \rangle\, , \\
		 \pounds_- |j, m \rangle &= \sqrt{(j + m)(j - m + 1)} \, |j, m-1\rangle\,.
	\end{aligned}
	\end{equation}
	Two representations of $\lab{SO}(3)$ can be combined into a single representation with definite total and azimuthal angular momentum using the Clebsch-Gordan coefficients,
	\begin{equation}
		|j, m; \ell, s \rangle = \sum_{m_\ell = - \ell}^{\ell} \sum_{m_s = -s}^{s} |(s\, m_s)\, (\ell\, m_\ell)\rangle \langle (s\, m_s)\, (\ell\, m_\ell) | j, m; \ell, s \rangle\,.
	\end{equation}
	This yields a well-defined prescription for generating tensor spherical harmonics of any rank. That is, given a tensor with an arbitrary number of spatial indices, we can define the basis of tensors $|s, m_s \rangle$ by first solving 
	\begin{equation}
	\begin{aligned}
		\pounds_+\hskip 2pt \chi^{s,s}_{ijk\dots} &= 0 \, ,\\ \pounds_z\hskip 2pt \chi^{s,s}_{ijk\dots} &= s\hskip 2pt \chi^{s, s}_{ijk\dots}\,,
		\end{aligned}
	\end{equation}
  and then using $\pounds_-$, with the normalization convention above, to define the other $|s, m_s\rangle$. Note that these conditions immediately imply that $\pounds^2 \chi^{s, s}_{ijk\dots} = s(s+1) \chi^{s,s}_{ijk\dots}$.
	We may then combine $|s, m_s\rangle$ with a \emph{scalar} representation of $|\ell, m_\ell\rangle$, i.e.~the scalar spherical harmonics $Y_{\ell\hspace{0.75pt} m_\ell}(\theta, \phi)$,
	  to generate the tensor spherical harmonics
  \begin{equation}
    T_{ijk\dots}^{j,m; \ell, s}(\theta, \phi) = \sum_{m_\ell = -\ell}^{\ell} \sum_{m_s = -s}^{s} \langle (s\, m_s)\, (\ell\, m_\ell) | j, m; \ell, s \rangle\, Y_{\ell \hspace{0.75pt} m_\ell}(\theta, \phi)\, \chi^{s,m_s}_{ijk\dots}\,. \label{eq:tensorSphericalHarmonics}
  \end{equation}
  This procedure yields a tensor representation of $\lab{SO}(3)$. Because our ultimate goal is to expand vector fields in $\mathbbm{R}^3$, the $T_{ijk\dots}^{j, m; \ell,s}$ must actually form a representation of $\lab{O}(3)$, i.e. the group $\lab{SO}(3)$ together with inversions. Since the scalar spherical harmonics have definite parity $Y_{\ell m_\ell}(\pi-\theta, \phi+\pi) = (-1)^{\ell} Y_{\ell m_\ell}(\theta, \phi)$, we can choose $\chi^{s, m_s}_{ijk\dots}$ to also have definite parity, so that (\ref{eq:tensorSphericalHarmonics}) defines a tensor representation of $\lab{O}(3)$. Below, we will explicitly construct a family of vector spherical harmonics. In that case, the choice of parity determines whether (\ref{eq:tensorSphericalHarmonics}) yields \emph{vector} or \emph{pseudo-vector} spherical harmonics.

	\subsection{Vector Spherical Harmonics} 
	\label{app:vsh} 

To form a basis of vector spherical harmonics, 
we first construct the basis of vectors $\chi_{1, m_s}^{i}$ by the defining relations
    \begin{equation}
    \begin{aligned}
      \pounds_{+} \chi_{1, 1}^i &= 0 \, , \\ \pounds_z \chi_{1, 1}^i &= \chi_{1, 1}^i \,.
      \end{aligned}
    \end{equation}
    These equations have the general solution
    \begin{equation}
      \chi_{1,1}^{i} \,\partial_i = \frac{e^{i \phi}}{\sqrt{2}}\Big[F_{r}(r) \sin \theta\, \partial_r + F_{\phi}(r)(\partial_\theta + i \cot \theta \,\partial_\phi) + F_{\theta}(r)(\cos \theta \, \partial_\theta + i \csc \theta\, \partial_\phi)\Big]\,,
    \end{equation}
    where the $F_{i}$ are, in general, undetermined functions of $r$. Imposing that $\chi_{1,1}^i$ transforms like a \emph{vector} under inversions, leads to $F_{\phi} = 0$. Taking, instead, $F_r = F_\theta = 0$ would yield a basis of \emph{pseudo-vector} spherical harmonics.
We may then define $\sqrt{2} \chi_{1, 0}^i = \pounds_- \chi_{1,1}^i$ and $\sqrt{2}\chi_{1, \sminus 1}^{i} = \pounds_{-} \chi_{1,0}^i$, to find the complete basis
    \begin{equation}
      \begin{aligned}
        \chi_{1,1}^{i}\,\partial_i &= \frac{e^{i \phi}}{\sqrt{2}}\Big[F_{r}(r) \sin \theta\, \partial_r  + F_{\theta}(r)(\cos \theta \, \partial_\theta + i \csc \theta\, \partial_\phi)\Big] \, ,\\
        \chi_{1, 0}^{i} \, \partial_i &= - F_{r}(r) \cos \theta \,\partial_r + F_\theta(r) \sin \theta \, \partial_\theta\, , \\
        \chi_{1, \sminus 1}^{i}\, \partial_i &= \frac{e^{-i \phi}}{\sqrt{2}}\Big[\minus F_{r}(r) \sin \theta\, \partial_r  + F_{\theta}(r)(\minus \cos \theta \, \partial_\theta + i \csc \theta\, \partial_\phi)\Big]\, .
      \end{aligned}
      \label{eq:vecBasis}
    \end{equation}
In flat space, we can take $F_r = -1$ and $F_\theta = -1/r$, so that these basis vectors are covariantly constant, $\nabla_\mu \chi_{1, m_s}^\nu = 0$.\footnote{If we can find basis vectors that are covariantly constant, then it is possible to unambiguously disentangle spin and orbital angular momentum, so that $\ell$ can be a good quantum number.}  These harmonics then reduce to those commonly used in the literature; see e.g.~\cite{Thorne:1980ru}. This is impossible in the Kerr geometry. It will instead be more convenient to choose $F_r$ and $F_\theta$ to simplify the boundary conditions of the vector field at the horizon and at spatial infinity, as discussed in \S\ref{app:boundaryConditions}. 	For each choice of $F_{r}(r)$ and $F_{\theta}(r)$, we then have a set of \emph{vector spherical harmonics}
		\begin{equation}
			Y^i_{\ell, j m} = \sum_{m_s = -1}^{1} \langle (1\, m_s)\, (\ell\, m - m_s) | j, m \rangle \,Y_{\ell, m - m_s}(\theta,\phi)\, \chi_{1, m_s}^{i} \,, \label{eq:vectorSphericalHarmonics}
		\end{equation}
		which have definite angular momentum,
		\begin{equation}
		\begin{aligned}
			\pounds^2 Y^i_{\ell, j m} &= j(j+1) Y^i_{\ell, j m} \, ,\\ \pounds_z Y^i_{\ell, j m} &= m Y^i_{\ell,j m}\,,
			\end{aligned}
		\end{equation}
    and by construction satisfy
    \beq
    \begin{aligned}
      \pounds^2 \Big(e^{-i \omega t} G(r) Y^i_{\ell, j m} \Big) &= j(j+1) e^{-i \omega t} G(r) Y^i_{\ell, j  m}\,, \\
       \pounds_z \Big(e^{-i \omega t} G(r) Y^i_{\ell, j m} \Big) &= m e^{-i \omega t} G(r) Y^i_{\ell,j  m}\, .
    \end{aligned}	
    \eeq
The Clebsch-Gordan coefficients are only non-vanishing for $j \geq 0$, $|m| \leq j$, and $j+1 \geq \ell \geq |j-1|$.
   Any smooth spatial vector field can be decomposed into these harmonics,
      \begin{equation}
        V^i(t, r, \theta, \phi) = \sum_{j = 0}^{\infty} \,\sum_{\ell = |j -1|}^{j +1} \,\sum_{m = -j}^{j}  V_{\ell, j m}(t, r) \,Y^i_{\ell, jm}(\theta, \phi)\, .
      \end{equation} 
    By construction, the harmonics (\ref{eq:vectorSphericalHarmonics}) have definite parity, and acquire a factor of $(-1)^{\ell+1}$ under the transformation $(\theta, \phi) \to (\pi - \theta, \phi + \pi)$. 
   A general odd-parity vector field $E^i$ 
can thus be expanded as
   \begin{equation}
   	E^i = \sum_{j, \ell, m} E_{\ell, jm}(t,r) Y^i_{\ell, j m}(\theta, \phi)\,,
   \end{equation}
   where $j$ runs over all positive integers, 
   $\ell$ is even and takes values between $j\pm 1$, 
   and $m$ runs from $-j$ to $j$. Similarly, a general even-parity vector field $B^i$ can be expanded as
   \begin{equation}
   	B^i = \sum_{j, \ell, m} B_{\ell, jm}(t, r) \,Y^i_{\ell,j m}(\theta, \phi) \, ,
   \end{equation}
where $j$ runs over all non-negative integers, $\ell$ runs over all odd numbers between $j\pm 1$, and $m$ runs from $-j$ to~$j$. 
In both cases, if the field has definite azimuthal angular momentum $m_z$, the sum over $m$ is restricted to $m_z$ and hence can be dropped.

    \vskip 4pt
    A major benefit of this basis of vector fields (\ref{eq:vecBasis}) is that it also acts simply on the scalar spherical harmonics. We may define the operators
      \begin{align}
        \mathcal{D}_\pm Y_{\ell m_\ell} &= \frac{e^{\pm i \phi}}{\sqrt{2}} \left(\pm \cos \theta \, \partial_\theta + i \csc \theta \, \partial_\phi\right)Y_{\ell m_\ell}(\theta, \phi) \label{eq:dpm} \\
      &= \sqrt{\frac{(\ell+1)^2 (\ell\mp m_\ell)(\ell\mp m_\ell-1)}{2(2 \ell + 1)(2 \ell-1)}} Y_{\ell-1, m_\ell \pm 1} + \sqrt{\frac{\ell^2(\ell\pm m_\ell +1)(\ell\pm m_\ell +2)}{2(2\ell+3)(2 \ell+1)}} Y_{\ell+1, m_\ell \pm 1}\,, \nonumber \\[6pt]
      \mathcal{D}_0 Y_{\ell m_\ell}  &=\sin \theta \,\partial_\theta Y_{\ell m_\ell }(\theta, \phi) \label{eq:d0} \\[2pt]
      &=  \sqrt{\frac{\ell^2 (\ell+m_\ell + 1)(\ell-m_\ell +1)}{(2\ell+1)(2\ell+3)}} Y_{\ell+1, m_\ell} -\sqrt{\frac{(\ell+1)^2 (\ell+m_\ell)(\ell-m_\ell)}{(2 \ell+1)(2 \ell-1)}}Y_{\ell-1, m_\ell}\, , \nonumber
      \end{align}
    so that the angular legs of each of (\ref{eq:vecBasis}) can be rewritten using these operators,
    \begin{equation}
      \begin{aligned}
        \chi_{1,1}^i \,\partial_i &= \frac{1}{\sqrt{2}} e^{i \phi} \sin \theta F_r \,\partial_r + F_\theta \,\mathcal{D}_+ \,,\\
        \chi_{1, 0}^i \,\partial_i &= -F_r \cos\theta \, \partial_r + F_\theta \,\mathcal{D}_0 \, ,\\
        \chi_{1, \sminus 1}^i \,\partial_i &= \minus \frac{1}{\sqrt{2}}e^{-i \phi} F_r \sin \theta \, \partial_r + F_\theta\, \mathcal{D}_-\,.
      \end{aligned}
    \end{equation}
    While not strictly necessary, it will prove extremely convenient in our decomposition of the Proca equation to write all angular derivatives in terms of $\pounds^2$, $\pounds_z$, $\mathcal{D}_\pm$, and $\mathcal{D}_0$, as derivatives like $\partial_\theta Y_{\ell, m}$ are not easily expressible in terms of other spherical harmonics.

    \vskip 4pt
    Finally, it will also be useful to construct a basis of one-form harmonics $\chi_i^{1, m_s}\, \ud x^i$, which are dual to the vector fields (\ref{eq:vecBasis}), in the sense that
    \begin{equation}
      \bar{\chi}^{1, m_s}_i \chi_{1, m_s'}^{i} = \delta^{m_s}_{m_s'} \quad\text{and} \quad \bar{\chi}^{1 m_s}_{i} \chi_{1, m_s}^{j} = \delta_i^j\,,
    \end{equation} 
    where the bar denotes complex conjugation.
    Explicitly, these form fields are
    \beq
    \begin{aligned}
          \chi^{1, 1}_{i} \,\ud x^i &= \frac{1}{\sqrt{2}}e^{i\phi}\left(F_r^{-1} \, \sin \theta \, \ud r + F_\theta^{-1}\left( \cos \theta \, \ud \theta + i \sin \theta\, \ud \phi\right)\right) , \\
          \chi^{1,0}_{i} \,\ud x^i &= -F_r^{-1} \cos \theta\, \ud r + F_\theta^{-1} \cos \theta \, \ud \theta\,, \\
          \chi^{1,\sminus 1}_i \, \ud x^i &= \frac{1}{\sqrt{2}} e^{-i \phi} \left( \minus F_r^{-1} \sin \theta \, \ud r + F_\theta^{-1}\left( - \cos \theta \, \ud \theta + i \sin \theta \, \ud \phi\right)\right).
      \end{aligned} 
      \eeq
      We can then write the set of {\it one-form spherical harmonics} as
      \begin{equation}
        Y_i^{\ell, j m} = \sum_{m_s = -1}^{1} \langle (1\, m_s)\, (\ell\, m - m_s) | j, m \rangle \,Y_{\ell, m - m_s}(\theta,\phi) \,\chi^{1, m_s}_{i}\,, \label{eq:formSphericalHarmonics}
      \end{equation}
   which satisfy the orthonormality condition,
      \begin{equation}
        \int_{\lab{S}^2}\!\ud \Omega\,\,  \bar{Y}^{\ell, jm}_{i} (\theta,\phi) \,Y^i_{\ell', j'm'}(\theta, \phi) = \delta_{j'}^{j} \delta^\ell_{\ell'} \delta^{m}_{m'}\, ,
      \end{equation}
      where $\int_{\lab{S}^2} \!\ud \Omega = \int_{0}^{2\pi}\!\ud \phi\, \int_{0}^{\pi}\!\sin \theta\, \ud \theta$ denotes integration over the sphere.

\newpage
\section{Details of the Analytical Treatment}
\label{app:details}  

In Section~\ref{sec:analytic}, we derived the quasi-bound state spectra for massive scalar and vector fields on the Kerr geometry, using the method of match asymptotic expansion. In this appendix, we provide some of the technical details that we left out in Section~\ref{sec:analytic}. 

\subsection{Separable Ansatz for Vectors}

In the coordinates (\ref{equ:Kerr2}), the polarization tensor $B^{\mu \nu}$ in (\ref{eqn: Proca ansatz}) reads 
\beq
\begin{aligned}
B^{\mu \nu} =  \hskip 3pt & 
\frac{1}{\Sigma q_r} \left( 
\begin{array}{cccc}
0 & i q_a \lambda^{-1} r & 0 & 0 \\
- i q_a \lambda^{-1} \hskip 1pt r & \Delta & 0 & - i \alpha \hskip 0.5pt \tilde{a} \lambda^{-1} r \\ 
0 & 0 & 0 & 0 \\
0 & i \alpha \hskip 0.5pt \tilde{a} \lambda^{-1} \hskip 1pt r & 0 & - \alpha^2 \tilde{a}^2 \Delta^{-1} 
\end{array} \right) + \frac{q_a}{ \Delta  q_r q_\theta } \left( 
\begin{array}{cccc}
-1 + \alpha^2 \tilde{a}^2 \lambda^{-2} \hskip 5pt &  0 \hskip 5pt & \hskip 5pt  0 &  \hskip 5pt \alpha  \hskip 0.5pt \tilde{a}  \lambda^{-2}   \\
0  &  0  \hskip 5pt & \hskip 5pt 0 & 0 \\
0  & 0  \hskip 5pt & \hskip 5pt 0 & 0 \\
\alpha \hskip 0.5pt  \tilde{a} \lambda^{-2}  & 0 \hskip 5pt & \hskip 5pt 0 & 0
\end{array} \right)  \\[4pt]
& + \frac{1}{\Sigma q_\theta} \left( 
\begin{array}{cccc}
- 2\alpha^3   \tilde{a}^2 \Delta^{-1} r \hskip 1pt \sin^2 \theta   &  \hskip 5pt 0  & \hskip 5pt  i \alpha^2 \tilde{a}^2 \lambda^{-1} \cos \theta \sin \theta & \hskip 5pt  - 2 \alpha^2 \tilde{a} \Delta^{-1} r \\
0 & \hskip 5pt  0 & \hskip 5pt  0 &\hskip 5pt  0 \\
- i \alpha^2 \tilde{a}^2 \lambda^{-1} \cos \theta \sin \theta & \hskip 5pt  0 & \hskip 5pt  1 & \hskip 5pt  - i \alpha \tilde{a} \lambda^{-1} \cot \theta \\
- 2 \alpha^2 \tilde{a} \Delta^{-1} r & \hskip 5pt  0 & \hskip 5pt  i \alpha \tilde{a} \lambda^{-1} \cot \theta & \hskip 5pt  \csc^2 \theta \end{array} \right) \, , \label{eqn:BtensorFull}
\end{aligned}
\eeq
where the following quantities have been introduced
\begin{equation}
\begin{aligned}
q_r & \equiv 1 + \lambda^{-2} r^2 \, , \\
q_\theta & \equiv 1 - \alpha^2 \tilde{a}^2  \lambda^{-2} \cos^2 \theta \, ,\\
q_a & \equiv r^2 + \alpha^2 \tilde{a}^2  \, .
\end{aligned}
\end{equation}
Expanding $B^{\mu \nu}$ to leading order in $\alpha$, which is equivalent to taking the flat-space limit, leads to the result (\ref{eqn: ansatz a0 explicit}).

\vskip 4pt
In \cite{Dolan:2018dqv}, the following special limit was taken\footnote{To accommodate scenarios with non-vanishing black hole spin, we take $\alpha \to 0$, instead of $\tilde{a} \to 0$ as in~\cite{Dolan:2018dqv}.}
\begin{align}
\lim_{\alpha \to 0} \left( \alpha \tilde{a}  \lambda^{-1} \right) = \chi  \, ,  \label{eqn: Dolan magnetic mode limit}
\end{align}
with $\chi =  m \pm 1$, and the separation constant $\lambda$ formally vanishes. It was then found that the ansatz~(\ref{eqn: Proca ansatz}) in this limit captures a subset of the magnetic-type modes of the vector field.  We will now investigate these special solutions in more detail. The vector field in this case becomes
\begin{align}
A^\mu_{\indlab{0}} = \frac{1}{r^2\left( 1 - \chi^2 \cos^2 \theta \right)}\begin{pmatrix}
\hphantom{1} 0\hphantom{1}  & \hphantom{1} 0\hphantom{1} & 0 & 0 \\
0 & 0 & 0 & 0 \\
 0 & 0 & 1 & -i \chi \cot \theta \\
 0 & 0 & i \chi \cot \theta & \csc^{2} \theta \\
\end{pmatrix}  \!\begin{pmatrix} \hphantom{|}\partial_t \hphantom{|}\\ \partial_r \\ \partial_\theta \\ \partial_\phi \end{pmatrix} \!Z_0 \, , \label{eqn: Dolan magnetic mode A}
\end{align}
where only the angular gradient terms are non-vanishing. The leading-order angular equation then is
\beq
\begin{aligned}
\frac{\ud^2 S_0}{\d \theta^2} \,+ & \left( \cot \theta +  2 \tan \theta +  \frac{\tan \theta}{ \chi \cos \theta - 1} - \frac{\tan \theta}{\chi \cos \theta + 1} \right) \frac{\d S_0}{\d \theta} \\
& \hskip 5pt + \left( m \chi - \frac{m^2}{\sin^2 \theta} + \frac{m \sec \theta}{\chi \cos \theta-1} + \frac{m \sec \theta}{\chi \cos \theta+ 1}\right)S_0 = 0 \, . \label{eqn: Dolan magnetic mode angular}
\end{aligned}
\eeq
Although (\ref{eqn: Dolan magnetic mode angular}) is not the general Legendre equation, the associated Legendre functions $P_{\pm (m \pm 1), m}$ are solutions to this equation, with the upper (lower) sign denoting solutions with positive (negative) $m$.
Substituting these solutions into (\ref{eqn: Dolan magnetic mode A}), and taking the appropriate scale factors $\{1, r, r \sin \theta \}$ into account, we find that 
\beq
A^i_{\indlab{0}} \propto Y^i_{j, j m} \, ,   \label{eqn: Dolan magnetic mode B}
\eeq
with $ j = |m|$. The relation in (\ref{eqn: Dolan magnetic mode B}) shows that the limit  (\ref{eqn: Dolan magnetic mode limit}) recovers a special type of magnetic mode, subject to the restriction $ j = |m|$. It remains an open problem to obtain separated equations for all magnetic modes in the Kerr background, and it would be nice to understand if the other magnetic modes can also be recovered from the ansatz (\ref{eqn: Proca ansatz}) or to show conclusively that this is not possible.

\subsection{Higher-Order Corrections}

In \S\ref{sec: Scalar Relativistic Corrections}, we showed that the equations of motion at higher orders in $\alpha$ can be solved iteratively by treating the lower-order solutions as source terms; cf.~\eqref{eqn:Jdef}. In the following, we discuss in more detail how these inhomogeneous equations  can be solved through the method of variation of parameters.

\vskip 4pt
Imposing the relevant boundary conditions at the horizon and at infinity, the near and far-zone solutions of the scalar field are 
\beq
\begin{aligned}
R^{\rm near}_i(z) & = \mathcal{C}_{i}^{\rm near} n_c(z) - n_c(z) \int^z_0 \d t \, \frac{J_i^z(t) n_d(t)}{W_n(t)} + n_d(z) \int^z_0 \d t \, \frac{J_i^z(t) n_c(t)}{W_n(t)} \, , \\
R^{\rm far}_i(x) & = \mathcal{C}_{i}^{\rm far} f_c(x)  + f_c(x) \int^\infty_x \d t \, \frac{J_i^x(t) f_d(t)}{W_f(t)} - f_d(x) \int^\infty_x \d t \, \frac{J_i^x(t) f_c(t)}{W_f(t)} \, ,  \label{eqn:VOP2}
\end{aligned}
\eeq
where we defined the homogeneous solutions of the leading-order equations of motion as
\beq
\begin{aligned}
n_c(z) & \equiv  \left( \frac{z}{z+1} \right)^{i P_+ } {}_2 F_1(-\ell, \ell+1, 1-2 i P_+, 1+z)  \, , \\
n_d(z) & \equiv z^{i P_+ }( z + 1)^{ i P_+} 
  {}_2F_1\left( - \ell + 2 i P_+, \ell + 1 + 2 i P_+, 1 + 2 i P_+, 1 + z \right) \, , \\[4pt]
f_c(x) & \equiv e^{-x/2} x^{\ell}\, U( \ell+1 -\nu_0, 2+2\ell, x) \, , \\[4pt]
f_d(x) & \equiv e^{-x/2} x^{\ell}\, {}_1 F_1( \ell+1 -\nu_0, 2+2\ell, x) \, ,
\end{aligned}
\eeq
and introduced the Wronskian $W_{n} \equiv W\left[ n_c, n_d \right] = n_c n^\prime_d -n_d n^\prime_c $ (and similarly for $W_{f}$), with the prime denoting a derivative with respect to the argument of the function. In general, the integrals in~(\ref{eqn:VOP2}) cannot be solved analytically. However, to perform the matched asymptotic expansions, we are only interested in the asymptotic expansions of these integrals.  To avoid clutter, we will work with the $z$ and $x$-coordinates for the near and far-zone solutions, instead of converting them to the matching coordinate $\xi$. Recall that the limit $\alpha \to 0$  of the asymptotic expansions in terms of $\xi$ is equivalent to taking the limits $z\to \infty$ and $x \to 0$. 

\vskip 4pt
We first consider the integrals for the far-zone solutions. It is convenient to rewrite these integrals in the following general form 
\begin{align}
\int^\infty_x \d t \, \frac{J_i^x(t) f (t)}{W_f(t)} \equiv \int^\infty_x \d t \, e^{-t} H_i^x(t) \, ,
\end{align}
where the exponential factor regulates all divergences in the limit $t \to \infty$. To obtain the asymptotic series in the limit $x \to 0$, we Taylor expand $H_i^x(t)$ for  $t\to 0$. The result can be organized in the following way~\cite{bender1999advanced}
\beq
\begin{aligned}
\int^\infty_x \d t \, e^{-t} H_i^x(t)  = & \int^\infty_x \d t  \, e^{-t} \sum_{k=N_{\rm min}}^{-1} \sum_{m=0}^\infty  a_{k,m} \hskip 1pt t^k \log^m(t) \\
& +  \int^\infty_0 \d t  \, e^{-t} \Bigg[ H_i^x(t) - \sum_{k=N_{\rm min}}^{-1} \sum_{m=0}^\infty  a_{k,m} \hskip 1pt t^k \log^m(t)\Bigg]   \\
& -  \int^x_0 \d t  \, e^{-t}  \sum_{k=0}^{\infty} \sum_{m=0}^\infty  a_{k,m} \hskip 1pt t^k \log^m(t)\, , \label{eqn:xintegral}
\end{aligned}
\eeq
where $a_{k, m}$ are constants, and $N_{\rm min}$ is the smallest power that appears in the series. The first line in (\ref{eqn:xintegral}) consists of terms that converge as    $t \to \infty$, the second line is a constant, while the third line consists of terms that diverge as $t \to \infty$, but converge as $t \to 0$. This is why we have rewritten the integral limits $\int^\infty_x \to \int^\infty_0 - \int^x_0$ in the last line. Written in this form, (\ref{eqn:xintegral}) can be integrated term by term to give the asymptotic series as $ x \to 0$.

\vskip 4pt
Similarly, in the near zone, we encounter integrals of the form 
\begin{align}
\int_0^z \d t \, \frac{J^z_i (t) n(t)}{W_n(t)} \equiv \int_0^z \d t \, e^{-1/t} \hskip 1pt H_i^z(t) \, ,
\end{align}
where the exponential factor regulates potential divergences for $t \to 0$. To obtain the asymptotic expansion as $z \to \infty$, we Taylor expand $H^z_i(t)$ for $t \to \infty$. The integral  then becomes 
\beq
\begin{aligned}
\int_0^z \d t \, e^{-1/t} \hskip 1pt H_i^z(t)  = & \int_0^z \d t \, e^{-1/t} \hskip 1pt \sum^{N_{\rm max}}_{k=-1} \sum_{m=0}^\infty  a_{k,m} \hskip 1pt t^k \log^m(t)  \\
& + \int_0^\infty \d t \, e^{-1/t} \hskip 1pt \Bigg[ H_z(t) -  \sum^{N_{\rm max}}_{k=-1} \sum_{m=0}^\infty  a_{k,m} \hskip 1pt t^k \log^m(t)  \Bigg]  \\
& -\int_z^\infty \d t \, e^{-1/t} \hskip 1pt
\sum^{-2}_{k=-\infty} \sum_{m=0}^\infty  a_{k,m} \hskip 1pt t^k \log^m(t) \, , \label{eqn:zintegral}
\end{aligned}
\eeq
where $N_{\rm max}$ is the maximum power of the integrand (which varies depending on the form of~$J^z_i $). The reasoning behind the organization of (\ref{eqn:zintegral}) is similar to that for (\ref{eqn:xintegral}): the first line converges as $t \to 0$, the second line is a constant, and the third line is convergent as $t \to \infty$. To evaluate the integrals in \eqref{eqn:zintegral}, it is convenient to perform the coordinate transformation $s \equiv 1/t$, such that they become directly analogous to the integrals in (\ref{eqn:xintegral}).

\subsection{Ordinary Perturbation Theory} 

In the main text, we have argued that the $\ell > 0$ modes can be treated by extrapolating the far-zone solutions towards the outer horizon, $x \to 0$, and solve the spectra using ordinary perturbation theory.
Here, we will provide further details of that claim, We will also present results for the higher-order electric angular problem that were omitted in~\S\ref{sec:ProcaLO}.

\vskip 4pt
To compute the higher-order eigenvalues, we take the inner product of the general equations of motion (\ref{eqn:Jdef}) with the zeroth-order eigenstates 
\beq
0 = \int w \, X^*_0 \, \square^{\indlab{0}}  X_i = \int w \, X^*_0 \, J^X_i \, , \label{eqn:ExpEval}
\eeq
where $w$ is the Sturm-Louiville weight factor and the integral is performed with respect to the relevant coordinate. The left-hand side vanishes after using the hermiticity of $\square^{\indlab{0}}$ and the leading-order equations of motion. The eigenvalues, which are contained in $J_i^X$, are then computed at every order by performing this integral. At each order in $\alpha$, we may back-substitute the eigenvalues and equations of motion of the previous order to simplify $J_i^X$. 

\vskip 4pt
The angular equations for the scalar field are trivially solved, since $J_i^\theta = 0$ up to the order of interest, cf.~(\ref{eqn:ScalarHigherAngular}). For the vector field, on the other hand, the higher-order angular equations contain additional $\theta$-dependent terms on the right-hand side of (\ref{equ:ProcaS}), which induce new cross couplings in the angular eigenstates, cf.~(\ref{eqn: Electric angular eigenstate order2}). The coefficients in (\ref{eqn: Electric angular eigenstate order2}) are
\beq
\begin{aligned}
b_{j-2} & =  - \left[ \frac{(j^2 - m^2)[(j-1)^2 - m^2]}{(2j-3)(2j-1)^2(2j+1)} \right]^{1/2} \frac{(j+1 - \lambda_0)}{\lambda_0^2(2j-1)} \, ,  \\[4pt]
b_{j+2} & = - \left[\frac{[(j+1)^2 - m^2] [(j+2)^2 - m^2] }{(2j+1)(2j+3)^2(2j+5)} \right]^{1/2}  \!\!\frac{( j + \lambda_0)}{\lambda_0^2(2j+3)} \, , 
\end{aligned}
\eeq
and  $c_{j \pm 2} = 2 m \lambda_0^{-5} \, b_{j \pm 2} $. These expressions are valid for both $\lambda_0 = \lambda_0^\pm$. Up to order $\alpha^3$, the angular eigenvalues for the $j=\ell \pm 1$ modes are  
\begin{align}
\lambda = \lambda_0 - \frac{\alpha \tilde{a} m}{\lambda_0} &- \left(\frac{\lambda_0}{2 n^2( 2 \lambda_0 - 1)}-\frac{\tilde{a}^2 (\lambda_0 + 1) (\lambda_0^2 - m^2)}{\lambda_0^3(2 \lambda_0 + 1)} \right)\alpha^2  \nonumber \\
&+ \left(\frac{1}{n^2 (2 \lambda_0 - 1)} + \frac{\tilde{a}^2 ( 2 + \lambda_0)(\lambda_0^2 - m^2)}{\lambda_0^5 (2 \lambda_0 + 1)}\right) \tilde{a}\es m \es \alpha^3\, . \label{eqn:lambdaPlusMinus}
\end{align}
These results are substituted into the radial equations to solve for the energy eigenvalues at higher orders.

\vskip 4pt
For the radial equations, we restrict ourselves to the far-zone radial equations 
and impose regular boundary conditions at $x=0$. The latter assumption allows us to remove derivative terms in $J_i^x$ through integrations by parts. For a scalar field, the leading-order far-zone solution can be written as
\beq
R^{\text{far}}_0(x) =  \sqrt{\frac{(n-\ell-1)!}{2 n (n+\ell)!}} \, e^{-x/2} x^\ell L^{(2\ell+1)}_{n-\ell-1}(x) \, , \label{eqn: scalar radial far LO 2}
\eeq
where $L_k^{(\rho)}$ is the associated Laguerre polynomial. 
The overall coefficient has been fixed by requiring the integral of the 
square of the mode function to be unity. 
Each term in the inner product (\ref{eqn:ExpEval}) then has the following generic form
\beq
\langle x^{-s} \rangle \equiv \frac{(n-\ell-1)!}{2 n (n+\ell)!} \int_0^\infty \d x \, e^{-x} x^{2\ell+2-s}\, \left( L^{(2\ell+1)}_{n-\ell-1} \right)^2 \, , \label{eqn:farexpval}
\eeq
where $J_i^x \supset x^{-s}$, with positive integer $s$. We may then compute the energy eigenvalues through the following identities
\beq
\begin{alignedat}{2}
\left\langle \frac{1}{x} \right\rangle & = \frac{1}{2n} \, , &&  \qquad
\left\langle \frac{1}{x^2} \right\rangle = \frac{1}{2n (2\ell+1)} \, , \\
\left\langle \frac{1}{x^3} \right\rangle & = \frac{1}{2\ell (2\ell+1) (2\ell+2)}   \, , && \qquad  \left\langle \frac{1}{x^4} \right\rangle   = \frac{3 n^2-\ell ^2-\ell }{n  (2 \ell -1)(2 \ell) (2 \ell +1)(2\ell +2) (2 \ell +3)} \, , 
\end{alignedat}
 \label{eqn:Identities1}
\eeq
which are valid for all $\ell$. The fact that $\langle x^{-3} \rangle$ and $\langle x^{-4} \rangle$ diverge for $\ell =0$ reflects the sensitivity of these modes to the near region of the black hole. In general, the presence of these terms indicates a breakdown of ordinary perturbation theory, and calculating the eigenvalues requires the more rigorous matched asymptotic expansion. However, in the case of a scalar field, these terms have coefficients that are proportional to $m$, which vanish for $\ell=0$, so that the naively divergent terms do not contribute to the spectrum. Ordinary perturbation theory, with the assumption of a regular boundary condition at the horizon, is therefore sufficient for the scalar spectrum.

\vskip 4pt
Although the far-zone radial equations of the scalar and vector fields are different, the inner product (\ref{eqn:farexpval}) is also valid for the vector case. This is because the additional power of $x$ that appears in the radial solution for a vector field cancels, as the Sturm-Liouville weight factor is now $1$, instead of $x^2$ for the scalar case. The identities (\ref{eqn:Identities1}) are therefore also valid for the vector field. 
Unlike in the scalar case, however, the coefficients of the $\langle x^{-3} \rangle$ and $\langle x^{-4} \rangle$  terms do not vanish for the vector $\ell=0$ modes, so ordinary perturbation theory is expected to yield divergent results. Indeed, these divergences can be written as divergent boundary terms in the integral (\ref{eqn:farexpval}). Remarkably, however, if these boundary terms are discarded, we obtain results that agree with those obtained through the more rigorous matched asymptotic expansion. Since an implicit assumption in deriving the rules (\ref{eqn:Identities1}) is precisely that any boundary terms are negligible, substituting them into the divergent operators can still provide finite results. Indeed, we find that the coefficients of the $\langle x^{-3} \rangle$ and $\langle x^{-4} \rangle$ terms are precisely such that the divergences for $\ell=0$ cancel in the sum. As a consequence of this cancellation, ordinary perturbation theory gives the correct results even for the fine and hyperfine splittings of vector spectrum. 
Interestingly, two wrongs \emph{can} make a right.

\newpage
  \section{Details of the Numerical Treatment} 
  \label{app:numericalDetails}
  
  A large part of the analysis presented in Section~\ref{sec:numeric} was focused on achieving accurate numeric results for the quasi-bound state spectrum without a separable ansatz. In \S\ref{sec:nosep}, we presented a schematic outline for how one translates the unseparated Proca equation and Lorenz constraint into a nonlinear eigenvalue problem that can be readily solved on a computer. In the interest of pedagogy, we kept technical details there to a minimum, and will instead provide them in this appendix. 

  \vskip 4pt
  We first detail the decomposition of the Proca equation into its temporal, radial, and angular components using the vector spherical harmonics defined in \S\ref{app:vsh}, and describe how to choose these harmonics such that the quasi-bound state boundary conditions are satisfied. We then detail the construction of the finite-dimensional matrix, whose nonlinear eigenvalues determine the bound state spectrum. Finally, in \S\ref{app:cheb}, we provide a brief introduction to Chebyshev interpolation with an emphasis on its convergence properties and numeric implementation.  

\subsection{Decomposition of the Proca Equation} 
\label{app:decompProca}

Our numerical analysis relies on a $1+1+2$ decomposition of the Proca equation into its temporal, radial and angular components.  While such a decomposition is straightforward for a scalar field, the vector index complicates things. It will therefore be useful to first rewrite the Proca equation as a set of coupled scalar equations, reminiscent of the decomposed Klein-Gordon equation (\ref{eq:startingPoint}), with operators that act simply on the scalar spherical harmonics.
  
  \vskip 4pt
  To this end, we introduce a basis of vector fields  $e\indices{_a^\mu} \partial_\mu$ and dual form fields $f\indices{^a_\mu} \, \ud x^\mu$, where $a = 0, 1, 2, 3$. These bases are dual to one another in the sense that
    \begin{equation}
    \begin{aligned}
      \bar{e}\indices{_a^\mu} f\indices{^b_\mu} &= \delta^a_b \,,\\
       \bar{e}\indices{_a^\mu} f\indices{^a_\nu}  &= \delta_\nu^\mu \,, \label{eqn:Tetrad}
      \end{aligned}
    \end{equation}
    where the bar denotes complex conjugation. Taking the vector fields to have definite angular momentum, we may write
      \begin{align}
          e\indices{_0^\mu} \partial_\mu &= F_t(r) \, \partial_t\,, \quad \ e\indices{_1^\mu} \partial_\mu = \chi_{1, 1}^i \, \partial_i \, , \qquad \ e\indices{_2^\mu} \,\partial_\mu = \chi_{1, 0}^i \, \partial_i\,, \qquad e\indices{_3^\mu} \,\partial_\mu = \chi_{1, \sminus 1}^{i} \,\partial_i\,, \label{eq:vecFieldDefs} \\
       f\indices{^0_\mu} \ud x^\mu &= F_t^{-1} \,\ud t\, ,\quad f\indices{^1_\mu}\ud x^\mu = \chi_i^{1,1}\,\ud x^i\,, \quad f\indices{^2_\mu} \ud x^\mu  = \chi_i^{1, 0} \, \ud x^i \, , \quad f\indices{^3_\mu} \ud x^\mu = \chi_{i}^{1 ,\sminus 1} \, \ud x^i\,. \label{eq:formFieldDefs}
\end{align}
where we have used the forms and vectors defined in the previous section.  By design, these forms are well-behaved under the Kerr isometries,
    \begin{equation}
      \pounds_t f\indices{^a_\mu} = 0\,, \quad \pounds_z f\indices{^0_\mu} = 0\,, \quad \text{and} \quad \pounds_z f\indices{^i_\mu} = (2 - i) f\indices{^i_\mu}\,, \quad \text{where} \quad i = 1, 2, 3\, ,
    \end{equation}
    and are eigenstates of the angular momentum operator, $\pounds^2 f\indices{^0_\mu} = 0$ and $\pounds^2 f\indices{^i_\mu} = 2 f\indices{^i_\mu}$. Similar relations hold for the basis of vectors, $e\indices{_a^\mu}$. Finally, it is useful to note that
    \begin{align}
      \partial_\mu &= \bar{f}\indices{^a_\mu}e\indices{_a^\nu}\partial_\nu \nonumber \\[2pt]
      &=  i \bar{f}\indices{^{\,0}_\mu} F_t \pounds_t + \delta_\mu^r \partial_r +  \bar{f}\indices{^{\,1}_\mu} F_\theta \mathcal{D}_+ + \bar{f}\indices{^{\,2}_\mu} F_\theta \mathcal{D}_0 + \bar{f}\indices{^{\,3}_\mu} F_\theta \mathcal{D}_-\,.
    \end{align}
    Our goal is to rewrite the Proca equation,\footnote{At a technical level, the Proca equation with a lowered index is much simpler than with the upper index and so we focus exclusively on this form.} $\nabla^2 A_\mu =A_\mu$,
    and the Lorenz condition, $\nabla^\mu A_\mu = 0$, into a form that is (roughly) a system of coupled scalar differential equations.  
    We assume that $A_\mu$ is in a state of definite frequency and azimuthal angular momentum,
    \begin{equation}
    \begin{aligned}
      \pounds_t A_\mu &= - \omega A_\mu \, , \\ \pounds_z A_\mu &= +m A_\mu\,,
      \end{aligned}
    \end{equation}
    and write   $A_\mu = A_a f\indices{^a_\mu}$.
       
    \vskip 4pt
    We begin by noting that, for any scalar $\Phi$, the purely radial derivative operator $\partial_r(\Delta \partial_r)$ can be rewritten using the Laplacian $\nabla^2$ and isometry generators $\pounds^2$, $\pounds_t$ and $\pounds_z$:
    \begin{equation}
        \partial_r\left(\Delta \partial_r \Phi\right) = \left[\Sigma \nabla^2 + \pounds^2 - \left(\Sigma + \frac{2 \alpha r(r^2 + \alpha^2 \tilde{a}^2)}{\Delta}\right)\pounds_t^2 - \frac{\alpha^2 \tilde{a}^2}{\Delta} \pounds_z^2 - \frac{4 \alpha^2 \tilde{a} r}{\Delta} \pounds_t \pounds_z\right] \Phi\,.
    \end{equation}
    For the Klein-Gordon equation, this almost immediately yields the decomposition into temporal, radial, and angular degrees of freedom we are after. The Proca equation requires a bit more work, but eventually can be written as
      \begin{align}
        0\, =\, \ & \bar{e}\indices{_b^\mu}\Delta^{-1} \Sigma \left(\nabla^2 - 1\right)\left(A_a f\indices{^a_\mu}\right) \nonumber \\
          =  \ & \bigg[\frac{1}{\Delta} \partial_r(\Delta \partial_r) - \frac{1}{\Delta} \left(\pounds^2 + \alpha^2 \tilde{a}^2 \left(1 - \omega^2\right) \cos^2 \theta \right) \nonumber \\
        & - (1 - \omega^2) + \frac{P_+^2}{(r - r_+)^2} + \frac{P_-^2}{(r -r_-)^2} - \frac{A_+}{(r - r_+)(r_+ - r_-)} + \frac{A_-}{(r -r_-)(r_+ - r_-)}\bigg] A_b \nonumber \\[4pt]
        &+ \mathcal{S}\indices{_b^a} A_a + \mathcal{Q}\indices{_b^a} \pounds_z A_a + \mathcal{R}\indices{_b^a} \partial_r A_a + \mathcal{P}\indices{_b^a} \mathcal{D}_+ A_a + \mathcal{Z}\indices{_b^a} \mathcal{D}_0 A_a + \mathcal{M}\indices{_b^a} \mathcal{D}_- A_a\,, \label{eq:procaEqDecomp}
      \end{align} 
      where we have introduced the following `mixing matrices':
          \begin{equation}
        \begin{aligned}
          \mathcal{S}\indices{_b^a} &= \frac{1}{\Delta} \bar{e}\indices{_b^\mu}\Big[\Sigma \nabla^2 f\indices{^a_\mu} - \frac{\alpha^2 \tilde{a}^2}{\Delta} \pounds_z^2 f\indices{^a_\mu} + \frac{4 \alpha^2 \tilde{a} \omega r}{\Delta} \pounds_z f\indices{^a_\mu} - 2 i \omega F_t \Sigma \,\bar{f}\indices{^{\,0}_\rho} \nabla^\rho f\indices{^a_\mu} \Big] \, ,\\[0.6ex]
          \mathcal{Q}\indices{_b^a} &=  - \frac{2 \alpha^2 \tilde{a}^2}{\Delta^2} \bar{e}\indices{_b^\mu} \pounds_z f\indices{^a_\mu}\, , \qquad  \quad \ \mathcal{R}\indices{_b^a} = 2 \Sigma \Delta^{-1}\,  \bar{e}\indices{_b^\mu} \nabla^\rho f\indices{^a_\mu} \delta^r_\rho\, , \\[0.6ex]
          \mathcal{P}\indices{_b^a} &= 2\Sigma \Delta^{-1}\,  \bar{e}\indices{_b^\mu} \nabla^\rho f\indices{^a_\mu} \bar{f}\indices{^{\,1}_\rho} F_\theta\, ,  \quad \mathcal{Z}\indices{_b^a} = 2\Sigma\Delta^{-1}\,  \bar{e}\indices{_b^\mu} \nabla^\rho f\indices{^a_\mu} \bar{f}\indices{^{\,2}_\rho} F_\theta\, , \\[0.75ex]
          \mathcal{M}\indices{_b^a} &= 2 \Sigma \Delta^{-1}\,  \bar{e}\indices{_b^\mu} \nabla^\rho f\indices{^a_\mu} \bar{f}\indices{^{\,3}_\rho} F_\theta\,.
        \end{aligned} \label{eq:mixingMatrices}
      \end{equation}
      Similarly, the Lorenz condition can be written as
      \begin{equation}
        0 \,=\, \mathcal{T}^0 A_0 + \mathcal{S}^i A_i + \mathcal{R}^i \partial_r A_i + \mathcal{P}^i \mathcal{D}_+ A_i + \mathcal{Z}^i \mathcal{D}_0 A_i + \mathcal{M}^i \mathcal{D}_- A_i\,, \label{eq:lorenzEqDecomp}
      \end{equation}
      where we have defined the `mixing vectors':
      \begin{equation}
        \begin{aligned}
          \mathcal{S}^i &= \nabla^\mu f\indices{^i_\mu} - i \omega F_t f\indices{^i_\mu} g^{\mu \lambda} \bar{f}\indices{^{\,0}_\lambda}\,, \qquad \mathcal{R}^i = f\indices{^i_\mu} g^{\mu r}\,, \qquad \mathcal{P}^i = f\indices{^i_\mu} g^{\mu \lambda} \bar{f}\indices{^{\,1}_\lambda} F_\theta\,, \\[0.8ex]
           \mathcal{Z}^i &= f\indices{^i_\mu} g^{\mu \lambda} \bar{f}\indices{^{\,1}_\lambda} F_\theta\,, \qquad  \mathcal{M}^i = f\indices{^i_\mu} g^{\mu \lambda} \bar{f}\indices{^{\,1}_\lambda} F_\theta \,, \qquad \mathcal{T}^0 = i F_t^{-1}\big(m g^{t \phi} - \omega g^{tt}\big).
        \end{aligned} \label{eq:mixingVectors}
      \end{equation}
      These mixing matrices and vectors encode how the temporal and spatial components couple to one another and how the Kerr geometry distinguishes scalar and vector fields.

  \subsection{Boundary Conditions} 
  \label{app:boundaryConditions}

    So far, the radial functions $F_r$, $F_\theta$, $F_t$ have not been specified. 
    We are free to choose these functions in such a way that all of the components $A_a$ 
scale in the same way as $r \to r_+$ and $r \to \infty$. This will make it easy to impose the ingoing boundary conditions at the outer horizon and the decaying boundary conditions at spatial infinity.

\vskip 4pt
    Let us first concentrate on the behavior at the event horizon, which will be easiest to analyze using the standard form of the Proca equation $\nabla^2 A_\mu = A_\mu$ instead of the decomposition (\ref{eq:procaEqDecomp}). We demand that $A_\mu$ is an eigenstate of both frequency $\pounds_t A_\mu = - \omega A_\mu$ and angular momentum, $\pounds_z A_\mu = m A_\mu$, and so the coefficients $A_\mu$ can all be written as $A_\mu = A_\mu(r, \theta)\,e^{-i \omega t + i m \phi}$. We can then solve the Lorenz condition for $A_0(r, \theta)$ and eliminate it from the Proca equation to attain equations of motion that only involve the $A_i$. While the system of equations is quite complicated, for what follows, we only need to consider the asymptotic behavior of these equations as $r \to r_+$. We will thus use $(\,\dots)$ to denote functions that are constant in $r$ (though they generally depend on $\theta$, $r_\pm$, $m$ and $\omega$) and  whose precise form are irrelevant to the discussion.

\vskip 4pt
    With the benefit of hindsight, we take
    \begin{equation}
      A_r = \left(\frac{r - r_-}{r- r_+}\right) \tilde{A}_r\,, \quad A_\theta = (r - r_-) \tilde{A}_\theta\,, \quad A_\phi = (r - r_-) \tilde{A}_\phi \,.
    \end{equation}
    With these factors peeled off, the near-horizon behavior for the $r$, $\theta$, $\phi$ components of the Proca equation are 
      \begingroup \allowdisplaybreaks
      \begin{align}
        0 \,=\,\ & \partial_r^2 \tilde{A}_r + \frac{1}{r - r_+} \partial_r \tilde{A}_r + \frac{P_+^2}{(r -r_+)^2} \tilde{A}_r + \frac{(\,\dots)}{r - r_+} \partial_\theta \tilde{A}_\theta + \frac{(\,\dots)}{r - r_+} \partial_\theta^2 \tilde{A}_r + (\,\dots) \partial_r \tilde{A}_\theta  \nonumber \\
        & +  \frac{(\,\dots)}{r - r_+} \partial_\theta \tilde{A}_r+ \frac{(\,\dots)}{r -r_+} \tilde{A}_\phi  + \frac{(\,\dots)}{r-r_+} \tilde{A}_\theta + \cdots \, ,\\
        0 \,=\, \ & \partial_r^2 \tilde{A}_\theta +\frac{1}{r - r_+} \partial_r \tilde{A}_\theta +  \frac{P_+^2}{(r - r_+)^2} \tilde{A}_\theta  + \frac{(\,\dots)}{r - r_+} \partial_\theta \tilde{A}_r +\frac{(\,\dots)}{r - r_+} \partial_\theta \tilde{A}_\theta   \nonumber \\
        &+ \frac{(\,\dots)}{r - r_+} \partial_\theta^2 \tilde{A}_\theta +  \frac{(\,\dots)}{r - r_+} \tilde{A}_\phi+ (\,\dots) \partial_r \tilde{A}_r +(\,\dots) \tilde{A}_r + \cdots\, ,\\
        0 \,=\ \, & \partial_r^2 \tilde{A}_\phi + \frac{1}{r - r_+} \partial_r \tilde{A}_\phi + \frac{P_+^2}{(r - r_+)^2} \tilde{A}_\phi + \left(\,\dots\right)\left(\partial_r^2 \tilde{A}_r + \frac{1}{r - r_+} \partial_r \tilde{A}_r + \frac{P_+^2}{(r -r_+)^2}\tilde{A}_r \right) \nonumber \\
        &+ \frac{(\,\dots)}{r -r_+} \tilde{A}_\theta + (\,\dots) \partial_r \tilde{A}_\theta + (\,\dots) \partial_r \partial_\theta \tilde{A}_\theta  + (\,\dots) \partial_\theta \tilde{A}_r + \frac{(\,\dots)}{r - r_+} \partial_\theta \tilde{A}_\theta \nonumber \\
        &+ \frac{(\,\dots)}{r - r_+} \partial_\theta \tilde{A}_\phi + (\,\dots) \partial_\theta^2 \tilde{A}_\theta + \frac{(\,\dots)}{(r - r_+)} \partial_\theta^2 \tilde{A}_\phi + \cdots\, .
      \end{align}
      \endgroup
      We see that most singular terms imply that $\tilde{A}_r$, $\tilde{A}_\theta$, $\tilde{A}_\phi$ all go as $(r - r_+)^{\pm i P_+}$ as $r \to r_+$. In a similar way, we can also examine the Lorenz condition as $r \to r_+$, 
      \begin{equation}
        A_0 = (\,\dots)(r -r_+) \tilde{A}_r + (r-r_+) \partial_r \tilde{A}_r + (\,\dots)(r-r_+) \partial_\theta \tilde{A}_\theta + (\,\dots)(r-r_+) \tilde{A}_\theta + (\,\dots) \tilde{A}_\phi\,,
      \end{equation}
      and we see that $A_0$ also approaches $(r - r_+)^{\pm i P_+}$ as $r \to r_+$. With these scalings in mind, we take the radial functions to be
      \begin{equation}
        F_t = 1\,, \quad F_r = \frac{r - r_+}{r - r_-}\,, \quad F_\theta = \frac{1}{r - r_-}\,, \label{eq:decompRadialFuncs}
      \end{equation} 
      so that $A_a \sim (r- r_+)^{\pm i P_+}\left(1 + \mathcal{O}(r -r_+)\right)$ as $r \to r_+$ for each $a$.

\vskip 4pt
      To analyze how the coefficients $A_a$ behave as $r \to \infty$, we turn to the decomposed Proca equation (\ref{eq:procaEqDecomp}), where the mixing matrices are computed with the radial functions (\ref{eq:decompRadialFuncs}). 
      Ignoring the mixing matrices, for a moment, (\ref{eq:procaEqDecomp}) predicts that the $A_a$ scale in the same way as the scalar solution~(\ref{eqn: scalar radial infinity}):
      \begin{equation}
        A_a \sim r^{-1 - \nu + 2 \alpha^2/\nu} e^{-\sqrt{1 - \omega^2}(r - r_+)}\left(1 + \mathcal{O}(r^{-1})\right) .
      \end{equation}
      This conclusion will not be affected by including the mixing matrices, as long as they 
       decay faster at spatial infinity than $r^{-1}$ and, by explicit computation, we confirm that $\mathcal{S}\indices{_0^0} = 0$, $\{\mathcal{S}\indices{_0^i}, \mathcal{S}\indices{_i^0}, \mathcal{R}\indices{_0^0}, \mathcal{R}\indices{_i^j}\}$ scale as $r^{-2}$, $\{\mathcal{S}\indices{_i^j}, \mathcal{R}\indices{_0^i}, \mathcal{R}\indices{_i^0}, \mathcal{P}\indices{_0^0}, \mathcal{P}\indices{_i^j}, \mathcal{Z}\indices{_0^0}, \mathcal{Z}\indices{_i^j}, \mathcal{M}\indices{_0^0}, \mathcal{M}\indices{_i^j}\}$ scale as $r^{-3}$, while the rest decay as $r^{-4}$.

\vskip 4pt
      We have chosen the radial functions (\ref{eq:decompRadialFuncs}), so that the components $A_a$ all have the same asymptotic behavior as $r \to r_+$ and $r \to \infty$. We then peel off this asymptotic behavior by writing
      \begin{equation}
        A_a(r) = \left(\frac{r - r_+}{r - r_-}\right)^{i P_+} \!\!\!\left(r - r_-\right)^{-1+\nu-2 \alpha^2/\nu} e^{-\alpha(r - r_+)/\nu} B_a(r)\,, \label{eq:asympBehaviorVec}
      \end{equation}
      and work with the functions $B_a$, which, for the modes of interest, approach constants as $r \to r_+$ and $r \to \infty$. In the numerics, this allows us to impose the boundary conditions for the quasi-bound states by simply expanding $B_a$ in a set of functions that approach constant values at the boundaries.

    \subsection{Constructing the Matrix Equation}
      To convert (\ref{eq:procaEqDecomp}) into a finite-dimensional matrix equation, we first write \eqref{eq:asympBehaviorVec} as
       \begin{equation}
        A_a(r) = \left(\frac{r - r_+}{r - r_-}\right)^{i P_+} \!(r - r_-)^{-1 + \nu - 2 \alpha^2/\nu}e^{-\alpha(r - r_+)/\nu} B_a(\zeta(r)) \equiv F(r) B_a(\zeta(r))\,,
      \end{equation}
    which is similar to the scalar case (\ref{eq:pseudoSpecPeelOff}), with $\zeta(r)$ a map from $[r_+, \infty)$ to $[-1, 1]$. The spatial components $i = 1, 2, 3$ of the Proca equation (\ref{eq:procaEqDecomp}) can then be written as
      \begin{equation}
        \mathcal{D}\indices{_i^k}[B_k] + \mathcal{D}\indices{_i^0}[B_0] = 0\,, \label{eq:procaSchematic}
      \end{equation}
      where
      \begin{align}
        \mathcal{D}\indices{_i^k}[B_k] \,=\,\ & \Bigg(\partial_\zeta^2 + \frac{1}{\zeta'(r)}\left(\frac{1}{r - r_+} + \frac{1}{r - r_-} + \frac{2 F'(r)}{F(r)} + \frac{\zeta''(r)}{\zeta'(r)}\right) \partial_\zeta \Bigg) B_i\nonumber \\
        & - \frac{1}{\zeta'(r)^2 \Delta} \left(\pounds^2 + \alpha^2 \tilde{a}^2 \left(1 - \omega^2\right) \cos^2 \theta \right)B_i + \frac{1}{\zeta'(r)^2} \Bigg(\frac{F'/F}{r-r_+} + \frac{F'/F}{r - r_-} + \frac{F''}{F} \nonumber
\\
        & - (1 - \omega^2) + \frac{P_+^2}{(r - r_+)^2} + \frac{P_-^2}{(r -r_-)^2} - \frac{A_+}{(r - r_+)(r_+ - r_-)} + \frac{A_-}{(r -r_-)(r_+ - r_-)}\Bigg) B_i \nonumber \\[4pt]
        &+ \tilde{\mathcal{S}}\indices{_i^k} B_k + \tilde{\mathcal{Q}}\indices{_i^k}\, \pounds_z B_k + \tilde{\mathcal{R}}\indices{_i^k}\, \partial_\zeta B_k + \tilde{\mathcal{P}}\indices{_i^k}\, \mathcal{D}_+ B_k + \tilde{\mathcal{Z}}\indices{_i^k}\, \mathcal{D}_0 B_k + \tilde{\mathcal{M}}\indices{_i^k} \,\mathcal{D}_- B_k\, ,     \\[14pt]
        \mathcal{D}\indices{_i^0}[B_0] \,=\,\ & \tilde{\mathcal{S}}\indices{_i^0} B_0 + \tilde{\mathcal{Q}}\indices{_i^0} \pounds_z B_0 + \tilde{\mathcal{R}}\indices{_i^0} \partial_\zeta B_i + \tilde{\mathcal{P}}\indices{_i^0} \,\mathcal{D}_+ B_0 + \tilde{\mathcal{Z}}\indices{_i^0} \,\mathcal{D}_0 B_0 + \tilde{\mathcal{M}}\indices{_i^0} \,\mathcal{D}_- B_0\, .
      \end{align}
 In the above, we have defined the transformed mixing matrices 
      \begin{equation}
        \begin{aligned}
        \tilde{\mathcal{S}}\indices{_b^a} &= \frac{1}{\zeta'^{\hskip 1pt 2}} \mathcal{S}\indices{_b^a} + \frac{F'}{ \zeta'^{\hskip 1pt 2} F} \mathcal{R}\indices{_b^a}\,, \quad \tilde{\mathcal{R}}\indices{_b^a} = \mathcal{R}\indices{_b^a}/\zeta'\,, \quad \tilde{\mathcal{Q}}\indices{_b^a} = \mathcal{Q}\indices{_b^a}/\zeta'^{\hskip 1pt 2} \,,\\[1ex]
         \tilde{\mathcal{P}}\indices{_b^a} &= \mathcal{P}\indices{_b^a}/\zeta'^{\hskip 1pt 2} \,,\quad \tilde{\mathcal{Z}}\indices{_b^a} = \mathcal{Z}\indices{_b^a}/\zeta'^{\hskip 1pt 2} \,,\quad \tilde{\mathcal{M}}\indices{_b^a} = \mathcal{M}\indices{_b^a}/\zeta'^{\hskip 1pt 2}\,,
      \end{aligned}
      \end{equation}
      where $\zeta' = \partial_r \zeta$.  Similarly, the Lorenz condition can be written as
      \begin{equation}
        \mathcal{D}\indices{_0^0}[B_0] + \mathcal{D}\indices{_0^i}[B_i] = 0\,,\label{eq:lorenzConstraintSchematic}
      \end{equation}
      where
      \begin{align}
    \mathcal{D}\indices{_0^0}[B_0] &= \mathcal{T}^0 B_0\,, \\
        \mathcal{D}\indices{_0^i}[B_i] &=\tilde{\mathcal{S}}^i  B_i + \tilde{\mathcal{R}}^i \partial_\zeta B_i + \mathcal{P}^i \mathcal{D}_+ B_i + \mathcal{Z}^i \mathcal{D}_0 B_i + \mathcal{M}^i \mathcal{D}_- B_i\,,
      \end{align}
      with $ \tilde{\mathcal{S}}^i = \mathcal{S}^i + F' \mathcal{R}^i/F$ and $\tilde{\mathcal{R}}^i = \zeta' \, \mathcal{R}^i$.
  
\vskip 4pt
      We then expand the  temporal component of the field into scalar harmonics and the spatial components into one-form harmonics, 
      \begin{align}
       B_0 &= \sum_{j'} B_{j' m} Y_{j' m}\, , \\
        B_i &= \sum_{\ell', j'} B_{\ell',j' m} \,Y_i^{\ell',j' m} \, .
      \end{align}
      As discussed in \S\ref{app:vsh}, the summation ranges of $j'$ and $\ell'$ depend on the parity of the mode we are solving for. 
      We then project the Proca equation (\ref{eq:procaSchematic}) onto the vector harmonics and the Lorenz constraint (\ref{eq:lorenzConstraintSchematic}) onto the scalar harmonics, to obtain
      \begin{align}
        \sum_{\ell',j'} \mathcal{D}^{\lab{ss}}_{\ell j\,|\,\ell' j'}[B_{\ell', j' m}(\zeta)] + \sum_{j'} \mathcal{D}^{\lab{st}}_{\ell j\, |\, j'}[B_{j'm}(\zeta)] &= 0\,, \label{eq:procaProject} \\
          \sum_{\ell',j'} \mathcal{D}^{\lab{ts}}_{j\, |\, \ell' j'}[B_{\ell', j' m}(\zeta)] + \sum_{j'} \mathcal{D}^{\lab{tt}}_{j\, | \, j'}[B_{j' m}(\zeta)] &= 0\,. \label{eq:lorenzProject}
      \end{align}
      Explicitly, we find
      \begin{equation}
        \begin{aligned}
          \mathcal{D}^\lab{ss}_{\ell j\, |\, \ell' j'}[B_{\ell',j' m}(\zeta)] &= \int_{\lab{S}^2}\!\ud \Omega\,\, \bar{Y}^{i}_{\ell,j m} \, \mathcal{D}\indices{_i^k}\big[B_{\ell',j'm}(\zeta) Y^{\ell',j' m}_k\big]\, , \\
          \mathcal{D}^{\lab{st}}_{\ell j\,|\,j'}[B_{j'm}(\zeta)] & = \int_{\lab{S}^2}\!\ud \Omega\, \, \bar{Y}^{i}_{\ell,jm} \, \mathcal{D}\indices{_i^0}\big[B_{j'm}(\zeta) Y_{j' \,m}\big]\, , \\
          \mathcal{D}^{\lab{t s}}_{j\,|\,\ell'j'}[B_{\ell',j'm}(\zeta)] &= \int_{\lab{S}^2}\!\ud \Omega \,\, \bar{Y}_{jm}\,\mathcal{D}\indices{_0^k}\big[B_{\ell', j' m}(\zeta) Y^{\ell',j' m}_{k}\big]\, , \\
          \mathcal{D}^{\lab{tt}}_{j\,|\,j'}[B_{j'm}(\zeta)] &= \int_{\lab{S}^2}\!\ud \Omega\,\,\bar{Y}_{jm}\,\mathcal{D}\indices{_0^0}\big[B_{j'm}(\zeta) Y_{j'm}\big]\,.
        \end{aligned}\label{eq:sphericalMatrices}
      \end{equation}
      Because we defined the $f\indices{^a_\mu}$ to have definite angular momentum, the harmonics (\ref{eq:vectorSphericalHarmonics}) and \ref{eq:formSphericalHarmonics}) take simple forms in this tetrad basis. For instance, the components of the one-form harmonics are
      \begin{equation}
        Y^{\ell, j m}_{i} = \langle (1 \, m_i) \, ( \ell\,\, m - m_i) | j \, m \rangle \,Y_{\ell, m- m_i}(\theta, \phi)\,,
      \end{equation}
      where $m_i = 2 - i$. In practice, the overlaps (\ref{eq:sphericalMatrices}) can be computed efficiently, as each term in the rewritten Proca equation (\ref{eq:procaEqDecomp}) operates very simply on these harmonics.

\vskip 4pt
    A vector field
     is a solution of 
       the Proca equation and the Lorenz constraint if and only if (\ref{eq:procaProject}) and (\ref{eq:lorenzProject}) are satisfied for all $j$ and $\ell$. Following the scalar case in \S\ref{sec:nosep}, we may approximate the radial functions as
      \begin{equation}
        B_{\ell',j'm}(\zeta) = \sum_{k = 0}^{N} B_{\ell',j'm}(\zeta_k) \,p_k(\zeta) \qquad \text{and} \qquad B_{j' m} = \sum_{k = 0}^{N} B_{j' m}(\zeta_k) \,p_k(\zeta)\,,
      \end{equation}
      where $p_k(\zeta)$ are the cardinal polynomials associated to the points $\{\zeta_k\}$. By substituting these approximations into (\ref{eq:procaProject}) and (\ref{eq:lorenzProject}), sampling each equation at the $\{\zeta_n\}$, and truncating the angular expansion, we convert this system of equations into a finite-dimensional matrix equation. This matrix, whose structure is depicted in Fig.\,\ref{fig:vectorMatrixStructure}, can then be passed to a nonlinear eigenvalue solver to determine the bound state spectrum.

  \subsection{Chebyshev Interpolation} \label{app:cheb} 

    Our numerical techniques rely on approximating the scalar and vector field configurations using finite and discrete sets of data. This discretization is a necessary step towards approximating the Klein-Gordon and Proca equations---i.e.~partial \emph{differential} equations---as finite-dimensional matrix equations. 
    How we discretize matters immensely, and a careless choice can cause the numerics to fail outright. This section
     explains both what Chebyshev interpolation is and why we choose it, and provides various technical details needed for the numerics outlined in \S\ref{sec:Separable}. For an excellent introduction to the subject, see \cite{Trefethen:2013ata}.

    \vskip 4pt
    Any smooth function $f(\zeta)$ on the interval $\zeta \in [-1, 1]$ has a unique representation in terms of Chebyshev polynomials,
    \begin{equation}
      f(\zeta) = \sum_{k = 0}^{\infty} a_k T_k(\zeta)\,,\label{eq:chebExpApp}
    \end{equation}
    where the  Chebyshev polynomials of  degree $k$ are  $T_k(\cos \theta) = \cos k \theta$. 
    The Chebyshev coefficients are then given by
    \begin{equation}
      a_k = \frac{2}{\pi} \int_{-1}^{1} \!\ud \zeta\, \frac{T_k(\zeta)\, f(\zeta)}{\sqrt{1 - \zeta^2}}\, ,\label{eq:chebCoefficients}
    \end{equation}
    for $k \geq 1$, while the right-hand side is multiplied by a factor of $1/2$ when $k = 0$. If we define an expanded function on the unit circle, 
    $F(z) = F(1/z) = f(\zeta)$, with $\zeta =  (z+z^{-1})/2$, then the Chebyshev expansion (\ref{eq:chebExpApp}) is nothing more than the unique Laurent series for $F(z)$ and (\ref{eq:chebCoefficients}) is simply a translation of the Cauchy integral formula.

    \vskip 4pt
    With this expansion in hand, we may define a degree $N$ polynomial approximation to $f(\zeta)$ by truncating the sum,
    \begin{equation}
      \tilde{f}_{N}(\zeta) = \sum_{k = 0}^{N} a_k T_k(\zeta)\,. \label{eq:truncApprox}
    \end{equation}
    While this approximation is guaranteed to become exact as $N \to \infty$, its accuracy at finite $N$ depends on how quickly the Chebyshev coefficients decay. This, in turn, depends on the analytic structure of $f(\zeta)$ and thus $F(z)$. Since $f(\zeta)$ is analytic on the interval $\zeta \in [-1, 1]$, the expanded function $F(z)$ is necessarily analytic within an annulus about the unit circle, the largest of which we denote $\rho^{-1} \leq |z| \leq \rho$. The Laurent series for $F(z)$ diverges outside of this annulus, and so its size determines the asymptotic behavior of the Chebyshev coefficients, i.e.~$a_k < \mathcal{C} \rho^{-k}$~as~$k \to \infty$ for some constant $\mathcal{C}$. Since $|T_k(\zeta)| \leq 1$ on the interval, the truncation error scales as
    \begin{equation}
      |f(\zeta)  - \tilde{f}_{N}(\zeta)| \leq \sum_{k = N+1}^{\infty} \left|a_k T_k(\zeta)\right| \leq \frac{\mathcal{C} \rho^{-N}}{\rho - 1}\,, \quad \text{as}\quad N \to \infty\,. \label{eq:convergenceProps}
    \end{equation}
    The Chebyshev approximation (\ref{eq:truncApprox}) thus converges to $f(\zeta)$ exponentially quickly, at a rate set by the singularity closest to the unit circle $|z| = 1$. If $F(z)$ is entire, the approximation converges even faster.
   
\begin{figure}
      \begin{center}
        \includegraphics{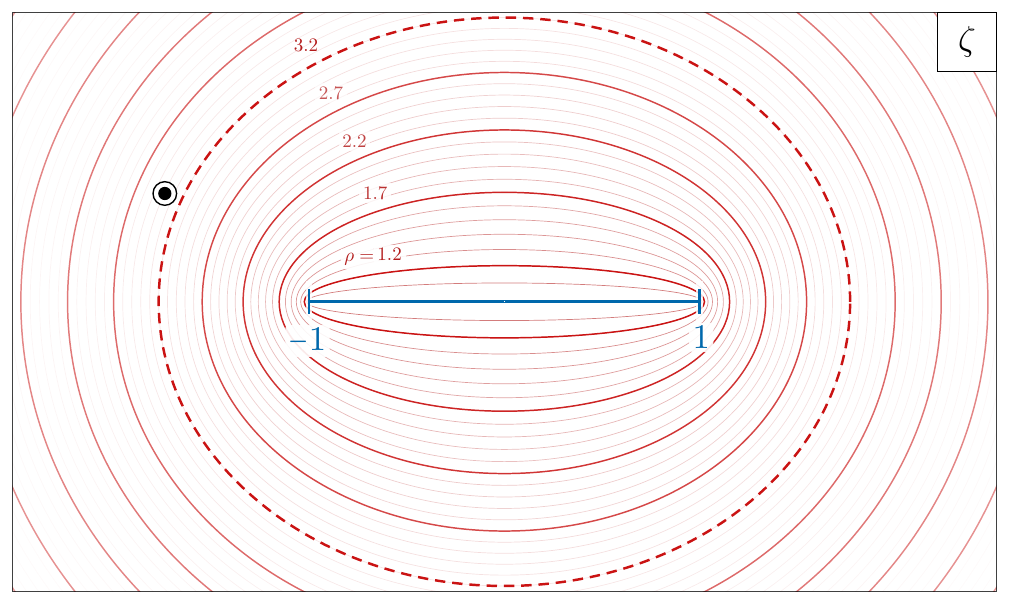}
        \caption{The interval $\zeta \in [-1, 1]$ and Bernstein ellipses of various sizes. The distance between the closest singularity of the function $f(\zeta)$ (pictured as \mySingularity) and the finite interval determines how quickly the Chebyshev expansion and interpolation converge. As illustrated here, truncation errors roughly decay as $(3.2)^{-N}$. \label{fig:bernstein}}
      \end{center} 
    \end{figure}

    \vskip 4pt
    Of course, we are mainly interested in the analytic structure of $f(\zeta)$, and not its expanded counterpart $F(z)$. However, we can relate the two by noting that $\zeta = (z+z^{-1})/2$ maps a circle of radius $\rho$ into an ellipse with foci at $\zeta = \pm 1$ defined by
    \begin{equation}
      \zeta(\theta) = \frac{1}{2}\left(\rho + \frac{1}{\rho}\right) \cos \theta + \frac{i}{2} \left(\rho - \frac{1}{\rho}\right) \sin \theta\,. \label{eq:bernsteinEllipse}
    \end{equation}
    This is called the Bernstein ellipse of radius $\rho$, depicted in Fig.~\ref{fig:bernstein}.  From the above logic, the size of the largest such ellipse inside which $f(\zeta)$ is analytic then determines how quickly the Chebyshev expansion (\ref{eq:chebExpApp}) converges, and thus the number of terms needed to approximate $f(\zeta)$ to a desired accuracy.

    \vskip 4pt
   In practice, we will use a different degree $N$ polynomial approximation, called the \emph{interpolant}
    \begin{equation}
      f_{N}(\zeta) = \sum_{k = 0}^{N} c_k T_k(\zeta)\,,
    \end{equation}
    that will inherit the convergence properties of the projection (\ref{eq:truncApprox}), yet is more convenient and numerically robust. The $N+1$ coefficients $c_k$ are defined by requiring that $f_{N}(\zeta_n) = f(\zeta_n)$ at a set of $N+1$ interpolation points $\{\zeta_n\}$. We may rewrite this interpolant in terms of the degree~$N$~\emph{Lagrange polynomials} 
    \begin{equation}
      f_{N}(\zeta) = \sum_{k = 0}^{N} f(\zeta_k) \,p_k(\zeta)\,, \label{eq:approxLagrange}
    \end{equation}
    which are defined with respect to the interpolation points as 
    \begin{equation}
      p_{k}(\zeta_n) = \delta_{nk}\,. \label{eqn:CardialPDef}
    \end{equation}
    Explicitly, the Lagrange polynomials are given by
    \begin{equation}
      p_n(\zeta) = \frac{\prod_{k \neq n} (\zeta - \zeta_k)}{\prod_{k \neq n} (\zeta_n - \zeta_k)} = \frac{p(\zeta)\, w_n}{\,(\zeta - \zeta_n)}\,, \label{eq:lagrangePolynomials}
    \end{equation}
    where we have defined both the degree $N+1$ \emph{node polynomial}
    \begin{equation}
      p(\zeta) = \prod_{k = 0}^{N} (\zeta - \zeta_n)\,
    \end{equation}
    and the \emph{weights} $w_n^{-1} = p'(\zeta_n)$.

    \vskip 4pt
    Whether or not this interpolant accurately approximates $f(\zeta)$ away from the points $\{\zeta_n\}$ depends crucially on how the interpolation points are distributed on the interval. One can show that the difference between the exact function and its approximation is given by \cite{Trefethen:2013ata,Boyd:2001cfp}
    \begin{equation}
      f(\zeta) - f_N(\zeta) = \frac{f^{(N+1)}(\zeta)\, p(\zeta)}{(N+1)!}\,.
    \end{equation}
    If the interpolation points are such that the maximum value of $|p(\zeta)|$ on the interval does not grow with its degree, then,  for large enough $N$, this difference is guaranteed to be small everywhere on the interval. Given that the Chebyshev polynomial $T_{N+1}(\zeta)$ is both extremely simple in its trigonometric form and is bounded on the interval by $\pm 1$ for all $N$, we will take $p(\zeta) = T_{N+1}(\zeta)/2^{N}$ and sample it at its zeros
    \begin{equation}
      \zeta_{k} = \cos \left(\frac{\pi(2 k + 1)}{2(N+1)}\right),\qquad k = 0, \dots, N\,, \label{eq:chebNodes}
    \end{equation}  
    with corresponding weights
    \begin{equation}
      w_k = \sin\left(\frac{2 \pi k(N+2) + \pi}{2(N+1)}\right), \qquad k = 0, \dots, N\,.
    \end{equation}
    The interpolant inherits the convergence properties (\ref{eq:convergenceProps}) of the truncation if we interpolate with the \emph{Chebyshev nodes} (\ref{eq:chebNodes}),\footnote{In fact, any set of interpolation points that are distributed according to the equilibrium distribution, $\rho(\zeta) = (\pi \sqrt{1 - \zeta^2})^{-1}$, will share the convergence properties of the truncation (\ref{eq:truncApprox}). See \cite{Trefethen:2013ata} for more details.} and so (\ref{eq:approxLagrange}) converges to $f(\zeta)$ at rate set by the largest Bernstein ellipse.

    \vskip 4pt
    In practice, evaluating the interpolant (\ref{eq:approxLagrange}) away from the interpolation points using (\ref{eq:lagrangePolynomials}) is a numerical disaster \cite{Berrut:2004bli,Trefethen:2011six,Trefethen:2013ata}. Fortunately, the \emph{second barycentric form},
    \begin{equation}
      p_n(\zeta) = \frac{\lambda_n}{\zeta - \zeta_n}\,\Bigg/\sum_{k = 0}^{N} \frac{\lambda_k}{\zeta - \zeta_k}\,,
    \end{equation}
    is numerically robust~\cite{Higham:2004num,Berrut:2004bli} and computationally efficient. This formula can be used to compute the derivative matrices $p_k'(\zeta_n)$ and $p_k''(\zeta_n)$ that appear throughout the text in a stable way. Explicitly, we take
    \begin{align}
      p^{\hskip 1pt \prime}_k(\zeta_n) &= \begin{dcases}
        \frac{w_k/w_n}{\zeta_n - \zeta_k} & n \neq k \\
        -\sum_{k\neq n} p^{\hskip 1pt \prime}_k(\zeta_n) & n = k
      \end{dcases} \, , \label{eq:cardDeriv} \\[4pt]
      p^{\hskip 1pt \prime \prime}_k(\zeta_n) &= \begin{dcases}
        2 p^{\hskip 1pt \prime}_{k}(\zeta_{n})p^{\hskip 1pt \prime}_{n}(\zeta_{n}) - \frac{2 p^{\hskip 1pt \prime}_{k}(\zeta_{n})}{\zeta_{n} - \zeta_k} & n \neq k \\
        -\sum_{k \neq n} p^{\hskip 1pt \prime \prime}_k(\zeta_n) & n = k 
      \end{dcases}\,. \label{eq:cardDDeriv}
    \end{align}
    Numerically, it is extremely important \cite{Baltensperger:2000imp} to set the diagonal elements of these differentiation matrices such that the sum of all rows vanish identically. This forces the differentiation matrices to annihilate constant functions on the interval and can have an outsized effect on accuracy.

\pagebreak

\newpage
\section{Notation and Conventions}
\label{app:notation}

\begin{center}
\renewcommand*{\arraystretch}{1.08}
\begin{longtable}{c p{10cm} c}
\toprule
\multicolumn{1}{c}{\textbf{Symbol}} &
\multicolumn{1}{l}{\textbf{Meaning}} &
\multicolumn{1}{c}{\textbf{Reference}} \\
\midrule
\endfirsthead
\multicolumn{3}{c}
{} \\
\toprule
\multicolumn{1}{c}{\textbf{Symbol}} &
\multicolumn{1}{l}{\textbf{Meaning}} &
\multicolumn{1}{c}{\textbf{Reference}} \\
\midrule
\endhead
\bottomrule
\endfoot
\bottomrule
\endlastfoot
$M$	  & 	Mass of a black hole	&	(\ref{equ:Kerr}) 	\\ 
$a$	  & Spin of a black hole	&  (\ref{equ:Kerr})	\\  
$\tilde{a}$	  & Dimensionless spin,  $\tilde{a} \equiv a/M$ 	& (\ref{equ:Kerr2}) 	\\  
$r_g$ & Gravitational radius, $r_g \equiv GM/c^2$  &  (\ref{eqn:alphaDef})  \\
$\Omega_H$	  & Angular velocity at the horizon, $\Omega_H \equiv a/(2M r_+)$	&  (\ref{equ:Kerr})	\\  
$\mu$ & Mass of a bosonic field		 & (\ref{eqn:alphaDef})	\\ 
$\lambda_c$ & Compton wavelength  of a bosonic field, $\lambda_c \equiv \hbar/(\mu c)$ & (\ref{eqn:alphaDef})  \\
$\alpha$ & Gravitational fine structure constant, $\alpha \equiv r_g / \lambda_c$  & (\ref{eqn:alphaDef})  \\
$r_c$ & Typical Bohr radius, $r_c = (\mu \alpha)^{-1}$  
&  \eqref{equ:BohrRadius} \\
$r_{\pm}$ & Outer (inner) horizon, $r_\pm \equiv M \pm \sqrt{M^2 - a^2}$ &  (\ref{equ:Kerr}) \\
$\tilde{r}_{\pm}$ & Dimensionless horizon, $\tilde{r}_\pm \equiv r_\pm/M$ &  (\ref{equ:Kerr}) \\
$\hat{r}_{\pm}$ & Additional poles for the electric modes, $\hat{r}_\pm \equiv \pm i \lambda$ &  (\ref{eqn:ProcaR}) \\
$\check{r}_{\pm}$ & Approximate outer (inner) horizons of the magnetic modes &  (\ref{eqn:MagneticRadial}) \\
$k$ & Killing vector field &  \S\ref{sec:TensorKerr}  \\ 
$\pounds$ & Lie derivative &  \S\ref{sec:TensorKerr}  \\ 
\midrule
$\Phi$ & Ultralight scalar field &  (\ref{equ:KG})  \\ 
$A_\mu$ & Ultralight vector field &  (\ref{equ:Proca})  \\ 
$\pounds^2$ & Total angular momentum operator, $\pounds^2 \equiv \pounds^2_x + \pounds^2_y + \pounds^2_z$ &  (\ref{eq:quadCasimir})  \\ 
$R$ & Radial function in separable ansatz  &  (\ref{eqn:ScalarAnsatz})\hskip 1pt,\hskip 1pt(\ref{eqn: Proca ansatz}) \\ 
$S$ & Angular function in separable ansatz &  (\ref{eqn:ScalarAnsatz})\hskip 1pt,\hskip 1pt(\ref{eqn: Proca ansatz})  \\ 
$c^2$ & Spheroidicity parameter, $c^2 \equiv \alpha^2 \tilde{a}^2 ( \omega^2 - \mu^2)$ &   (\ref{eqn: spheroidal harmonic equation}) \\ 
$B^{\mu \nu}$ & Polarization tensor for electric ansatz &   (\ref{eqn: Proca ansatz}) \\ 
\midrule
$x$ & Far-zone coordinate, $x \equiv 2 \sqrt{\mu^2 - \omega^2} \left( r-r_+ \right)$ &  (\ref{eqn: x variable})  \\ 
$y$ & Intermediate-zone coordinate, $ y \equiv \mu ( r-r_+ )$ &  (\ref{eqn: y variable}) \\
$z$ & Near-zone coordinate, $z \equiv (r-r_+) / (r_+ - r_-)$ &  (\ref{eqn: z variable})  \\ 
$\xi$ & Matching coordinate &  (\ref{eqn:IntCoord})\hskip 1pt,\hskip 1pt(\ref{eqn:IntCoordVector})  \\ 
$J$ & Source terms of inhomogeneous equations &  (\ref{eqn:Jdef})  \\ 
\midrule
$\omega$ & Frequency eigenvalue& \S\ref{sec:TensorKerr} \\ 
$E$ & Energy eigenvalue, $E \equiv {\rm Re} \, \omega $ & (\ref{eqn:generalspectrum}) \\ 
$\Gamma$ & Instability rate, $\Gamma \equiv {\rm Im} \, \omega $& (\ref{eqn:generalspectrum})  \\ 
$\nu$ & Rescaled energy eigenvalue, $\nu \equiv \mu \alpha / \sqrt{\mu^2 - \omega^2}$ &  (\ref{eqn:generalspectrum})\hskip 1pt,\hskip 1pt(\ref{eqn:nudef}) \\ 
$n$ & Principal quantum number, $n \geq \ell+1$ & \S\ref{sec:TensorKerr} \\ 
$\ell$ & Orbital angular momentum quantum number & \S\ref{sec:TensorKerr} \\ 
$j$ & Total angular momentum quantum number & \S\ref{sec:TensorKerr}\\ 
$m$ & Azimuthal angular momentum quantum number  &  \S\ref{sec:TensorKerr} \\ 
$\Lambda$ & Angular eigenvalue  &  (\ref{eqn: spheroidal harmonic equation})\hskip 1pt,\hskip 1pt(\ref{equ:ProcaS}) \\ 
$\lambda$ & Electric angular eigenvalue  &  (\ref{eqn:Bdef}) \\ 
$\lambda_0^\pm$ & Leading order electric angular eigenvalues for $j = \ell \pm 1$  &  (\ref{eqn:lambda0sol})\hskip 1pt,\hskip 1pt(\ref{eqn:lambda0solL}) \\ 
\midrule
$\mathcal{M}_{nk}$ & Nonlinear eigenvalue matrix of the radial sector &  (\ref{eq:nlev})  \\ 
$\mathcal{A}_{nk}$ & Nonlinear eigenvalue matrix of the angular sector &  (\ref{eq:angMatSol})  \\ 
$\zeta$ & Rescaled coordinate, from $r=(r_+, \infty) $ to $\zeta = (-1,1)$ &  (\ref{equ:z1})\hskip 1pt,\hskip 1pt(\ref{equ:z2})  \\ 
$\zeta_n$ & Chebyshev nodes &  (\ref{eq:chebPoints})  \\ 
$p_k$ & Cardinal polynomials, $p_k(\zeta_n) = \delta_{nk}$ &  (\ref{eqn:CardialPDef})  \\ 
$\rho$ & Size of the Bernstein ellipse &  (\ref{eq:bernsteinEllipse})  \\ 
$e\indices{_a^\mu} $ & Tetrad vector field components &  \S\ref{app:decompProca}  \\ 
$f\indices{^a_\mu} $ & Tetrad one-form field components &  \S\ref{app:decompProca}  \\ 
$A_a$ & $A_\mu$'s components in the tetrad basis, $A_\mu = A_a f\indices{^a_\mu}$ &  \S\ref{app:decompProca} \\ 
\midrule
$P_{j m}$ & Associated Legendre polynomial &  \cite{NIST:DLMF}  \\ 
$Y_{j m}$ & Scalar spherical harmonic &  \cite{NIST:DLMF}  \\ 
$Y^i_{\ell, j m}$ & Pure-orbital vector spherical harmonic &  \cite{Thorne:1980ru}  \\ 
${}_2 F_1$ & Hypergeometric function &  \cite{NIST:DLMF}  \\ 
${}_1 F_1$ & Confluent hypergeometric function of the first kind &  \cite{NIST:DLMF}  \\ 
$U$ & Confluent hypergeometric function of the second kind &  \cite{NIST:DLMF}  \\ 
$L^{(\rho)}_k$ & Associated Laguerre polynomial &  \cite{NIST:DLMF}  \\ 
$\Gamma$ & Gamma function &  \cite{NIST:DLMF}  \\ 
$\psi$ & Digamma function & \cite{NIST:DLMF}  \\ 
$\gamma_E$ & Euler-Mascheroni constant & \cite{NIST:DLMF}  \\ 
$T_n$ & Chebyshev polynomial &  \cite{NIST:DLMF}  \\  
\end{longtable}
\end{center}

\newpage
\phantomsection
\addcontentsline{toc}{section}{References}
\bibliographystyle{utphys}
\bibliography{Spectrum-Refs}

\providecommand{\href}[2]{#2}\begingroup\raggedright\begin{thebibliography}{10}

\bibitem{Cowan:1992xc}
C.~Cowan, F.~Reines, F.~Harrison, H.~Kruse, and A.~McGuire, ``{Detection of the
  Free Neutrino: A Confirmation},''
\href{http://dx.doi.org/10.1126/science.124.3212.103}{{\em Science} {\bfseries
  124} (1956) 103--104}.

\bibitem{Fukuda:1998mi}
{Y.~Fukuda (Super-Kamiokande Collaboration)}, ``{Evidence for Oscillation of
  Atmospheric Neutrinos},''
  \href{http://dx.doi.org/10.1103/PhysRevLett.81.1562}{{\em Phys. Rev. Lett.}
  {\bfseries 81} (1998) 1562--1567},
\href{http://arxiv.org/abs/hep-ex/9807003}{{\ttfamily arXiv:hep-ex/9807003
  [hep-ex]}}.

\bibitem{Zeldovich:1971a}
Y.~{Zel'Dovich}, ``{Generation of Waves by a Rotating Body},'' {\em JETP
  Letters} {\bfseries 14} (1971) 180.

\bibitem{Zeldovich:1972spj}
Y.~{Zel'Dovich}, ``{Amplification of Cylindrical Electromagnetic Waves
  Reflected from a Rotating Body},'' {\em Sov. Phys. JETP} {\bfseries 35}
  (1972) 1085.

\bibitem{Arvanitaki:2009fg}
A.~Arvanitaki, S.~Dimopoulos, S.~Dubovsky, N.~Kaloper, and J.~March-Russell,
  ``{String Axiverse},''
  \href{http://dx.doi.org/10.1103/PhysRevD.81.123530}{{\em Phys. Rev. D}
  {\bfseries 81} (2010) 123530},
\href{http://arxiv.org/abs/0905.4720}{{\ttfamily arXiv:0905.4720 [hep-th]}}.

\bibitem{Arvanitaki:2010sy}
A.~Arvanitaki and S.~Dubovsky, ``{Exploring the String Axiverse with Precision
  Black Hole Physics},''
  \href{http://dx.doi.org/10.1103/PhysRevD.83.044026}{{\em Phys. Rev. D}
  {\bfseries 83} (2011) 044026},
\href{http://arxiv.org/abs/1004.3558}{{\ttfamily arXiv:1004.3558 [hep-th]}}.

\bibitem{Detweiler:1980uk}
S.~Detweiler, ``{Klein-Gordon Equation and Rotating Black Holes},''
\href{http://dx.doi.org/10.1103/PhysRevD.22.2323}{{\em Phys. Rev.} {\bfseries
  D22} (1980) 2323--2326}.

\bibitem{Dolan:2007mj}
S.~Dolan, ``{Instability of the Massive Klein-Gordon Field on the Kerr
  Spacetime},'' \href{http://dx.doi.org/10.1103/PhysRevD.76.084001}{{\em Phys.
  Rev.} {\bfseries D76} (2007) 084001},
\href{http://arxiv.org/abs/0705.2880}{{\ttfamily arXiv:0705.2880 [gr-qc]}}.

\bibitem{Yoshino:2012kn}
H.~Yoshino and H.~Kodama, ``{Bosenova Collapse of Axion Cloud around a Rotating
  Black Hole},'' \href{http://dx.doi.org/10.1143/PTP.128.153}{{\em Prog. Theor.
  Phys.} {\bfseries 128} (2012) 153--190},
\href{http://arxiv.org/abs/1203.5070}{{\ttfamily arXiv:1203.5070 [gr-qc]}}.

\bibitem{Dolan:2012yt}
S.~Dolan, ``{Superradiant Instabilities of Rotating Black Holes in the Time
  Domain},'' \href{http://dx.doi.org/10.1103/PhysRevD.87.124026}{{\em Phys.
  Rev.} {\bfseries D87} (2013) 124026},
\href{http://arxiv.org/abs/1212.1477}{{\ttfamily arXiv:1212.1477 [gr-qc]}}.

\bibitem{Okawa:2014nda}
H.~Okawa, H.~Witek, and V.~Cardoso, ``{Black Holes and Fundamental Fields in
  Numerical Relativity: Initial Data Construction and Evolution of Bound
  States},'' \href{http://dx.doi.org/10.1103/PhysRevD.89.104032}{{\em Phys.
  Rev.} {\bfseries D89} (2014) 104032},
\href{http://arxiv.org/abs/1401.1548}{{\ttfamily arXiv:1401.1548 [gr-qc]}}.

\bibitem{Yoshino:2014}
H.~Yoshino and H.~Kodama, ``{Gravitational Radiation from an Axion Cloud around
  a Black Hole: Superradiant Phase},''
  \href{http://dx.doi.org/10.1093/ptep/ptu029}{{\em PTEP} {\bfseries 2014}
  (2014) 043E02},
\href{http://arxiv.org/abs/1312.2326}{{\ttfamily arXiv:1312.2326 [gr-qc]}}.

\bibitem{Baumann:2018vus}
D.~Baumann, H.~S. Chia, and R.~A. Porto, ``{Probing Ultralight Bosons with
  Binary Black Holes},''
  \href{http://dx.doi.org/10.1103/PhysRevD.99.044001}{{\em Phys. Rev.}
  {\bfseries D99} (2019) 044001},
\href{http://arxiv.org/abs/1804.03208}{{\ttfamily arXiv:1804.03208 [gr-qc]}}.

\bibitem{Brito:2014wla}
R.~Brito, V.~Cardoso, and P.~Pani, ``{Black Holes as Particle Detectors:
  Evolution of Superradiant Instabilities},''
  \href{http://dx.doi.org/10.1088/0264-9381/32/13/134001}{{\em Class. Quant.
  Grav.} {\bfseries 32} (2015) 134001},
\href{http://arxiv.org/abs/1411.0686}{{\ttfamily arXiv:1411.0686 [gr-qc]}}.

\bibitem{Arvanitaki:2014wva}
A.~Arvanitaki, M.~Baryakhtar, and X.~Huang, ``{Discovering the QCD Axion with
  Black Holes and Gravitational Waves},''
  \href{http://dx.doi.org/10.1103/PhysRevD.91.084011}{{\em Phys. Rev. D}
  {\bfseries 91} (2015) 084011},
\href{http://arxiv.org/abs/1411.2263}{{\ttfamily arXiv:1411.2263 [hep-ph]}}.

\bibitem{Arvanitaki:2016qwi}
A.~Arvanitaki, M.~Baryakhtar, S.~Dimopoulos, S.~Dubovsky, and R.~Lasenby,
  ``{Black Hole Mergers and the QCD Axion at Advanced LIGO},''
  \href{http://dx.doi.org/10.1103/PhysRevD.95.043001}{{\em Phys. Rev. D}
  {\bfseries 95} (2017) 043001},
\href{http://arxiv.org/abs/1604.03958}{{\ttfamily arXiv:1604.03958 [hep-ph]}}.

\bibitem{Brito:2017wnc}
R.~Brito {\em et~al.}, ``{Stochastic and Resolvable Gravitational Waves from
  Ultralight Bosons},''
  \href{http://dx.doi.org/10.1103/PhysRevLett.119.131101}{{\em Phys. Rev.
  Lett.} {\bfseries 119} (2017) 131101},
\href{http://arxiv.org/abs/1706.05097}{{\ttfamily arXiv:1706.05097 [gr-qc]}}.

\bibitem{Brito:2017zvb}
R.~Brito {\em et~al.}, ``{Gravitational Wave Searches for Ultralight Bosons
  with LIGO and LISA},''
  \href{http://dx.doi.org/10.1103/PhysRevD.96.064050}{{\em Phys. Rev. D}
  {\bfseries 96} (2017) 064050},
\href{http://arxiv.org/abs/1706.06311}{{\ttfamily arXiv:1706.06311 [gr-qc]}}.

\bibitem{Witek:2012tr}
H.~Witek, V.~Cardoso, A.~Ishibashi, and U.~Sperhake, ``{Superradiant
  Instabilities in Astrophysical Systems},''
  \href{http://dx.doi.org/10.1103/PhysRevD.87.043513}{{\em Phys. Rev.}
  {\bfseries D87} (2013) 043513},
\href{http://arxiv.org/abs/1212.0551}{{\ttfamily arXiv:1212.0551 [gr-qc]}}.

\bibitem{Pani:2012vp}
P.~Pani, V.~Cardoso, L.~Gualtieri, E.~Berti, and A.~Ishibashi, ``{Black Hole
  Bombs and Photon Mass Bounds},''
  \href{http://dx.doi.org/10.1103/PhysRevLett.109.131102}{{\em Phys. Rev.
  Lett.} {\bfseries 109} (2012) 131102},
\href{http://arxiv.org/abs/1209.0465}{{\ttfamily arXiv:1209.0465 [gr-qc]}}.

\bibitem{Pani:2012bp}
P.~Pani, V.~Cardoso, L.~Gualtieri, E.~Berti, and A.~Ishibashi, ``{Perturbations
  of Slowly Rotating Black Holes: Massive Vector Fields in the Kerr Metric},''
  \href{http://dx.doi.org/10.1103/PhysRevD.86.104017}{{\em Phys. Rev.}
  {\bfseries D86} (2012) 104017},
\href{http://arxiv.org/abs/1209.0773}{{\ttfamily arXiv:1209.0773 [gr-qc]}}.

\bibitem{Endlich:2016jgc}
S.~Endlich and R.~Penco, ``{A Modern Approach to Superradiance},''
  \href{http://dx.doi.org/10.1007/JHEP05(2017)052}{{\em JHEP} {\bfseries 05}
  (2017) 052},
\href{http://arxiv.org/abs/1609.06723}{{\ttfamily arXiv:1609.06723 [hep-th]}}.

\bibitem{Baryakhtar:2017ngi}
M.~Baryakhtar, R.~Lasenby, and M.~Teo, ``{Black Hole Superradiance Signatures
  of Ultralight Vectors},''
  \href{http://dx.doi.org/10.1103/PhysRevD.96.035019}{{\em Phys. Rev.}
  {\bfseries D96} (2017) 035019},
\href{http://arxiv.org/abs/1704.05081}{{\ttfamily arXiv:1704.05081 [hep-ph]}}.

\bibitem{East:2017ovw}
W.~East and F.~Pretorius, ``{Superradiant Instability and Backreaction of
  Massive Vector Fields around Kerr Black Holes},''
  \href{http://dx.doi.org/10.1103/PhysRevLett.119.041101}{{\em Phys. Rev.
  Lett.} {\bfseries 119} (2017) 041101},
\href{http://arxiv.org/abs/1704.04791}{{\ttfamily arXiv:1704.04791 [gr-qc]}}.

\bibitem{East:2017mrj}
W.~East, ``{Superradiant Instability of Massive Vector Fields around Spinning
  Black Holes in the Relativistic Regime},''
  \href{http://dx.doi.org/10.1103/PhysRevD.96.024004}{{\em Phys. Rev.}
  {\bfseries D96} (2017) 024004},
\href{http://arxiv.org/abs/1705.01544}{{\ttfamily arXiv:1705.01544 [gr-qc]}}.

\bibitem{East:2018glu}
W.~East, ``{Massive Boson Superradiant Instability of Black Holes: Nonlinear
  Growth, Saturation, and Gravitational Radiation},''
  \href{http://dx.doi.org/10.1103/PhysRevLett.121.131104}{{\em Phys. Rev.
  Lett.} {\bfseries 121} (2018) 131104},
\href{http://arxiv.org/abs/1807.00043}{{\ttfamily arXiv:1807.00043 [gr-qc]}}.

\bibitem{Cardoso:2018tly}
V.~Cardoso, {\'O}.~Dias, G.~Hartnett, M.~Middleton, P.~Pani, and J.~Santos,
  ``{Constraining the Mass of Dark Photons and Axion-Like Particles Through
  Black Hole Superradiance},''
  \href{http://dx.doi.org/10.1088/1475-7516/2018/03/043}{{\em JCAP} {\bfseries
  1803} (2018) 043},
\href{http://arxiv.org/abs/1801.01420}{{\ttfamily arXiv:1801.01420 [gr-qc]}}.

\bibitem{Dolan:2018dqv}
S.~Dolan, ``{Instability of the Proca Field on Kerr Spacetime},''
  \href{http://dx.doi.org/10.1103/PhysRevD.98.104006}{{\em Phys. Rev.}
  {\bfseries D98} (2018) 104006},
\href{http://arxiv.org/abs/1806.01604}{{\ttfamily arXiv:1806.01604 [gr-qc]}}.

\bibitem{Paper2}
D.~Baumann, H.~S. Chia, R.~A. Porto, and J.~Stout, {\em {to appear.}}

\bibitem{Hannuksela:2018izj}
O.~Hannuksela, K.~Wong, R.~Brito, E.~Berti, and T.~Li, ``{Probing the Existence
  of Ultralight Bosons with a Single Gravitational-Wave Measurement},''
  \href{http://dx.doi.org/10.1038/s41550-019-0712-4}{{\em Nat. Astron.}
  {\bfseries 3} (2019) 447--451},
\href{http://arxiv.org/abs/1804.09659}{{\ttfamily arXiv:1804.09659
  [astro-ph.HE]}}.

\bibitem{Zhang:2018kib}
J.~Zhang and H.~Yang, ``{Gravitational Floating Orbits around Hairy Black
  Holes},'' \href{http://dx.doi.org/10.1103/PhysRevD.99.064018}{{\em Phys.
  Rev.} {\bfseries D99} (2019) 064018},
\href{http://arxiv.org/abs/1808.02905}{{\ttfamily arXiv:1808.02905 [gr-qc]}}.

\bibitem{Berti:2019wnn}
E.~Berti, R.~Brito, C.~Macedo, G.~Raposo, and J.~Rosa, ``{Ultralight Boson
  Cloud Depletion in Binary Systems},''
  \href{http://dx.doi.org/10.1103/PhysRevD.99.104039}{{\em Phys. Rev.}
  {\bfseries D99} (2019) 104039},
\href{http://arxiv.org/abs/1904.03131}{{\ttfamily arXiv:1904.03131 [gr-qc]}}.

\bibitem{Zhang:2019eid}
J.~Zhang and H.~Yang, ``{Dynamic Signatures of Black Hole Binaries with
  Superradiant Clouds},''
\href{http://arxiv.org/abs/1907.13582}{{\ttfamily arXiv:1907.13582 [gr-qc]}}.

\bibitem{Krtous:2018bvk}
P.~Krtous, V.~Frolov, and D.~Kubiznak, ``{Separation of Maxwell Equations in
  Kerr--NUT--(A)dS Spacetimes},''
  \href{http://dx.doi.org/10.1016/j.nuclphysb.2018.06.019}{{\em Nucl. Phys.}
  {\bfseries B934} (2018) 7--38},
\href{http://arxiv.org/abs/1803.02485}{{\ttfamily arXiv:1803.02485 [hep-th]}}.

\bibitem{Frolov:2018ezx}
V.~Frolov, P.~Krtous, D.~Kubiznak, and J.~Santos, ``{Massive Vector Fields in
  Rotating Black-Hole Spacetimes: Separability and Quasinormal Modes},''
  \href{http://dx.doi.org/10.1103/PhysRevLett.120.231103}{{\em Phys. Rev.
  Lett.} {\bfseries 120} (2018) 231103},
\href{http://arxiv.org/abs/1804.00030}{{\ttfamily arXiv:1804.00030 [hep-th]}}.

\bibitem{Rosa:2011my}
J.~Rosa and S.~Dolan, ``{Massive Vector Fields on the Schwarzschild Spacetime:
  Quasi-Normal Modes and Bound States},''
  \href{http://dx.doi.org/10.1103/PhysRevD.85.044043}{{\em Phys. Rev.}
  {\bfseries D85} (2012) 044043},
\href{http://arxiv.org/abs/1110.4494}{{\ttfamily arXiv:1110.4494 [hep-th]}}.

\bibitem{Brill:1972xj}
D.~Brill, P.~Chrzanowski, M.~Pereira, E.~Fackerell, and J.~Ipser, ``{Solution
  of the Scalar Wave Equation in a Kerr Background by Separation of
  Variables},''
\href{http://dx.doi.org/10.1103/PhysRevD.5.1913}{{\em Phys. Rev.} {\bfseries
  D5} (1972) 1913--1915}.

\bibitem{Carter:1968}
B.~Carter, ``{Hamilton-Jacobi and Schr\"odinger Separable Solutions of
  Einstein's Equations},''
{\em Commun. Math. Phys.} {\bfseries 10} (1968) 280.

\bibitem{Andre:1995heun}
F.~Andr{\'e}~Ronveaux {\em et~al.}, {\em Heun's Differential Equations}.
\newblock Oxford University Press, 1995.

\bibitem{Frolov:2017kze}
V.~Frolov, P.~Krtous, and D.~Kubiznak, ``{Black Holes, Hidden Symmetries, and
  Complete Integrability},''
  \href{http://dx.doi.org/10.1007/s41114-017-0009-9}{{\em Living Rev. Rel.}
  {\bfseries 20} (2017) 6},
\href{http://arxiv.org/abs/1705.05482}{{\ttfamily arXiv:1705.05482 [gr-qc]}}.

\bibitem{Thorne:1980ru}
K.~Thorne, ``{Multipole Expansions of Gravitational Radiation},''
\href{http://dx.doi.org/10.1103/RevModPhys.52.299}{{\em Rev. Mod. Phys.}
  {\bfseries 52} (1980) 299--339}.

\bibitem{bender1999advanced}
C.~Bender and S.~Orszag, {\em Advanced Mathematical Methods for Scientists and
  Engineers I: Asymptotic Methods and Perturbation Theory}.
\newblock Springer, 1999.

\bibitem{holmes1998introduction}
M.~Holmes, {\em Introduction to Perturbation Methods}.
\newblock Texts in Applied Mathematics. Springer New York, 1998.

\bibitem{Starobinsky:1973aij}
A.~Starobinsky, ``{Amplification of Waves Reflected from a Rotating ``Black
  Hole"},'' {\em Sov. Phys. JETP} {\bfseries 37} (1973) 28.
[Zh. Eksp. Teor. Fiz. {\bf 64} (1973) 48].

\bibitem{Teukolsky:1973ha}
S.~Teukolsky, ``{Perturbations of a Rotating Black Hole. 1. Fundamental
  Equations for Gravitational, Electromagnetic and Neutrino Field
  Perturbations},''
\href{http://dx.doi.org/10.1086/152444}{{\em Astrophys. J.} {\bfseries 185}
  (1973) 635--647}.

\bibitem{Dias:2015nua}
{\'O}.~Dias, J.~Santos, and B.~Way, ``{Numerical Methods for Finding Stationary
  Gravitational Solutions},''
  \href{http://dx.doi.org/10.1088/0264-9381/33/13/133001}{{\em Class. Quant.
  Grav.} {\bfseries 33} (2016) 133001},
\href{http://arxiv.org/abs/1510.02804}{{\ttfamily arXiv:1510.02804 [hep-th]}}.

\bibitem{Gautschi:1967cat}
W.~Gautschi, ``{Computational Aspects of Three-Term Recurrence Relations},''
  {\em SIAM review} {\bfseries 9} (1967) 24--82.

\bibitem{Leaver:1985ax}
E.~Leaver, ``{An Analytic Representation for the Quasi-Normal Modes of Kerr
  Black Holes},''
\href{http://dx.doi.org/10.1098/rspa.1985.0119}{{\em Proc. Roy. Soc. Lond.}
  {\bfseries A402} (1985) 285--298}.

\bibitem{Dolan:2015eua}
S.~Dolan and D.~Dempsey, ``{Bound States of the Dirac Equation on Kerr
  Spacetime},'' \href{http://dx.doi.org/10.1088/0264-9381/32/18/184001}{{\em
  Class. Quant. Grav.} {\bfseries 32} (2015) 184001},
\href{http://arxiv.org/abs/1504.03190}{{\ttfamily arXiv:1504.03190 [gr-qc]}}.

\bibitem{Press:2007nr}
W.~Press, S.~Teukolsky, W.~Vetterling, and B.~Flannery, {\em Numerical Recipes:
  The Art of Scientific Computing}.
\newblock Cambridge University Press, 2007.

\bibitem{Cardoso:2005vk}
V.~Cardoso and S.~Yoshida, ``{Superradiant Instabilities of Rotating Black
  Branes and Strings},''
  \href{http://dx.doi.org/10.1088/1126-6708/2005/07/009}{{\em JHEP} {\bfseries
  07} (2005) 009},
\href{http://arxiv.org/abs/hep-th/0502206}{{\ttfamily arXiv:hep-th/0502206
  [hep-th]}}.

\bibitem{Guttel:2017cup}
S.~G{\"u}ttel and F.~Tisseur, ``{The Nonlinear Eigenvalue Problem},''
  \href{http://dx.doi.org/10.1017/S0962492917000034}{{\em Acta Numerica}
  {\bfseries 26} (2017) 1--94}.

\bibitem{NIST:DLMF}
{\em {\it NIST Digital Library of Mathematical Functions}}.
\newblock \url{http://dlmf.nist.gov/}.

\bibitem{Abramowitz:1965}
M.~Abramowitz and I.~A. Stegun, {\em Handbook of Mathematical Functions: with
  Formulas, Graphs, and Mathematical Tables}, vol.~55.
\newblock Courier Corporation, 1965.

\bibitem{Trefethen:2013ata}
L.~Trefethen, {\em Approximation Theory and Approximation Practice}, vol.~128.
\newblock Siam, 2013.

\bibitem{Boyd:2001cfp}
J.~Boyd, {\em {Chebyshev} and {Fourier} Spectral Methods}.
\newblock Dover Books on Mathematics. Dover Publications, Mineola, NY, 2nd~ed.,
  2001.

\bibitem{Berrut:2004bli}
J.-P. Berrut and L.~Trefethen, ``{Barycentric Lagrange Interpolation},'' {\em
  SIAM review} {\bfseries 46} (2004) 501--517.

\bibitem{Trefethen:2011six}
L.~Trefethen, ``{Six Myths of Polynomial Interpolation and Quadrature},'' 2011.

\bibitem{Higham:2004num}
N.~Higham, ``{The Numerical Stability of Barycentric Lagrange Interpolation},''
  {\em IMA Journal of Numerical Analysis} {\bfseries 24} (2004) 547--556.

\bibitem{Baltensperger:2000imp}
R.~Baltensperger, ``{Improving the Accuracy of the Matrix Differentiation
  Method for Arbitrary Collocation Points},'' {\em Applied Numerical
  Mathematics} {\bfseries 33} (2000) 143--149.

\end{thebibliography}\endgroup

\end{document}